\documentclass[10pt]{article}
\usepackage{dcolumn}
\usepackage{bm}
\usepackage{hyperref}
\usepackage{verbatim}       

\usepackage{color}
\usepackage{graphicx}
\usepackage{array}

\usepackage{amsthm}
\usepackage{amssymb}

\usepackage{amstext}
\usepackage{psfrag}
\usepackage{amsmath}
\usepackage{amsfonts}
\usepackage{units}
\usepackage{mathrsfs}

\newcommand{\p}{\partial}
\newcommand{\eu}{{\rm e}}
\newcommand{\dd}{{\rm d}}

\newtheorem{theorem}{Theorem}[section]
\newtheorem{corollary}[theorem]{Corollary}
\newtheorem{lemma}[theorem]{Lemma}
\newtheorem{proposition}[theorem]{Proposition}
\newtheorem*{theorem*}{Theorem}
\theoremstyle{definition}
\newtheorem{definition}[theorem]{Definition}
\theoremstyle{remark}
\newtheorem{remark}[theorem]{Remark}
\newtheorem{example}[theorem]{Example}

\newtheorem*{claim*}{Claim}  

\begin{document}

\title{\Large Volume and area theorems for cosmological spacetimes\\ via the Cauchy Busemann function}

\author{E. Minguzzi\footnote{Dipartimento di Matematica, Universit\`a degli Studi di Pisa,  Via
B. Pontecorvo 5,  I-56127 Pisa, Italy. E-mail:
ettore.minguzzi@unipi.it, ORCID:0000-0002-8293-3802}}

\date{}
\maketitle

\begin{abstract}
\noindent We study future timelike complete spacetimes admitting a compact Cauchy hypersurface, including spacetimes with an initial singularity. In a companion paper we showed that the graph of the Cauchy Busemann function $b$ of a compact Cauchy hypersurface is a Cauchy horizon of $M\times\mathbb{R}$, whose generators are the lifts of the $b$-calibrated rays, the observers who age maximally into the far future. Projecting the area theorem, we obtain a theorem for the volume measured by these observers. We show that $\nabla b$ is a special function of bounded variation, so that the volume theorem is an exact balance law: the divergence of the field of calibrated velocities is a measure, whose absolutely continuous part is the expansion $-\Box b$ and whose singular part is carried by the creases of $b$. In spacetimes with regular cosmological time $\tau$ a canonical Cauchy Busemann function selects a congruence of observers. If the future cosmological horizon is empty, the calibrated volume of the level sets of $\tau$ is finite, and under the timelike convergence condition it is non-decreasing; its increase is given exactly by the expansion and the creases, and its constancy on an interval forces local staticity. If the horizon is non-empty, the calibrated volume of these slices is infinite, and the same balance law and rigidity hold for the finite volumes of the events whose least asymptotic delay with respect to $\tau$ is bounded, with an additional flux of observers through the bound. These volumes have a limit independent of the bound and diverge at least linearly in the bound, with a lower asymptotic slope which is non-decreasing in time and, on smooth slices, at least the area of the horizon section, which obeys an area theorem. We also treat weighted volumes under Bakry--\'Emery conditions, and discuss the balance law as a geometric second law.
\end{abstract}

\newpage
\tableofcontents

\section{Introduction}

\subsection{Cauchy Busemann functions and the volume theorem}

This work studies the geometry of spacetimes $(M,g)$ which admit a compact Cauchy hypersurface and which are future timelike complete, namely such that every future inextendible timelike geodesic is future complete. As in the companion paper \cite{minguzzi26e}, in the title by a {\em cosmological spacetime} we simply mean a globally hyperbolic spacetime admitting a compact Cauchy hypersurface. This meaning is broader than the original one, due to Bartnik \cite{bartnik88}, according to which a cosmological spacetime is a globally hyperbolic spacetime with compact Cauchy hypersurfaces which satisfies the timelike convergence condition (TCC), $Ric(v,v)\ge 0$ for every timelike vector $v$. Future timelike completeness is compatible with an initial singularity and disregards only the future timelike incompleteness expected inside black holes, an approximation which is acceptable for a study at the cosmological scale; we refer to \cite{minguzzi26e} for a discussion.

Given a compact Cauchy hypersurface $\Sigma$, the future Cauchy Busemann function introduced in \cite{minguzzi26e} is the normalized limit $b=\lim_k[k-d(\cdot,\Sigma_k)]$ of the Lorentzian distances to the level sets $\Sigma_k$ of the distance from $\Sigma$. Its finiteness domain $\textrm{dom}\, b$, the future cosmological development, and the boundary $\p\,\textrm{dom}\, b$, the future cosmological horizon, do not depend on $\Sigma$. Through every point of $\textrm{dom}\, b$ passes a $b$-calibrated timelike ray, along which $b$ increases by the proper time, and these rays are the observers who age maximally into the far future. The product trick of \cite{minguzzi26e} shows that in $\tilde M=M\times\mathbb{R}$, $\tilde g=g+\dd z^2$, the graph of $b$ over $\textrm{dom}\, b$ is a Cauchy horizon $H$, whose generators are the lifts of the $b$-calibrated rays.

The area theorem for horizons \cite{chrusciel01,minguzzi14d} compares the areas of two sections of a horizon in terms of the integral of the expansion of the generators between them. Since the generators of $H$ are the lifts of the calibrated rays, the area theorem can be projected to $\textrm{dom}\, b$. It becomes a {\em volume theorem} (Theorem \ref{thmvol}): it compares the $n$-dimensional volumes, as measured by the calibrated observers, of the two bases of an $(n+1)$-dimensional tube whose sides are generated by segments of calibrated rays, in terms of the integral over the tube of the expansion $\theta^b=-\Box b$ of the semi-flow of the calibrated rays, and of the calibrated volume of its creases, namely of the points from which several calibrated rays start. We obtain it by translating the results of \cite[Sect.\ 2.6--2.7]{minguzzi14d} through a dictionary (Section \ref{secdict}). The volume theorem is in fact an exact balance law. In Gaussian coordinates the eikonal equation solved by $b$ is a Hamilton--Jacobi equation with a Hamiltonian which is uniformly convex on the bounded sets of momenta, and from a theorem of Bianchini and Tonon \cite{bianchini12} we deduce that $\nabla b$ is locally a special function of bounded variation (Theorem \ref{thmsbv}). Thus the Hessian of $b$ has no Cantor part, and the calibrated volume changes only through the expansion of the calibrated congruence and through its creases (Corollary \ref{corbal}).

\subsection{Canonical volumes}

The construction is of particular interest in spacetimes with an initial singularity, by which we mean spacetimes whose cosmological time $\tau$ \cite{andersson98} is regular; this is a condition of finite age, which by itself does not imply the blow up of the curvature or the past inextendibility of the spacetime. In these spacetimes no choice is involved. The Cauchy Busemann functions of the level sets $T_a=\tau^{-1}(a)$ differ by constants and define a canonical function $b^{\textnormal{\textsc{m}}}\ge\tau$ \cite{minguzzi26e}, hence a canonical congruence of observers, with almost everywhere defined unit tangent field $v=-\nabla b^{\textnormal{\textsc{m}}}$. The difference $w=b^{\textnormal{\textsc{m}}}-\tau\ge0$ is the least delay, with respect to the cosmological time, which an observer leaving the event will accumulate, and it is non-increasing along the calibrated rays. Our results in this setting can be summarized as follows.

\begin{theorem*}[Canonical volumes]
Let $(M,g)$ be future timelike complete with regular cosmological time $\tau$. For $a>0$ and $s\ge0$ let $N(a,s)$ be the calibrated volume, namely the volume measured by the calibrated observers, of the compact set $T_a\cap\{w\le s\}$ of the events of cosmological time $a$ with delay at most $s$. It is finite and it depends only on $(M,g)$. Moreover:
\begin{itemize}
\item[(i)] For $a_1<a_2$ the difference $N(a_2,s)-N(a_1,s)$ is the sum of three terms: the integral of the expansion $\theta^b=-\Box b^{\textnormal{\textsc{m}}}$ of the calibrated congruence over $\{a_1<\tau<a_2\}\cap\{w\le s\}$, twice the calibrated volume of the creases contained in this set, and the flux $\Phi(s)\ge0$ of the calibrated observers whose delay falls below $s$ between the two times.
\item[(ii)] Under the timelike convergence condition $\theta^b\ge0$, so $N(a,s)$ is non-decreasing in $a$. Its limit $V_\infty\in(0,+\infty]$ for $a\to+\infty$ does not depend on $s>0$, and it is also the limit of the volumes $N(a,\kappa a)$, $\kappa>0$, of the events whose delay is at most the fraction $\kappa$ of their cosmological time.
\item[(iii)] Under the timelike convergence condition, if $N(a_1,s)=N(a_2,s)$ for some $s>0$, then the set $\{a_1<\tau<a_2\}\cap\{w<s\}$ is locally isometric to a static product on which the calibrated observers are the static ones.
\item[(iv)] Under the timelike convergence condition, if the future cosmological horizon $\p\,\textrm{dom}\, b^{\textnormal{\textsc{m}}}$ is non-empty then $V_\infty=+\infty$, and once the slices $T_a$ meet the horizon $N(a,s)$ grows at least linearly in $s$, at a rate which is non-decreasing in $a$.
\end{itemize}
\end{theorem*}

These statements are proved in Theorems \ref{thmasym}, \ref{thmslope} and \ref{thmcanbal} and in Corollary \ref{correl}. Their scope should be stated precisely. The calibrated volume is the volume measured in the rest spaces of the calibrated observers. On the slices of the cosmological time it exceeds the volume induced by the metric by the Lorentz factor $\gamma\ge1$ of the calibrated observers with respect to the observers comoving with the cosmological time, and the same factor governs the decrease of the delay along the calibrated rays, $\dd w(v)=1-\gamma$; in particular, the divergence of the calibrated volume at the horizon is a divergence of $\gamma$, the induced volume of the slices being finite. The quantities $N(a,s)$ are determined by $(M,g)$, but not by the geometry up to the cosmological time $a$: the function $b^{\textnormal{\textsc{m}}}$ is defined through a limit at future infinity, so they depend on the whole future of the slice. By (i), their growth is not, in general, the expansion of a region comoving with the calibrated observers, as it includes the entry term $\Phi$, which can be the only one present (Example \ref{exflat}). Finally, what the geometry selects is a family of finite monotone quantities, equivalently the measures $\nu_a$ on $[0,+\infty)$ of which $N(a,\cdot)$ are the cumulative distribution functions, the {\em deficit spectra}, rather than a single number. The following table separates what is obtained from what is not.

\begin{center}
\small
\renewcommand{\arraystretch}{1.15}
\begin{tabular}{p{0.46\textwidth}p{0.46\textwidth}}
\hline
{\em Obtained} & {\em Not obtained}\\
\hline
A canonical congruence of observers and the finite quantities $N(a,s)$, non-decreasing in $a$ under the TCC, with an exact balance law and a rigidity statement. & A preferred finite total volume when the horizon is non-empty: then $V_\infty=+\infty$, and no weight which is a function of $w$ has a finite limit (Remark \ref{remtrunc}).\\
A finite $V_\infty$ excludes the horizon. & The converse: $V_\infty=+\infty$ also in spacetimes without horizon (Example \ref{exwarp2}).\\
A weight with finite weighted volumes which is determined by $(M,g)$ (Remark \ref{remspec}); finite volumes at bounded relative delay (Corollary \ref{correl}). & The finiteness for a universal positive weight, such as $\eu^{-w}$, which would follow from an upper bound on the growth of $N(a,s)$ in $s$.\\
The rate of growth of the truncated volumes at the horizon is non-decreasing in time and, for smooth slices, at least the area of the section of the horizon (Theorem \ref{thmslope}). & The equality of this rate with the area of the section, and the monotonicity of the finite part (Remark \ref{remfinite}).\\
\hline
\end{tabular}
\end{center}

The case in which the future cosmological horizon is empty, $\textrm{dom}\, b^{\textnormal{\textsc{m}}}=M$, is the cleanest. Then the calibrated volume $\mathcal{V}(a)$ of the slice $T_a$ is finite, its variation between two cosmological times is exactly the integral of the expansion of the calibrated congruence plus twice the calibrated volume of its creases, so that under the TCC it is non-decreasing, and its constancy on an interval forces the intervening region to be locally a static product, with the calibrated observers as static observers (Corollary \ref{corhf}). In local form, the field $v$ of calibrated velocities has a distributional divergence which is a measure, non-negative under the TCC, whose absolutely continuous part is the expansion and whose singular part is carried by the creases (Corollary \ref{cordiv}). This is the structure of a second law: a current $\eta v$, $\eta>0$, with non-negative production, together with a rigidity statement for the absence of production; moreover, the calibrated volume of a slice is the area of a section of the horizon $H$ of the product trick. In Section \ref{secphys} we propose that $\eta\,\mathcal{V}$ is a geometric entropy of the cosmological spacetime, and we compare it with other proposals involving the volume of Cauchy hypersurfaces. This entropy is distinct from the entropy of matter and from measures of gravitational clumping, as $\mathcal{V}$ grows also when the entropy of matter is constant, as for adiabatic radiation, and in flat spacetimes. Since $H$ lies in a spacetime of one more dimension, the constant $\eta$ involves the length of the auxiliary direction; its determination, together with a temperature and a first law, is the subject of ongoing work. When the horizon is non-empty, the finite quantities $N(a,s)$ satisfy the same balance law, with an additional flux of observers through the bound $s$ on the delay, which is not production.

\subsection{Summary of the results}

In more detail, our results are the following.
\begin{itemize}
\item[(a)] {\em Volume monotonicity.} Under the TCC the semi-flow of the calibrated rays is non-contracting \cite{minguzzi26e}, and the calibrated volume is non-decreasing, also when weighted by functions which are non-decreasing along the calibrated rays (Theorem \ref{corvol}); by the balance law its increase is given exactly by the expansion and by the creases. Weights which decrease along the rays are allowed, without the TCC, if a weighted Ricci curvature of Bakry--\'Emery type is non-negative in the calibrated directions, with any synthetic dimension $N\in(-\infty,0]\cup[n,+\infty]$ and a completeness condition taken from the $\epsilon$-range of Lu, Minguzzi and Ohta \cite{lu21,lu22} (Proposition \ref{propBE} and Corollary \ref{corBE}). The weights need not be $C^2$: it suffices that they are locally Lipschitz and semi-concave along the calibrated rays, the singular part of their second derivative having the favourable sign, and for locally semi-concave weights the condition can be imposed pointwise, as $Ric_N^\psi(v)\ge0$ almost everywhere, with $v=-\nabla b$ (Corollary \ref{corpoint}). For $N>n$ no completeness assumption beyond future timelike completeness is needed, while for $N=0$ and smooth weights the result is the area theorem for the horizon $H$ in a conformally rescaled metric. The equality of the calibrated volumes of the bases of a tube forces $b$ to be smooth on it, with vanishing d'Alembertian, and the tube to be locally isometric to a static product (Theorem \ref{thmvolrig}); this rests on an elliptic regularity property of the calibrating semi-flow, which holds without energy conditions (Proposition \ref{propellreg}).
\item[(b)] {\em Whole slices.} The calibrated volume of the slices $f^{-1}(t)\cap \textrm{dom}\, b$ of a Cauchy time function $f$ is infinite once they leave $\textrm{dom}\, b$ (Proposition \ref{propinfvol}). Monotonicity persists for their compact subsets $\{b-u\le s\}$, where $u$ grows at least as fast as the proper time along the calibrated rays; besides the expansion and the creases, their calibrated volume has a third non-negative source, the flux of the calibrated rays which enter the set (Theorem \ref{corvol}(c)). Natural choices of $u$ are steep temporal functions, the signed distance from a $C^{1,1}$ Cauchy hypersurface and, in spacetimes with an initial singularity, the cosmological time of Andersson, Galloway and Howard \cite{andersson98}, in which case these sets depend only on the spacetime (Remark \ref{remcosmvol}). Monotonicity also persists for the weighted calibrated volumes of the whole slices, with weights $F(b-u)$ vanishing at the future cosmological horizon, which are finite for suitable $F$ (Corollary \ref{corweight}). The decay of $F$ needed for finiteness is dictated by the growth of the areas of the sections of the level sets of $b$ by the slices, rather than by the rate at which $b$ diverges at the horizon.
\item[(c)] {\em Transport form and canonical volumes.} By means of the past semi-flow of the generalized characteristics of $b$ \cite{minguzzi26f}, the monotonicity is localized to the Borel subsets of the slices: under the TCC the calibrated volume of a portion of a slice does not decrease when the portion is transported to a later slice by the calibrated observers, and no tube with ruled sides is needed (Theorem \ref{thmtransp}). In spacetimes with an initial singularity the difference $w=b^{\textnormal{\textsc{m}}}-\tau\ge 0$ between the canonical Cauchy Busemann function and the cosmological time, namely the least asymptotic delay of the observers leaving a point, is non-increasing and tends to zero along the calibrated rays. The calibrated volume $N(a,s)$ of the events of cosmological time $a$ with $w\le s$ is finite and non-decreasing in $a$, and its limit for $a\to+\infty$, the {\em asymptotic calibrated volume} $V_\infty$, does not depend on $s>0$; every weighted volume $V_F$ converges to $F(0)V_\infty$, and $V_\infty=+\infty$ whenever the future cosmological horizon is non-empty (Theorem \ref{thmasym}). The areas of the portions $\{w\le s\}$ of the horospheres of $b^{\textnormal{\textsc{m}}}$ are also non-decreasing (Corollary \ref{corhoro}). These quantities depend only on $(M,g)$, and a weight $F$ with finite $V_F$ depending only on $(M,g)$ exists (Remark \ref{remspec}). The divergence of the calibrated volume at the future cosmological horizon is at least linear in the bound $s$: the calibrated volume of a slice is the area of the graph of $b$ over it, which rises to infinity over the section of the horizon, and the lower asymptotic slope of the truncated volumes is non-decreasing in time and, for smooth slices and almost every time, bounded from below by the area of this section (Proposition \ref{propslope} and Theorem \ref{thmslope}). Where the areas of the level sets of $b$ on the slice converge in the Ces\`aro sense to that of the section, the slope equals this area, and the area theorem for the horizon of (d) is the leading term of the volume monotonicity theorem. The horospheres do not approach the horizon, and their total area is non-decreasing (Corollary \ref{corhoro2}). Finally, we consider weights $F(w)$ with $F$ increasing, which decrease along the calibrated rays. They are semi-concave along the rays, because the cosmological time is semi-convex, and their weighted volumes are non-decreasing, without the TCC, if the weighted Ricci curvature of synthetic dimension $N=0$ is non-negative in the calibrated directions, $Ric_0^{\psi}(v)\ge0$ almost everywhere, where $\psi=\ln F(w)$ and $v=-\nabla b^{\textnormal{\textsc{m}}}$; if $F^{-1/n}$ is convex this follows from $Ric(v,v)+(\ln F)'(w)\,\textrm{Hess}\,\tau(v,v)\ge0$. In this case the expansion of the calibrating semi-flow is bounded from below by $(\ln F)'(w)(\dd\tau(v)-1)\ge 0$ (Theorem \ref{thmdefw}). The balance law and the rigidity statement for the canonical volumes, the volumes at bounded relative delay, which are invariant under the rescalings of the metric up to the obvious factor, and two examples, in which the canonical volumes grow only by expansion and only by entry, are given in Section \ref{secfurther}.
\item[(d)] {\em Area theorem for the future cosmological horizon.} The horizon $\p\,\textrm{dom}\, b$, a $C^0$ null hypersurface which does not depend on $\Sigma$, satisfies an area theorem under the TCC, or under the null convergence condition and future null completeness (Theorem \ref{thmhorarea}). Its proof adapts Hawking's argument for event horizons, the role of null infinity being played by the lifts to $\tilde M$ of the calibrated rays (Lemma \ref{lemstar} and Theorem \ref{thmhorexp}). Only future timelike completeness is needed, so the result applies to spacetimes with an initial singularity, where the slices of the cosmological time give an area function which depends only on the spacetime (Corollary \ref{corcanarea}).
\end{itemize}

\subsection{Organization}

The paper is organized as follows. Section \ref{secprel} recalls the results of \cite{minguzzi26e} which we need. Section \ref{secdict} introduces the dictionary between the horizon $H$ and the future cosmological development, Section \ref{secvolthm} proves the volume theorem and the SBV regularity which turns it into a balance law, and Section \ref{secmono} studies the monotonicity of the calibrated volume, the slices of Cauchy time functions, the rigidity under the TCC and the elliptic regularity of the calibrating semi-flow (Section \ref{secellreg}). Section \ref{sectransp} proves the transport form of the monotonicity theorem, and Section \ref{secinitsing} applies it to spacetimes with an initial singularity: it introduces the canonical volumes and the asymptotic calibrated volume, it studies the weights which increase with the asymptotic delay (Section \ref{secdefw}) and the divergence of the calibrated volume at the future cosmological horizon (Section \ref{secdiv}), and it ends with the balance law, the rigidity and the examples for the canonical volumes (Section \ref{secfurther}). Section \ref{secareahor} establishes the area theorem for the future cosmological horizon, Section \ref{secphys} discusses the interpretation as a geometric second law, and Section \ref{secconcl} contains conclusions and open problems.

\section{Preliminaries} \label{secprel}

\subsection{Conventions}

Reference for background causality theory is \cite{minguzzi18b} unless otherwise stated, and we use its terminology. In particular, Cauchy hypersurfaces are acausal by definition, a {\em time function} is a continuous function which is strictly increasing over causal curves, a {\em Cauchy time function} is a time function whose restriction to every inextendible causal curve has image $\mathbb{R}$, and $d$ denotes the Lorentzian distance. For $A,B\subset M$ we set $d(A,B)=\sup_{p\in A,q\in B} d(p,q)$ and $d(p,B)=d(\{p\},B)$. Given a closed set $S$, an {\em $S$-ray} is a future inextendible causal curve starting from $S$ such that for every point $q$ on it the segment from $S$ to $q$ has Lorentzian length $d(S,q)$.

Throughout the paper $(M,g)$ is a connected, time-oriented, globally hyperbolic spacetime of dimension $n+1$ which admits a compact Cauchy hypersurface $\Sigma$. We say that $(M,g)$ is {\em future timelike complete} if every future inextendible timelike geodesic is future complete, and {\em future null complete} if every future inextendible lightlike geodesic is future complete. These assumptions, the timelike convergence condition (TCC), $Ric(v,v)\ge 0$ for every timelike $v$, and the null convergence condition (NCC), $Ric(k,k)\ge 0$ for every lightlike $k$, are stated in each result in which they are used. The TCC implies the NCC by continuity.

\subsection{The Cauchy Busemann function}

For a Cauchy hypersurface $\Sigma$ let $d^+_\Sigma(q):=d(\Sigma,q)$ and $d^-_\Sigma(q):=d(q,\Sigma)$. The {\em signed distance} from $\Sigma$, $d_\Sigma:=d^+_\Sigma-d^-_\Sigma$, is finite and continuous \cite[Thm.\ 2.1]{minguzzi26e}, and it is rushing: $d_\Sigma(q)-d_\Sigma(p)\ge d(p,q)$ for $p\le q$ \cite[Prop.\ 2.4]{minguzzi19d}, \cite[Prop.\ 2.9]{minguzzi26e}. It is locally Lipschitz on $M\backslash\Sigma$, and on $M$ if $\Sigma$ is a $C^{1,1}$ spacelike hypersurface \cite[Prop.\ 2.10]{minguzzi26e}.

Let $\Sigma$ be a compact Cauchy hypersurface and assume that $(M,g)$ is future timelike complete. Then the level sets $\Sigma_\tau:=(d^+_\Sigma)^{-1}(\tau)$, $\tau>0$, are compact Cauchy hypersurfaces \cite[Thm.\ 2.4]{minguzzi26e}, and the {\em Cauchy Busemann function} of $\Sigma$ is \cite[Def.\ 3.2]{minguzzi26e}
\[
b(p):=\lim_{\tau\to+\infty}[\tau-d(p,\Sigma_\tau)]\in(-\infty,+\infty], \qquad p\in M .
\]
It is rushing and bounded from below by the signed distance \cite[Eqs.\ (4) and (3)]{minguzzi26e}:
\begin{align}
b(p)+d(p,q)&\le b(q), \qquad p\le q, \label{fcpb}\\
b&\ge d_\Sigma \quad \textrm{on } M. \label{cmlr}
\end{align}
Its finiteness domain $\textrm{dom}\, b$, the {\em future cosmological development}, is an open past set which meets every Cauchy hypersurface \cite[Thm.\ 5.12(ii), Prop.\ 3.7]{minguzzi26e}, and it contains the future $\Sigma$-rays, along which, once parametrized with proper time, $b(\sigma(t))=t=d_\Sigma(\sigma(t))$ \cite[Prop.\ 3.5]{minguzzi26e}. Its boundary $\p\, \textrm{dom}\, b$, the {\em future cosmological horizon}, is a $C^0$ future null hypersurface \cite[Prop.\ 5.15]{minguzzi26e}, namely a locally achronal topological hypersurface through every point of which passes a future inextendible lightlike geodesic contained in it. The sets $\textrm{dom}\, b$ and $\p\,\textrm{dom}\, b$ do not depend on $\Sigma$ \cite[Thm.\ 3.9]{minguzzi26e}. In \cite{minguzzi26e} the function $b$ is also denoted $b^+$, to distinguish it from its time dual $b^-$, the past Cauchy Busemann function, which requires past timelike completeness; here we use the notation $b^\pm$ only when $b^-$ also appears.

\subsection{The product trick}

Let $\tilde M:=M\times \mathbb{R}$, $\tilde g:=g+\dd z^2$, with the time orientation induced from that of $M$, and let $\pi:\tilde M\to M$ be the projection; causal relations and sets on $\tilde M$ are denoted with a tilde. A $C^1$ curve $(x,z)$ is causal if and only if $x$ is causal and $\vert\dot z\vert\le\sqrt{-g(\dot x,\dot x)}$, and timelike if and only if $x$ is timelike and $\vert\dot z\vert<\sqrt{-g(\dot x,\dot x)}$; in particular, for every causal curve $x$ of $M$ the lift $t\mapsto (x(t),z_0+\ell(x\vert_{[0,t]}))$ is a lightlike curve of $\tilde M$ \cite[Sect.\ 4]{minguzzi26e}. The spacetime $(\tilde M,\tilde g)$ is globally hyperbolic \cite[Lemma 4.1]{minguzzi26e}. Let $\tilde\Sigma:=\{(p,z): p\in\Sigma,\ z\le 0\}$. If $(M,g)$ is future timelike complete, then \cite[Prop.\ 4.3]{minguzzi26e}
\begin{align}
\tilde D^{-}(\tilde I^+(\tilde \Sigma))&=\{(p,z): p \in M,\ z< b(p)\}, \label{mxwe} \\
H:=\tilde H^{-}(\tilde I^+(\tilde \Sigma))&= \p \operatorname{epi}  b,\label{mxwh}
\end{align}
and $\tilde D^{-}(\tilde I^+(\tilde \Sigma))$ is an open future set, so that $H$ is an achronal boundary. Moreover, $b:M\to(-\infty,+\infty]$ is continuous, and $H$ is the graph of $b$ over $\textrm{dom}\, b$ \cite[Thm.\ 5.12(iv)]{minguzzi26e}.

\subsection{Calibrated rays and regularity}

A causal curve $x$ with image in $\textrm{dom}\, b$ is {\em $b$-calibrated} if $b(x(t_2))=b(x(t_1))+\ell(x\vert_{[t_1,t_2]})$ for $t_1<t_2$ \cite[Def.\ 5.2]{minguzzi26e}. Assume that $(M,g)$ is future timelike complete. From every point $p\in\textrm{dom}\, b$ starts a future inextendible $b$-calibrated curve; these curves are the projections of the generators of $H$ starting from $(p,b(p))$, and they are future complete timelike rays, the {\em $b$-calibrated rays} \cite[Prop.\ 5.6, Thm.\ 5.12(i),(iv)]{minguzzi26e}. The function $b$ is locally Lipschitz and locally semi-concave on $\textrm{dom}\, b$; more precisely, if $\alpha$ is a proper time parametrized calibrated ray starting from $p$, then for every $s>0$ the function $b_{p,s}(q):=b(p)+s-d(q,\alpha(s))$ is a smooth upper support function for $b$ at $p$ \cite[Thm.\ 5.12(iii),(vi)]{minguzzi26e}. For every compact set $K\subset \textrm{dom}\, b$, the unit initial tangents of the calibrated rays starting from $K$ form a compact subset of $TM$ \cite[Thm.\ 5.12(v)]{minguzzi26e}. The function $b$ is differentiable at $p\in \textrm{dom}\, b$ if and only if exactly one calibrated ray starts from $p$, and in this case $-\nabla b(p)$ is a future directed unit timelike vector, the initial tangent of this ray \cite[Lemma 5.16, Prop.\ 5.19]{minguzzi26e}; in particular, $b$ is differentiable at the interior points of the calibrated rays. The set of non-differentiability points of $b$ is $\mathcal{H}^{n+1}$-null \cite[Thm.\ 5.21(c)]{minguzzi26e}.

\subsection{Mean convexity under the TCC}

Assume the TCC and that $(M,g)$ is future timelike complete. Then $\Box b\le 0$ in the support sense on $\textrm{dom}\, b$, the upper support functions being the functions $b_{p,s}$ \cite[Thm.\ 6.9(a)]{minguzzi26e}, and $\Box b\le 0$ at every point at which $b$ is twice differentiable \cite[Cor.\ 6.11(a)]{minguzzi26e}. As $-\nabla b$ is the initial unit tangent of the calibrated rays, $-\Box b$ is the expansion of their semi-flow, which is thus non-contracting.

\subsection{Spacetimes with an initial singularity} \label{secinit}

Andersson, Galloway and Howard \cite{andersson98} introduced the {\em cosmological time function}
\[
\tau(q):=\sup\{d(p,q): p\in J^-(q)\},
\]
and called it {\em regular} if $\tau(q)<+\infty$ for every $q\in M$ and $\tau\to 0$ along every past inextendible causal curve \cite[Def.\ 1.1]{andersson98}. If $\tau$ is regular, it is a time function \cite[Thm.\ 1.2(ii)]{andersson98}, it is locally Lipschitz \cite[Thm.\ 1.2(v)]{andersson98}, and it is rushing, $\tau(q)\ge\tau(p)+d(p,q)$ for $p\le q$, which follows from the definition. Let $(M,g)$ be future timelike complete with regular cosmological time. Then the level sets $T_a:=\tau^{-1}(a)$, $a>0$, are compact Cauchy hypersurfaces, $d^+_{T_a}=\tau-a$ on $J^+(T_a)$, and the limit
\[
b^{\textnormal{\textsc{m}}}(p):=\lim_{r\to +\infty}[r-d(p,T_r)]\in(-\infty,+\infty], \qquad p\in M,
\]
exists and satisfies $b^{\textnormal{\textsc{m}}}=b^{T_a}+a$ for every $a>0$, where $b^{T_a}$ is the Cauchy Busemann function of $T_a$, and $b^{\textnormal{\textsc{m}}}\ge\tau$ \cite[Prop.\ 5.23(a),(b)]{minguzzi26e}. The function $b^{\textnormal{\textsc{m}}}$ depends only on $(M,g)$, and $\textrm{dom}\, b^{S}=\textrm{dom}\, b^{\textnormal{\textsc{m}}}$ for the Cauchy Busemann function $b^S$ of every compact Cauchy hypersurface $S$ \cite[Prop.\ 5.23(c)]{minguzzi26e}. Moreover, with $\tilde M_-:=M\times(-\infty,0]$, the past Cauchy horizon $\tilde H^-(\tilde I^+(\tilde M_-))$ is the graph of $b^{\textnormal{\textsc{m}}}$ over $\textrm{dom}\, b^{\textnormal{\textsc{m}}}$, and it is the image of the horizon $H$ associated with $T_a$ under the translation $(p,z)\mapsto(p,z+a)$ \cite[Prop.\ 5.23(e)]{minguzzi26e}.

In this paper, as in \cite{minguzzi26e}, by a {\em spacetime with an initial singularity} we mean a spacetime with regular cosmological time. This is a condition of finite age: every observer has existed for a finite proper time, bounded by $\tau$, and the bound tends to zero in the past. It does not imply that the curvature blows up, nor that the spacetime is past inextendible, as shown by the flat spacetime of Example \ref{exflat}.

The results of this paper also apply to $b^{\textnormal{\textsc{m}}}$, although $b^{\textnormal{\textsc{m}}}$ is a Cauchy Busemann function only up to a constant. Indeed, all the objects entering the statements below are invariant under the addition of a constant to $b$: the domain $\textrm{dom}\, b$, the calibrated curves, the gradient $\nabla b$ and hence the field $v$ and the form $\omega$ introduced in Section \ref{secdict}, the d'Alembertian $\Box b$, and the sets $S^k$. The horizon is translated by the isometry $(p,z)\mapsto (p,z+a)$ of $\tilde M$, which commutes with the projection to $M$, so the area theorem for the horizon of $T_a$ translates into the same volume theorem. Finally, the hypersurfaces $T_a$ are in general only Lipschitz, but the results below, as those of \cite{minguzzi26e} on which they rely, only use that $\Sigma$ is a compact Cauchy hypersurface.

\section{The dictionary} \label{secdict}

Throughout the rest of the paper $b$ is the Cauchy Busemann function of a compact Cauchy hypersurface $\Sigma$; in particular, the objects introduced below are defined only if $(M,g)$ is future timelike complete, an assumption which is recalled in each result.

Let $\Psi:\textrm{dom}\, b\to H$, $\Psi(p)=(p,b(p))$, which is a locally bi-Lipschitz homeomorphism with inverse $\pi\vert_H$, since $H$ is the graph of $b$ over $\textrm{dom}\, b$ and $b$ is locally Lipschitz there \cite[Thm.\ 5.12(iv),(vi)]{minguzzi26e}. Let $S\subset \textrm{dom}\, b$ be the set of non-differentiability points of $b$, which is $\mathcal{H}^{n+1}$-null \cite[Thm.\ 5.21(c)]{minguzzi26e}, and for $p\in \textrm{dom}\, b\backslash S$ let $v(p):=-\nabla b(p)$ be the initial tangent of the unique $b$-calibrated ray starting from $p$ \cite[Lemma 5.16]{minguzzi26e}. Let $\mu_g$ be the volume form of $g$ (a measure if $M$ is not oriented) and let
\[
\omega:=i_v\mu_g ,
\]
an $n$-form defined on $\textrm{dom}\, b\backslash S$. If $B$ is a $C^1$ spacelike hypersurface with future unit normal $N$, then $\omega\vert_B=-g(v,N)\, \mu_B$, where $\mu_B$ is the induced volume form, so $\omega\vert_B$ is the volume of $B$ as measured in the rest frames of the calibrated observers, larger than the induced volume by the Lorentz factor $-g(v,N)\ge 1$. On the horospheres, at the differentiability points, $v$ is normal and $\omega$ restricts to the induced volume form.

Fix an auxiliary Riemannian metric $h$ on $M$, and let $\mathcal{H}^n$ be the $n$-dimensional Hausdorff measure of its distance; as Hausdorff measures of different Riemannian metrics are locally comparable, the null sets of $\mathcal{H}^n$ do not depend on $h$. Let $B\subset \textrm{dom}\, b$ be an acausal locally Lipschitz hypersurface, namely, near each of its points, the image of a locally bi-Lipschitz map $F:U\to M$, $U\subset \mathbb{R}^n$ open, for instance a Lipschitz graph in a chart. By Rademacher's theorem, $F$ is differentiable almost everywhere on $U$, so $B$ has a tangent hyperplane $T_xB$ at $\mathcal{H}^n$-almost every $x\in B$ \cite[Sect.\ 3.1]{evans92}. If $\mathcal{H}^n(B\cap S)=0$, at $\mathcal{H}^n$-almost every $x\in B$ both $T_xB$ and $v(x)$ are defined and we can set
\[
\rho_B(x):=\vert \omega_x(e_1,\dots,e_n)\vert ,
\]
where $e_1,\dots,e_n$ is an $h$-orthonormal basis of $T_xB$. The function $\rho_B$ is the density of the measure induced by $\omega$ on $B$ with respect to $\mathcal{H}^n$.

\begin{definition} \label{defcalvol}
Let $B\subset \textrm{dom}\, b$ be an acausal locally Lipschitz hypersurface such that $\mathcal{H}^n(B\cap S)=0$, and let $\varphi\ge 0$ be a Borel function on $B$. The {\em calibrated volume} of $B$ with weight $\varphi$ is
\[
\textrm{Vol}_b(B,\varphi):=\int_B\varphi\, \rho_B\, \dd\mathcal{H}^n \in [0,+\infty] .
\]
We write $\textrm{Vol}_b(B):=\textrm{Vol}_b(B,1)$.
\end{definition}

The integral is well defined and does not depend on $h$. Indeed, $\rho_B$ is Borel measurable and locally bounded, since $v$ is continuous on $\textrm{dom}\, b\backslash S$ \cite[Thm.\ 5.21(a)]{minguzzi26e} and takes values in a compact subset of $TM$ over every compact subset of $\textrm{dom}\, b$ \cite[Thm.\ 5.12(v)]{minguzzi26e}. Moreover, if $F:U\to B$ is a locally bi-Lipschitz parametrization of an open subset of $B$, the area formula \cite[Sect.\ 3.3]{evans92} gives
\[
\int_{F(U)}\varphi\,\rho_B\,\dd\mathcal{H}^n=\int_U \varphi(F(y))\,\big\vert \omega_{F(y)}\big(\p_1F(y),\dots,\p_nF(y)\big)\big\vert\, \dd y ,
\]
since the Jacobian of $F$ is the $h$-volume of $\p_1F,\dots,\p_nF$, and $\vert\omega(\p_1F,\dots,\p_nF)\vert$ is $\rho_B$ times this volume. The right-hand side does not involve $h$, and it does not depend on the parametrization by the change of variables formula for bi-Lipschitz maps of $\mathbb{R}^n$ \cite[Sect.\ 3.3]{evans92}. In particular $\textrm{Vol}_b(B,\varphi)<+\infty$ if $\varphi$ is bounded and $B$ is relatively compact in $\textrm{dom}\, b$.

We use the notation of \cite[Sect.\ 2.6--2.7]{minguzzi14d}. Let $n$ be a field of semitangents of $H$, whose normalization will be fixed in Lemma \ref{lemdict}. At the points where $H$ is differentiable, its tangent space is a null hyperplane whose $\tilde g$-orthogonal is spanned by $n$, so that $\tilde g$ induces a positive definite metric on the quotient bundle $TH/n$. The area form $\sigma=i_n\mu_H$ vanishes on $n$, hence it descends to an $n$-form on $TH/n$, and it is the volume form of this quotient metric. Equivalently, the restriction of $\sigma$ to a spacelike hyperplane $W\subset TH$, which projects isometrically onto $TH/n$, is the volume form of the metric induced by $\tilde g$ on $W$; in particular, it is the induced area form on the spacelike sections of $H$, and it does not depend on the normalization of $n$ \cite[Remarks 8, 9]{minguzzi14d}. The volume measure $\mu_H$ on $H$ is characterized by $i_n\mu_H=\sigma$ \cite[Eq.\ (14), Remark 9]{minguzzi14d}, so it depends on the normalization of $n$. Finally, $\theta_H\in L^1_{loc}(H,\mu_H)$ is the expansion, namely the density of the absolutely continuous part of the expansion measure. By \cite[Remark 10]{minguzzi14d}, if $n$ is replaced by $fn$ then $\mu_H$, $\theta_H$ and $\p_n$ are replaced by $\mu_H/f$, $f\theta_H$ and $f\p_n$, so that the measures $\theta_H\,\mu_H$ and $(\p_n\varphi)\,\mu_H$ do not depend on the normalization.

\begin{lemma} \label{lemdict}
Assume that $(M,g)$ is future timelike complete. For $p\in\textrm{dom}\, b$ let $U_p\subset T_pM$ be the set of initial unit tangents of the $b$-calibrated rays starting from $p$, so that $U_p=\{v(p)\}$ for $p\notin S$ \cite[Thm.\ 5.21(a)]{minguzzi26e}. Normalize the semitangents of $H$ at $\Psi(p)$ as $n=(u,1)$, $u\in U_p$; in particular $n=(v,1)$ on $\Psi(\textrm{dom}\, b\backslash S)$.
\begin{itemize}
\item[(a)] The generators of $H$ are the curves $t\mapsto \Psi(\alpha(t))=(\alpha(t),b(\alpha(0))+t)$, where $\alpha$ is a proper time parametrized $b$-calibrated ray, and $t$ is an affine parameter with $n=\dd/\dd t$. For every $k$, $\Psi$ maps $S^k$ onto the set $\Sigma^k(H)$ of points at which the semitangents span a vector space of dimension $k+1$.
\item[(b)] $\pi_*\mu_H=\mu_g$ on $\textrm{dom}\, b$. If $B\subset \textrm{dom}\, b$ is as in Definition \ref{defcalvol}, then the area measure $\vert\sigma\vert_{\Psi(B)}$ induced by $\sigma$ on $\Psi(B)$ satisfies $\pi_*\vert\sigma\vert_{\Psi(B)}=\rho_B\,\mathcal{H}^n$ on $B$, namely the area of $\Psi(B)$ is the calibrated volume of $B$. For every locally Lipschitz function $\varphi$ on $\textrm{dom}\, b$, $\p_n(\varphi\circ\pi)=\dd\varphi(v)$ almost everywhere.
\item[(c)] The function $\theta^b:=\theta_H\circ\Psi\in L^1_{loc}(\textrm{dom}\, b)$, which we call the {\em expansion of the calibrated congruence}, satisfies $\theta^b=-\Box b$ at every point at which $b$ is twice differentiable, hence almost everywhere. Where $b$ is $C^2$, $\theta^b=\textrm{div}\, v$.
\end{itemize}
\end{lemma}

\begin{proof}
(a). By (\ref{mxwh}) and \cite[Thm.\ 5.12(iv)]{minguzzi26e}, $H$ is an achronal boundary which coincides with the graph of $b$ over $\textrm{dom}\, b$. Let $\alpha$ be a proper time parametrized $b$-calibrated ray. Since $b(\alpha(t))=b(\alpha(0))+t$, its lift $t\mapsto \Psi(\alpha(t))=(\alpha(t),b(\alpha(0))+t)$ lies in $H$. As $\tilde g=g+\dd z^2$, the lift is a geodesic of $\tilde M$ with tangent $(\dot\alpha,1)$, which is lightlike since $\dot \alpha$ is unit timelike \cite[Sect.\ 4]{minguzzi26e}. Being a lightlike geodesic contained in the achronal set $H$, it is contained in a generator of $H$. Conversely, by \cite[Prop.\ 5.6]{minguzzi26e} every generator of $H$ projects to a $b$-calibrated ray, along which $b$, and hence $z$, increases by the proper time, so it is of the stated form. In particular $t$ is an affine parameter on the generators, and $\dd/\dd t=(\dot\alpha,1)$ is the normalized semitangent. The statement on $S^k$ was proved in the proof of \cite[Thm.\ 5.21]{minguzzi26e}.

(b). Let $p\in\textrm{dom}\, b\backslash S$, so that $H$ is differentiable at $\Psi(p)$, with tangent space $\dd\Psi_p(T_pM)=\{(X,-g(v,X)): X\in T_pM\}$ and normalized semitangent $n=(v,1)=\dd\Psi_p(v)$. For $X_1,\dots,X_n\in T_pM$ let $\tilde X_i:=\dd\Psi_p(X_i)=(X_i,-g(v,X_i))$. Then
\[
\tilde g(\tilde X_i,\tilde X_j)=g(X_i,X_j)+g(v,X_i)g(v,X_j)=g(PX_i,PX_j),
\]
where $P$ is the $g$-orthogonal projection on $v^\perp$. Thus $\dd\Psi_p$ induces an isomorphism of $T_pM/v$ onto $T_{\Psi(p)}H/n$ which is an isometry when $T_pM/v$ is endowed with the positive definite metric $g(P\cdot,P\cdot)$, that is, when it is identified with the spacelike hyperplane $v^\perp$. The volume form of $v^\perp$ is the restriction of $i_v\mu_g=\omega$. Since $\sigma$ is the volume form of the quotient metric on $TH/n$, we get, up to sign,
\[
\sigma(\tilde X_1,\dots,\tilde X_n)=\mu_g(v,PX_1,\dots,PX_n)=\omega(X_1,\dots,X_n),
\]
namely $\Psi^*\sigma=\omega$ at $p$, up to sign. Now let $B$ be as in Definition \ref{defcalvol}, and let $F:U\to B$ be a locally bi-Lipschitz parametrization of an open subset of $B$. Then $\Psi\circ F$ is a locally bi-Lipschitz parametrization of an open subset of $\Psi(B)$, and the measure $\vert\sigma\vert_{\Psi(B)}$ induced by $\sigma$ on $\Psi(B)$, namely its area measure, is given on $\Psi(F(U))$ by the density $\vert\sigma(\p_1(\Psi\circ F),\dots,\p_n(\Psi\circ F))\vert$ with respect to the Lebesgue measure $\dd y$ on $U$ \cite[Sect.\ 3.3]{evans92}. For almost every $y\in U$, $F$ is differentiable at $y$ and $F(y)\notin S$, as $\mathcal{H}^n(B\cap S)=0$ and $F$ is locally bi-Lipschitz, so $\Psi$ is differentiable at $F(y)$ and $\p_i(\Psi\circ F)(y)=\dd\Psi_{F(y)}(\p_iF(y))$. By the previous identity,
\[
\big\vert\sigma\big(\p_1(\Psi\circ F),\dots,\p_n(\Psi\circ F)\big)\big\vert(y)=\big\vert\omega_{F(y)}\big(\p_1F(y),\dots,\p_nF(y)\big)\big\vert ,
\]
which, by the parametric expression following Definition \ref{defcalvol}, is the density of $\rho_B\,\mathcal{H}^n$ with respect to $\dd y$. Thus $\pi_*\vert\sigma\vert_{\Psi(B)}=\rho_B\,\mathcal{H}^n$ on $B$. Next, $\mu_H$ is determined at $\Psi(p)$ by $i_n\mu_H=\sigma$, and $\dd\pi(n)=v$, so $\pi_*\mu_H$ is the $(n+1)$-form $\lambda$ on $T_pM$ such that $i_v\lambda=\omega=i_v\mu_g$, which is $\mu_g$, as an $(n+1)$-form is determined by its contraction with a non-zero vector. Again by the area formula, $\pi_*\mu_H=\mu_g$ as measures, the set $S$ being null. Finally $\p_n(\varphi\circ\pi)=\dd\varphi(\dd\pi(n))=\dd\varphi(v)$ wherever $\varphi$ is differentiable at $p\notin S$.

(c). Let $b$ be twice differentiable at $p$, so that $p\notin S$, and let $\alpha$ be the calibrated ray from $p$, $\dot\alpha(0)=v$. We first construct a smooth solution $u$ of the eikonal equation near $p$ with the same second order expansion as $b$ at $p$. A necessary condition is $\nabla^2b(p)(v,\cdot)=0$, since differentiating $g(\nabla u,\nabla u)=-1$ gives $\nabla^2u(\nabla u,\cdot)=0$ for every such $u$, and $\nabla u(p)=\nabla b(p)=-v$. This condition is where the upper support function $b_{p,s}$ of \cite[Thm.\ 5.12(iii)]{minguzzi26e} enters. Since $b(\alpha(t))=b(p)+t$ for $t\ge 0$ and $\alpha$ is a geodesic, the second order expansion of $b$ at $p$ gives $\nabla^2b(p)(v,v)=0$, and similarly $\nabla^2b_{p,s}(p)(v,v)=0$, as $b_{p,s}(\alpha(t))=b(p)+t$ for $t<s$. As $b_{p,s}\ge b$ near $p$ with equality at $p$, the symmetric form $A:=\nabla^2b_{p,s}(p)-\nabla^2b(p)$ is positive semi-definite, and $A(v,v)=0$, so $A(v,\cdot)=0$. Since $b_{p,s}$ solves the eikonal equation, $\nabla^2b_{p,s}(p)(v,\cdot)=0$, and therefore $\nabla^2b(p)(v,\cdot)=0$. The function $b_{p,s}$ itself cannot play the role of $u$, since its Hessian at $p$ is in general strictly larger than that of $b$ on $v^\perp$.

Let $Y$ be a smooth spacelike hypersurface through $p$ with future unit normal $v$ at $p$, and let $u:=b(p)+f$, where $f$ is the signed distance function of $Y$, defined near $p$ through the normal exponential map of $Y$ and increasing in the future direction. Then $u$ is smooth near $p$, solves the eikonal equation, $\nabla u(p)=-v$, $\nabla^2u(p)(v,\cdot)=0$, and the restriction of $\nabla^2u(p)$ to $v^\perp$ is, up to sign, the second fundamental form of $Y$ at $p$ \cite[Sect.\ 1]{eschenburg89}. The latter can be prescribed arbitrarily: given a symmetric bilinear form $Q$ on $v^\perp$, in normal coordinates $(x^0,\dots,x^n)$ centered at $p$ with $\p_0=v$ the hypersurface $Y=\{x^0=\frac12 Q_{ij}x^ix^j\}$ has second fundamental form $\pm Q$ at $p$, since the Christoffel symbols vanish at $p$. Choosing $Q$ so that $\nabla^2u(p)$ and $\nabla^2b(p)$ agree on $v^\perp$, the previous step gives $\nabla^2u(p)=\nabla^2b(p)$, so $u$ has the same second order expansion as $b$ at $p$.

Thus the graph of $u$ is a smooth hypersurface $N$ of $\tilde M$ which has second order contact with $H$ at $\Psi(p)$. It is null, since its tangent hyperplanes are null: they are $\tilde g$-orthogonal to the gradient $(-\nabla u,1)$ of $z-u$, which is lightlike by the eikonal equation. Hence $\Psi(p)$ is an Alexandrov point of $H$ and, by \cite[Remark 12]{minguzzi14d}, $\theta_H(\Psi(p))$ is the expansion of $N$ at $\Psi(p)$ with respect to $n=(-\nabla u,1)$. To compute it we use \cite[Eq.\ (18)]{minguzzi14d}, which expresses the expansion of a null hypersurface as $\theta=\tilde\nabla_\mu n^\mu-\kappa$, where $n$ is extended to a neighborhood as a lightlike vector field with $\tilde\nabla_n n=\kappa n$. We extend $n$ by translation in the $z$ direction, namely we take the vector field $\tilde n(x,z):=(-\nabla u(x),1)$ defined on a neighborhood of $\Psi(p)$ in $\tilde M$. It is lightlike by the eikonal equation, and it is geodesic: as $\tilde g=g+\dd z^2$ is a product metric, $\tilde\nabla_{\tilde n}\tilde n=(\nabla_{\nabla u}\nabla u,0)=(\frac12\nabla g(\nabla u,\nabla u),0)=0$, so that $\kappa=0$. Moreover, for a product metric the divergence of a vector field $(X(x),c)$, with $X$ independent of $z$ and $c$ constant, is $\textrm{div}_g X$, since the volume form is $\mu_g\wedge \dd z$ and the $z$ component is constant. Thus the expansion of $N$ at $\Psi(p)$ is $\textrm{div}_{\tilde g}\tilde n=\textrm{div}_g(-\nabla u)(p)=-\Box u(p)=-\Box b(p)$. As $b$ is locally semi-concave, it is twice differentiable almost everywhere. Moreover, its distributional Hessian is a matrix valued Radon measure whose absolutely continuous part has, as density, the almost everywhere defined Hessian of $b$ \cite[Sect.\ 6.3, 6.4]{evans92}, so that $\theta^b=-\Box b\in L^1_{loc}(\textrm{dom}\, b)$. The last statement is the same computation with $u=b$.
\end{proof}

The equality cases of the area theorem are expressed in \cite{minguzzi14d} through the local graphing functions $h$ of $H$, which represent $H$ as a graph over a spacelike hypersurface of $\tilde M$ along the flow of a timelike vector field \cite[Sect.\ 2.2]{minguzzi14d}. Instead, $b$ represents $H$ as a graph along the spacelike direction $\p_z$. The two graphing functions are nevertheless related closely enough for the relevant regularity properties to coincide. Both $Dh$ and $\nabla b$ are locally of bounded variation, as $h$ is locally semi-convex and $b$ is locally semi-concave, and we recall that a function of bounded variation is {\em special} if its distributional derivative has no Cantor part \cite{ambrosio00}.

\begin{lemma} \label{lemreg}
Assume that $(M,g)$ is future timelike complete. Let $O\subset \textrm{dom}\, b$ be open.
\begin{itemize}
\item[(a)] The local graphing functions $h$ of $H$ belong to $W^{2,1}_{loc}$ over $\Psi(O)$, namely their distributional Hessians have no singular part there, if and only if $b\in W^{2,1}_{loc}(O)$, namely the distributional Hessian of $b$ has no singular part on $O$.
\item[(b)] The derivatives $Dh$ of the local graphing functions of $H$ are special functions of bounded variation over $\Psi(O)$ if and only if $Db$ (equivalently $\nabla b$) is a special function of bounded variation on $O$.
\end{itemize}
\end{lemma}

\begin{proof}
Let $(x^0,x)$, $x\in U\subset \mathbb{R}^{n+1}$, be the coordinates of a coordinate cylinder of \cite[Sect.\ 2.2]{minguzzi14d}, in which $H=\{x^0=h(x)\}$, and let $(y,z)$ be product coordinates of $\tilde M=M\times \mathbb{R}$ on the same open set, so that $y=Y(x^0,x)$ and $z=Z(x^0,x)$ for smooth functions $Y,Z$. The map $\Theta(x):=Y(h(x),x)$, which sends $x$ to the projection on $M$ of the point of $H$ over $x$, is the composition of the locally bi-Lipschitz parametrization $x\mapsto (h(x),x)$ of $H$ with $\pi\vert_H$, hence it is locally bi-Lipschitz. Moreover $b(\Theta(x))=Z(h(x),x)$. Differentiating almost everywhere, which is possible since $\Theta$ maps null sets to null sets,
\[
(Db\circ\Theta)\,(\p_0Y\, Dh+\p_xY)=\p_0Z\,Dh+\p_xZ ,
\]
where the derivatives of $Y,Z$ are evaluated at $(h(x),x)$. For $(x,t,\xi)\in U\times \mathbb{R}\times \mathbb{R}^{n+1}$ let
\begin{align*}
A(x,t,\xi)&:=\p_0Y(t,x)\otimes \xi+\p_xY(t,x),\\
G(x,t,\xi)&:=\big(\p_0Z(t,x)\,\xi+\p_xZ(t,x)\big)\,A(x,t,\xi)^{-1},
\end{align*}
so that, at every point $x$ at which $h$ is differentiable, $\Theta$ is differentiable with $D\Theta(x)=A(x,h(x),Dh(x))$. Since $Y$ and $Z$ are smooth and matrix inversion is smooth on invertible matrices, $G$ is smooth on the open set $\mathcal{W}$ where $A$ is invertible. Let $C\subset U$ be compact, let $L$ be a bi-Lipschitz constant of $\Theta$ on $C$, so that $\vert \Theta(x')-\Theta(x)\vert\ge \vert x'-x\vert/L$ for $x,x'\in C$, and let $\Lambda$ be a Lipschitz constant of $h$ on $C$. At every interior point $x$ of $C$ at which $h$ is differentiable, for every $w\in \mathbb{R}^{n+1}$,
\[
\vert A(x,h(x),Dh(x))\,w\vert=\vert D\Theta(x)\,w\vert=\lim_{\epsilon\to 0^+}\frac{\vert\Theta(x+\epsilon w)-\Theta(x)\vert}{\epsilon}\ge \frac{\vert w\vert}{L} .
\]
In particular $A(x,h(x),Dh(x))$ is invertible, so the previous identity reads $Db\circ\Theta=G(x,h,Dh)$ almost everywhere, and, for almost every $x\in C$, the point $(x,h(x),Dh(x))$ belongs to the set
\begin{align*}
K:=\{(x,t,\xi):\ & x\in C,\ \vert t\vert\le \max_C\vert h\vert,\ \vert\xi\vert\le \Lambda,\\
& \vert A(x,t,\xi)\,w\vert\ge \vert w\vert/L \textrm{ for all } w\in \mathbb{R}^{n+1}\},
\end{align*}
This set is closed, as the last inequality is preserved under limits, and bounded, hence compact. It is contained in $\mathcal{W}$, since its last condition forces $A(x,t,\xi)$ to be injective, hence invertible. Modifying $G$ outside a neighborhood of $K$ we may assume that $G$ is smooth with bounded derivatives on $\mathbb{R}^{2n+3}$, without changing $G(x,h,Dh)$ on $C$. Exchanging the roles of $h$ and $b$ we get similarly $Dh\circ\Theta^{-1}=G'(y,b,Db)$ almost everywhere, with $G'$ smooth with bounded derivatives on compact subsets.

(a). As $h$ and $b$ are locally Lipschitz, $h\in W^{2,1}_{loc}$ if and only if $Dh\in W^{1,1}_{loc}$, and similarly for $b$. If $Dh\in W^{1,1}_{loc}$, then $G(x,h,Dh)\in W^{1,1}_{loc}$ by the chain rule \cite[Sect.\ 4.2]{evans92}, and $Db=G(x,h,Dh)\circ \Theta^{-1}\in W^{1,1}_{loc}$, as $W^{1,1}_{loc}$ is invariant under locally bi-Lipschitz changes of variables \cite[Thm.\ 2.2.2]{ziemer89}. The converse follows in the same way from the expression of $Dh\circ\Theta^{-1}$.

(b). By the chain rule in $BV$ \cite[Thm.\ 3.96]{ambrosio00}, the Cantor part of the derivative of $G(x,h,Dh)$ is obtained by applying the differential of $G$ to the Cantor part of $D(Dh)$, so it vanishes if $Dh$ is special. Next, if $w$ is a special function of bounded variation and $\Phi$ is a locally bi-Lipschitz map, then $w\circ \Phi$ is of bounded variation \cite[Thm.\ 3.16]{ambrosio00}, and $\vert D(w\circ \Phi)\vert\le C\,(\Phi^{-1})_\#\vert Dw\vert$ locally. The measure $\vert Dw\vert$ is the sum of a measure absolutely continuous with respect to the Lebesgue measure and of a measure concentrated on a set of $\sigma$-finite $\mathcal{H}^{n}$ measure, and both properties are preserved by $(\Phi^{-1})_\#$. The Cantor part of $D(w\circ\Phi)$ is absolutely continuous with respect to $\vert D(w\circ\Phi)\vert$, singular with respect to the Lebesgue measure, and it vanishes on the sets of $\sigma$-finite $\mathcal{H}^{n}$ measure \cite[Sect.\ 3.9]{ambrosio00}, hence it vanishes. Thus $w\circ\Phi$ is special. Applying these facts with $w=G(x,h,Dh)$ and $\Phi=\Theta^{-1}$, and conversely with $G'(y,b,Db)$ and $\Theta$, gives (b).
\end{proof}

\section{The volume theorem} \label{secvolthm}

\begin{definition} \label{deftube}
Let $f:M\to\mathbb{R}$ be a locally Lipschitz time function and $t_1<t_2$. A {\em calibrated tube} is an open set $\Omega=U\cap f^{-1}((t_1,t_2))$, where $U\subset M$ is open, such that $\overline{\Omega}\subset \textrm{dom}\, b$ is compact and $\p U\cap f^{-1}([t_1,t_2])$ is contained in a $C^1$ hypersurface $W$ each point of which lies in the interior of a segment of $b$-calibrated ray contained in $W$. The {\em bases} of $\Omega$ are $B_i:=\overline{U}\cap f^{-1}(t_i)$, $i=1,2$, and its {\em side} is $\p U\cap f^{-1}([t_1,t_2])$. If $U=M$ the tube has no side, and if moreover $f$ is a Cauchy time function, the bases are compact Cauchy hypersurfaces contained in $\textrm{dom}\, b$.
\end{definition}

Thus the sides of a calibrated tube are generated by calibrated segments, and the calibrated rays can enter the tube only through $B_1$ and leave it only through $B_2$. The level sets of $f$ are achronal, so $\mathcal{H}^n(f^{-1}(t)\cap S)=0$ for every $t\in\mathbb{R}$ \cite[Thm.\ 5.21(c)]{minguzzi26e}. Thus the calibrated volume of the bases is defined (Definition \ref{defcalvol}) and, $\Psi$ being locally bi-Lipschitz, the set of non-differentiability points of $H$ meets $\Psi(f^{-1}(t))$ in an $\mathcal{H}^n$-null set for every $t$. Hence the conclusion of \cite[Prop.\ 9]{minguzzi14d} for the locally Lipschitz time function $f\circ\pi$ of $\tilde M$, which is stated there for almost every $t$, holds here for every $t$, as its proof uses only this property: no value of $t$ needs to be excluded.

\begin{theorem}[Volume theorem] \label{thmvol}
Assume that $(M,g)$ is future timelike complete. Let $\Omega$ be a calibrated tube, let $\varphi$ be a positive locally Lipschitz function on a neighborhood of $\overline\Omega$, and let $v:=-\nabla b$, defined at the differentiability points of $b$, hence almost everywhere, where it is the initial unit tangent of the unique $b$-calibrated ray starting from the point \cite[Lemma 5.16]{minguzzi26e}. Then
\begin{equation} \label{volin}
\int_\Omega\big[\varphi\,\theta^b+\dd\varphi(v)\big]\,\mu_g\le \textrm{Vol}_b(B_2,\varphi)-\textrm{Vol}_b(B_1,\varphi),
\end{equation}
with equality if and only if $b\in W^{2,1}_{loc}(\Omega)$, namely the distributional Hessian of $b$ has no singular part on $\Omega$, for instance if $b$ is $C^{1,1}$ on $\Omega$. More precisely,
\begin{equation} \label{volin2}
\int_\Omega\big[\varphi\,\theta^b+\dd\varphi(v)\big]\,\mu_g+2\int_{S^1\cap\Omega}\varphi\,\vert\omega_+\vert\le \textrm{Vol}_b(B_2,\varphi)-\textrm{Vol}_b(B_1,\varphi),
\end{equation}
with equality if and only if $\nabla b$ is a special function of bounded variation on $\Omega$. Here $S^1$ is the set of points from which exactly two calibrated rays start, with initial tangents $v_\pm$, and $\omega_\pm=i_{v_\pm}\mu_g$; at $\mathcal{H}^n$-almost every point of the countably $\mathcal{H}^n$-rectifiable set $S^1$, the approximate tangent space $T$ is the $g$-orthogonal of $v_+-v_-$, it contains the timelike vector $v_++v_-$, and $\omega_+\vert_T=-\omega_-\vert_T$, so that $\vert\omega_+\vert=\vert\omega_-\vert$ on $S^1$.
\end{theorem}

For $\varphi=1$, (\ref{volin}) reads $\textrm{Vol}_b(B_2)-\textrm{Vol}_b(B_1)\ge \int_\Omega\theta^b\,\mu_g$, the analog of the area theorem, while (\ref{volin2}) adds the calibrated volume of the creases, counted twice. The inequalities (\ref{volin}) and (\ref{volin2}), though not their equality cases, extend to non-negative locally Lipschitz functions $\varphi$: it suffices to apply them to $\varphi+\delta$, $\delta>0$, and to let $\delta\to 0$, all terms being finite and affine in $\delta$.

\begin{proof}
Let $D:=\Psi(\Omega)$, an open relatively compact subset of $H$. Its boundary $\Psi(\p\Omega)$ is contained in $\Psi(B_1\cup B_2\cup W)$. The sets $f^{-1}(t_i)$ are boundaries of the future sets $f^{-1}((t_i,+\infty))$, hence locally Lipschitz hypersurfaces \cite[Prop.\ 6.3.1]{hawking73}, so $\p\Omega$ has finite $\mathcal{H}^n$ measure and so does $\p D$, $\Psi$ being locally bi-Lipschitz. Thus $D$ has finite perimeter \cite[Sect.\ 5.11]{evans92}. We apply \cite[Thm.\ 13]{minguzzi14d} to the past horizon $H$, the set $D$ and the function $\varphi\circ\pi$. By Lemma \ref{lemdict}(b),(c) and the independence of the left-hand side of \cite[Eq.\ (28)]{minguzzi14d} from the normalization of $n$, its left-hand side is the left-hand side of (\ref{volin}). Since $f\circ\pi$ increases along the generators, $\Psi(B_2)$ is the future boundary and $\Psi(B_1)$ the past boundary of $D$. By the observation preceding the theorem and \cite[Prop.\ 9]{minguzzi14d}, the area functional on $\Psi(B_i)$ is $\int_{\Psi(B_i)}(\varphi\circ\pi)\,\sigma$, which by Lemma \ref{lemdict}(b) is $\textrm{Vol}_b(B_i,\varphi)$. The side gives no contribution by \cite[Prop.\ 10]{minguzzi14d}, since $\Psi(W)$ is a $C^1$ hypersurface every point of which is internal to a generator contained in it: indeed, $b$ is differentiable at the interior points of the calibrated rays \cite[Prop.\ 5.19]{minguzzi26e} with continuous differential on the set of differentiability points \cite[Thm.\ 5.21(a)]{minguzzi26e}, so $b\vert_W$ is $C^1$. This proves (\ref{volin}), and its equality case follows from that of \cite[Thm.\ 13]{minguzzi14d} and Lemma \ref{lemreg}(a). Inequality (\ref{volin2}) and its equality case follow in the same way from \cite[Thm.\ 14]{minguzzi14d} and Lemma \ref{lemreg}(b), as $\Psi(S^1)=\Sigma^1(H)$ by Lemma \ref{lemdict}(a); the affine span of $U_p$ is a line which meets the unit hyperboloid in at most two points, so exactly two calibrated rays start from each point of $S^1$. The description of the approximate tangent space $T$ of $S^1$ follows from the structure of $Dh$ as a function of bounded variation recalled in \cite[Sect.\ 2.3]{minguzzi14d}, after Theorem 5 there, where $h$ is a local graphing function of $H$ in a coordinate cylinder $(x^0,x)$. By \cite[Thms.\ 6(4), 7]{minguzzi14d}, the non-differentiability points of $h$ form the approximate discontinuity set of $Dh$, and, as the sets $\Sigma^k(H)$, $k\ge 2$, are $\mathcal{H}^n$-null, $\Psi(S^1)=\Sigma^1(H)$ corresponds, up to an $\mathcal{H}^n$-null set, to the jump set $J_{Dh}$. At its points $Dh$ has one-sided approximate limits $Dh^\pm$ with respect to a unit normal $\nu$ orthogonal to the approximate tangent space of $J_{Dh}$ \cite[Thm.\ 6(4)]{minguzzi14d}. They are distinct limits of gradients of $h$ at nearby differentiability points, so they correspond to the two semitangents $(v_\pm,1)$. As the distributional Hessian of $h$ is symmetric, its jump part $(Dh^+-Dh^-)\otimes\nu\, \mathcal{H}^n\llcorner J_{Dh}$ \cite[Eq.\ (11)]{minguzzi14d} is symmetric, so $Dh^+-Dh^-$ is parallel to $\nu$, hence orthogonal to the approximate tangent space of $J_{Dh}$. Thus the two one-sided tangent hyperplanes $\{\xi^0=Dh^\pm\cdot \xi\}$ of $H$ agree on the vectors approximately tangent to $J_{Dh}$, and the approximate tangent space of $\Psi(S^1)$ is contained in both. These one-sided tangent hyperplanes are the null hyperplanes orthogonal to the semitangents $(v_\pm,1)$, so, as $\tilde g((v_\pm,1),(X,a))=g(v_\pm,X)+a$, they are the graphs of $-g(v_\pm,\cdot)$ over $T_pM$, so $g(v_+,X)=g(v_-,X)$ for $X\in T$, namely $T\subset (v_+-v_-)^\perp$, with equality by dimension. Next, for $X\in T$ the lifted vector $(X,-g(v_+,X))=(X,-g(v_-,X))$ is the same for both signs, so the computation in the proof of Lemma \ref{lemdict}(b), performed with $v_+$ or with $v_-$, shows that the area form of the lifted section, evaluated on lifted vectors, is $\vert\omega_+\vert$ and also $\vert\omega_-\vert$ on $T$. Since $T$ has dimension $n$, the restrictions $\omega_\pm\vert_T$ are multiples $c_\pm\,\epsilon_T$ of a fixed volume form $\epsilon_T$ of $T$, and the previous observation gives $\vert c_+\vert=\vert c_-\vert$, which is positive since the lifted section is spacelike. If it were $c_+=c_-$, then $\mu_g(v_+-v_-,\cdot)\vert_T=(\omega_+-\omega_-)\vert_T=0$, which is impossible: as $T=(v_+-v_-)^\perp$ and $v_+-v_-$ is spacelike, $v_+-v_-\notin T$, so $T$ and $v_+-v_-$ span $T_pM$ and $\mu_g(v_+-v_-,\cdot)$ does not vanish on $T$. Therefore $c_+=-c_-$, namely $\omega_+\vert_T=-\omega_-\vert_T$, in agreement with \cite[Remark 8]{minguzzi14d}. Equivalently $\mu_g(v_++v_-,\cdot)\vert_T=0$, which, by the same argument, means $v_++v_-\in T$.
\end{proof}

In particular the crease set $S^1$ is, in the measure theoretic sense, a timelike hypersurface: the two calibrated observers leaving one of its points move apart symmetrically with respect to it.

\begin{example}
For the warped products of \cite[Ex.\ 5.18]{minguzzi26e}, $b=t-t_0$, $v=\p_t$, $S=\emptyset$, $\omega=a^n\mu_h$ on the slices $S_t$, and $\theta^b=n\dot a/a$. Every set $U=I\times A$, $A\subset S$ with $C^1$ boundary, gives a calibrated tube for $f=t$, and (\ref{volin}) holds with equality, as $\frac{\dd}{\dd t}(a^n)\textrm{vol}_h(A)=\int_A n\frac{\dot a}{a}\,a^n\mu_h$.
\end{example}

The equality case of (\ref{volin2}) always holds. The reason is that, in Gaussian coordinates, the eikonal equation satisfied by $b$ is a Hamilton--Jacobi equation whose Hamiltonian is uniformly convex on the bounded sets of momenta, and for the viscosity solutions of these equations the gradient has no Cantor part, by a theorem of Bianchini and Tonon \cite{bianchini12}, which extends to Hamiltonians depending on time and space a result of Bianchini, De Lellis and Robyr \cite{bianchini11}. For a general horizon the graphing functions solve an eikonal equation whose Hamiltonian is positively homogeneous of degree one in the momenta, hence not uniformly convex, and the theorem does not apply; the horizon $H$ of the product trick is special in this respect.

\begin{theorem}[SBV regularity] \label{thmsbv}
Assume that $(M,g)$ is future timelike complete. Then $\nabla b$ is locally a special function of bounded variation on $\textrm{dom}\, b$. As a consequence, for every open set $O\subset\textrm{dom}\, b$, $b\in W^{2,1}_{loc}(O)$ if and only if $\mathcal{H}^n(S\cap O)=0$.
\end{theorem}

\begin{proof}
Let $p\in\textrm{dom}\, b$, let $Y$ be a smooth spacelike hypersurface through $p$, and let $(t,x)$ be Gaussian normal coordinates of $Y$ near $p$, so that $g=-\dd t^2+h_{ij}(t,x)\dd x^i\dd x^j$, with $\p_t$ future directed and $h_{ij}$ positive definite, on a coordinate neighborhood $\mathcal{U}$, relatively compact in $\textrm{dom}\, b$, on which $p$ has coordinates $(0,0)$. At the differentiability points of $b$ the gradient $\nabla b$ is past directed and $g(\nabla b,\nabla b)=-1$ \cite[Lemma 5.16]{minguzzi26e}, namely $\p_tb>0$ and $-(\p_tb)^2+h^{ij}\p_ib\,\p_jb=-1$. Thus the function $u(s,x):=b(-s,x)$ satisfies, at each of its differentiability points,
\begin{equation} \label{hjeq}
\p_su+\mathsf{H}(s,x,D_xu)=0,\qquad \mathsf{H}(s,x,p):=\sqrt{1+h^{ij}(-s,x)\,p_ip_j}\, ,
\end{equation}
where $\mathsf{H}$ is smooth and convex in the momentum $p\in\mathbb{R}^n$. Moreover, $u$ is locally Lipschitz and locally semi-concave, as $b$ is \cite[Thm.\ 5.12(iii),(vi)]{minguzzi26e} and these properties are preserved by diffeomorphisms.

{\em Step 1: $u$ is a viscosity solution of (\ref{hjeq}).} This is a well known property of the semi-concave functions which solve almost everywhere a Hamilton--Jacobi equation with a Hamiltonian convex in the momenta \cite[Ch.\ 5]{cannarsa04}, and we give the argument. We use the definition of viscosity solution of \cite[Def.\ 2.8]{bianchini12}, with smooth test functions. If $u-\phi$ has a local maximum (minimum) at a point, then $(\p_s\phi,D_x\phi)$ belongs to the superdifferential (subdifferential) of $u$ at the point. The superdifferential of a semi-concave function at a point is the convex hull of its reachable gradients, namely of the limits of the gradients at differentiability points converging to the point \cite[Thm.\ 2.5(iii)]{bianchini12}. By continuity, every reachable gradient $(\sigma_k,p_k)$ at $(s,x)$ satisfies $\sigma_k+\mathsf{H}(s,x,p_k)=0$, so, if $(\sigma,p)=\sum_k\lambda_k(\sigma_k,p_k)$ is a convex combination of reachable gradients, the convexity of $\mathsf{H}$ gives $\sigma+\mathsf{H}(s,x,p)\le\sum_k\lambda_k[\sigma_k+\mathsf{H}(s,x,p_k)]=0$. Thus $u$ is a viscosity subsolution. If the subdifferential of $u$ at a point is non-empty then, the superdifferential being non-empty by semi-concavity, $u$ is differentiable there, and (\ref{hjeq}) holds at the point; thus $u$ is a viscosity supersolution.

{\em Step 2: modification of the Hamiltonian.} Let $\delta>0$ be such that the closure of $Q:=(-\delta,\delta)\times\{\vert x\vert<\delta\}$ is contained in the image of $\mathcal{U}$ under the chart $(s,x)$, and let $R$ be larger than a Lipschitz constant of $u$ on $Q$ for the Euclidean distance. The elements $(\sigma,p)$ of the super- and subdifferentials of $u$ at the points of $Q$ satisfy $\vert p\vert<R$, so $u$ is a viscosity solution on $Q$ of every equation $\p_su+\tilde{\mathsf{H}}(s,x,D_xu)=0$ such that $\tilde{\mathsf{H}}=\mathsf{H}$ on $Q\times\{\vert p\vert\le R\}$. We construct such a Hamiltonian, defined on $\mathbb{R}\times\mathbb{R}^n\times\mathbb{R}^n$, which satisfies the assumptions (H1) and (H2) of \cite[Thm.\ 1.1]{bianchini12}, namely:
\begin{itemize}
\item[(H1)] $\tilde{\mathsf{H}}$ is of class $C^3$ with bounded second derivatives, and there are positive constants $a_0,b_0,c_0'$ such that $\tilde{\mathsf{H}}\ge-c_0'$, $\tilde{\mathsf{H}}(s,x,0)\le c_0'$ and $\vert \p_p\p_x\tilde{\mathsf{H}}\vert\le a_0+b_0\vert p\vert$;
\item[(H2)] there is $c>0$ such that $c^{-1}\,I\le \p^2_{pp}\tilde{\mathsf{H}}\le c\,I$ everywhere.
\end{itemize}

Let $\chi$ be a smooth function with compact support in the image of the chart, $0\le\chi\le1$, $\chi=1$ on $Q$, and let $\tilde h^{ij}(s,x):=\chi\,h^{ij}(-s,x)+(1-\chi)\,\delta^{ij}$, which is smooth, bounded with all its derivatives, and uniformly positive definite on $\mathbb{R}\times\mathbb{R}^n$. Let $\mathsf{H}_0:=(1+\tilde h^{ij}p_ip_j)^{1/2}$, so that $\mathsf{H}_0=\mathsf{H}$ over $Q$. By the Cauchy--Schwarz inequality for the scalar product $\tilde h^{ij}$, $\p^2_{pp}\mathsf{H}_0\ge \lambda\,(1+\tilde h^{ij}p_ip_j)^{-3/2}I$, where $\lambda>0$ is a lower bound for the eigenvalues of $\tilde h^{ij}$, so $\p^2_{pp}\mathsf{H}_0\ge c_0I$ for $\vert p\vert\le3R$, for some $c_0>0$. Let $\zeta:[0,+\infty)\to[0,1]$ be smooth and non-increasing, with $\zeta=1$ on $[0,2R]$ and $\zeta=0$ on $[3R,+\infty)$, let $\kappa:[0,+\infty)\to[0,2]$ be smooth, with $\kappa=0$ on $[0,R]$, $\kappa>0$ on $(R,+\infty)$ and $\kappa=2$ on $[2R,+\infty)$, and let $\beta(r):=\int_0^r\!\int_0^\rho\kappa(\varrho)\,\dd\varrho\,\dd\rho$. For $\Lambda>0$ let
\[
\tilde{\mathsf{H}}(s,x,p):=\zeta(\vert p\vert)\,\mathsf{H}_0(s,x,p)+\Lambda\,\beta(\vert p\vert) ,
\]
which is smooth, as $\zeta(\vert p\vert)=1$ and $\beta(\vert p\vert)=0$ near $p=0$, non-negative, and such that $\tilde{\mathsf{H}}(s,x,0)=1$ and $\tilde{\mathsf{H}}=\mathsf{H}_0$ for $\vert p\vert\le R$. Its derivatives with respect to $s$ and $x$, of every order, are bounded for $\vert p\vert\le3R$, as $\tilde h^{ij}$ is bounded with all its derivatives, and vanish for $\vert p\vert\ge3R$, where $\tilde{\mathsf{H}}=\Lambda\beta(\vert p\vert)$; in particular the second derivatives $\p_s^2\tilde{\mathsf{H}}$, $\p_s\p_x\tilde{\mathsf{H}}$, $\p^2_{xx}\tilde{\mathsf{H}}$, $\p_s\p_p\tilde{\mathsf{H}}$ and $\p_p\p_x\tilde{\mathsf{H}}$ are bounded, and the bound on $\p_p\p_x\tilde{\mathsf{H}}$ in (H1) holds with $b_0=0$. The function $p\mapsto\beta(\vert p\vert)$ is convex, as $\beta$ is convex and non-decreasing, and the eigenvalues of its Hessian are $\beta''(\vert p\vert)$ and $\beta'(\vert p\vert)/\vert p\vert$. On $\{\vert p\vert\ge2R\}$ they are bounded from above, and bounded from below by a positive constant $m$, as $\beta''=2$ and $\beta'(r)\ge\beta'(2R)>0$ there, with $\beta'(r)/r\to2$ for $r\to+\infty$. Thus $\p^2_{pp}\tilde{\mathsf{H}}\ge c_0I$ for $\vert p\vert\le2R$, where $\zeta=1$. For $2R\le\vert p\vert\le3R$ the Hessian of $\zeta(\vert p\vert)\mathsf{H}_0$ is bounded by a constant $C_1$ which does not depend on $(s,x)$, so $\p^2_{pp}\tilde{\mathsf{H}}\ge(\Lambda m-C_1)I\ge I$ provided $\Lambda\ge(C_1+1)/m$. Finally $\p^2_{pp}\tilde{\mathsf{H}}\ge\Lambda m\,I$ for $\vert p\vert\ge3R$. The upper bound on $\p^2_{pp}\tilde{\mathsf{H}}$ holds since the Hessian of $\zeta(\vert p\vert)\mathsf{H}_0$ is bounded and vanishes for $\vert p\vert\ge3R$, while the eigenvalues $\beta''(\vert p\vert)\le2$ and $\beta'(\vert p\vert)/\vert p\vert$ of the Hessian of $\beta(\vert p\vert)$ are bounded, as $\beta'(r)\le2r$. Thus (H2) holds, $\p^2_{pp}\tilde{\mathsf{H}}$ is bounded, and (H1) holds with $c_0'=1$.

{\em Step 3: conclusion.} The theorem of Bianchini and Tonon \cite[Thm.\ 1.1]{bianchini12} applies to viscosity solutions on an arbitrary open set $\Omega\subset[0,T]\times\mathbb{R}^n$, no initial condition being involved: the reduction to a Cauchy problem with bounded Lipschitz initial datum is part of its proof, through the locality property \cite[Prop.\ 3.20]{bianchini12}, \cite[Prop.\ 3.5]{bianchini11}. Their viscosity solutions are bounded and uniformly continuous \cite[Def.\ 2.8]{bianchini12}, a condition satisfied by $u$ on $Q$, which is Lipschitz on $Q$ and bounded, as $\overline Q$ is compact. We apply the theorem to the open set $\Omega:=Q+(\delta,0)\subset[0,2\delta]\times\mathbb{R}^n$, to the viscosity solution $(s,x)\mapsto u(s-\delta,x)$ and to the Hamiltonian $(s,x,p)\mapsto\tilde{\mathsf{H}}(s-\delta,x,p)$, which satisfies (H1) and (H2) with the same constants. Thus $D_xu$ and $\p_su$ are locally special functions of bounded variation on $Q$. Hence so are the partial derivatives of $b$ in the coordinates $(t,x)$ on a neighborhood of $p$, and the components $g^{\mu\nu}\p_\nu b$ of $\nabla b$, as the product of a special function of bounded variation with a smooth function is special. Since $p$ is arbitrary, and since the property does not depend on the chart (proof of Lemma \ref{lemreg}(b)), this proves the first statement.

Let $O\subset\textrm{dom}\, b$ be open. The gradient $\nabla b$ is continuous on $\textrm{dom}\, b\backslash S$ \cite[Thm.\ 5.21(a)]{minguzzi26e}, a set of full measure, so its approximate discontinuity set is contained in $S$. The distributional derivative of $\nabla b$ has no Cantor part, and its jump part is concentrated on the approximate discontinuity set and is absolutely continuous with respect to $\mathcal{H}^n$ \cite[Sect.\ 3.9]{ambrosio00}. Thus, if $\mathcal{H}^n(S\cap O)=0$, the derivative of $\nabla b$ is absolutely continuous with respect to the Lebesgue measure on $O$, namely $b\in W^{2,1}_{loc}(O)$. Conversely, if $b\in W^{2,1}_{loc}(O)$ then, by Lemma \ref{lemreg}(a), the local graphing functions $h$ of $H$ belong to $W^{2,1}_{loc}$ over $\Psi(O)$, so their Hessians have no jump part there. The jump part is $(Dh^+-Dh^-)\otimes\nu\,\mathcal{H}^n\llcorner J_{Dh}$ \cite[Eq.\ (11)]{minguzzi14d}, with $Dh^+\ne Dh^-$ on the jump set $J_{Dh}$, so $J_{Dh}$ is $\mathcal{H}^n$-null over $\Psi(O)$ and, as recalled in the proof of Theorem \ref{thmvol}, it corresponds to $\Psi(S^1)$ up to an $\mathcal{H}^n$-null set. As $\Psi$ is locally bi-Lipschitz, $\mathcal{H}^n(S^1\cap O)=0$, and $S\backslash S^1$ is $\mathcal{H}^n$-null, since $\Psi$ maps it into the union of the sets $\Sigma^k(H)$, $k\ge2$ (Lemma \ref{lemdict}(a)). Thus $\mathcal{H}^n(S\cap O)=0$.
\end{proof}

\begin{corollary}[Balance law for the calibrated volume] \label{corbal}
Assume that $(M,g)$ is future timelike complete. For every calibrated tube $\Omega$ and every non-negative locally Lipschitz function $\varphi$ on a neighborhood of $\overline\Omega$,
\begin{equation} \label{volbal}
\textrm{Vol}_b(B_2,\varphi)-\textrm{Vol}_b(B_1,\varphi)=\int_\Omega\big[\varphi\,\theta^b+\dd\varphi(v)\big]\,\mu_g+2\int_{S^1\cap\Omega}\varphi\,\vert\omega_+\vert .
\end{equation}
In particular, for positive $\varphi$, equality holds in (\ref{volin}) if and only if $\mathcal{H}^n(S\cap\Omega)=0$.
\end{corollary}

\begin{proof}
As $\overline\Omega$ is a compact subset of $\textrm{dom}\, b$, by Theorem \ref{thmsbv} $\nabla b$ is a special function of bounded variation on $\Omega$, so for positive $\varphi$ the identity is the equality case of (\ref{volin2}). For non-negative $\varphi$ we apply it to $\varphi+\delta$, $\delta>0$, and let $\delta\to0$, all terms being finite and affine in $\delta$. Comparing with (\ref{volin}), equality holds there if and only if $\int_{S^1\cap\Omega}\varphi\,\vert\omega_+\vert=0$, namely, for positive $\varphi$, if and only if $\mathcal{H}^n(S^1\cap\Omega)=0$, since the density of $\vert\omega_+\vert$ with respect to $\mathcal{H}^n$ on $S^1$ is positive almost everywhere (proof of Theorem \ref{thmvol}), and $S\backslash S^1$ is $\mathcal{H}^n$-null.
\end{proof}

Thus the calibrated volume is created or destroyed only by the expansion of the calibrated congruence and by its creases: the Hessian of $b$ has no diffuse singular part, and there is no other production of volume. In local form, let $\vert\omega_+\vert\llcorner S^1$ denote the Borel measure $E\mapsto\int_{S^1\cap E}\vert\omega_+\vert$ on $\textrm{dom}\, b$.

\begin{corollary}[Divergence of the calibrated velocity field] \label{cordiv}
Assume that $(M,g)$ is future timelike complete. The vector field $v=-\nabla b\in L^\infty_{loc}(\textrm{dom}\, b)$ has a distributional divergence which is a Radon measure on $\textrm{dom}\, b$, namely
\begin{equation} \label{divv}
\textrm{div}\, v=-\Box b=\theta^b\,\mu_g+2\,\vert\omega_+\vert\llcorner S^1 ,
\end{equation}
that is, $-\int\dd\varphi(v)\,\mu_g=\int\varphi\,\theta^b\,\mu_g+2\int_{S^1}\varphi\,\vert\omega_+\vert$ for every Lipschitz function $\varphi$ with compact support in $\textrm{dom}\, b$. Under the TCC, $\textrm{div}\, v\ge0$. Moreover, for every positive locally Lipschitz function $\phi$ on $\textrm{dom}\, b$, $\textrm{div}(\phi v)=\phi\,\textrm{div}\, v+\dd\phi(v)\,\mu_g$.
\end{corollary}

\begin{proof}
Let $K$ be the support of $\varphi$, and let $f$ be a smooth Cauchy temporal function and $t_1<t_2$ be such that $K\subset f^{-1}((t_1,t_2))$. Assume first $\varphi\ge0$. Equation (\ref{volc}) in the proof of Theorem \ref{corvol}(b) below does not use the TCC nor the sign of $\dd\varphi(v)$, and it holds for $\varphi$. Choose $c$ such that $K\subset\{b<c\}$. Then $F(b-c)\varphi=\varphi$, $F'(b-c)\varphi=0$, and $\varphi$ vanishes on $B_1\cup B_2$, so (\ref{volc}) reduces to the stated identity. The general case follows by linearity, writing $\varphi=\max\{\varphi,0\}-\max\{-\varphi,0\}$. Applying the identity to a non-negative $\varphi$ equal to $1$ on a compact set $C$ shows that $\int_{S^1\cap C}\vert\omega_+\vert<+\infty$, so the right-hand side of (\ref{divv}) is a Radon measure, as $\theta^b\in L^1_{loc}$. Its sign under the TCC follows from $\theta^b\ge0$. The last statement follows from the identity applied to $\phi\varphi$, as $\dd(\phi\varphi)=\phi\,\dd\varphi+\varphi\,\dd\phi$.
\end{proof}

The same method applies to the cosmological time of Andersson, Galloway and Howard (Section \ref{secinit}), whose singular set is studied in \cite{minguzzi26f}.

\begin{proposition}[SBV regularity of the cosmological time] \label{proptausbv}
Let $(M,g)$ be a spacetime with regular cosmological time $\tau$. Then $\nabla\tau$ is locally a special function of bounded variation on $M$. Thus the distributional Hessian of $\tau$ has no Cantor part, and its singular part is its jump part, which is concentrated on the set of non-differentiability points of $\tau$.
\end{proposition}

\begin{proof}
The function $\tau$ is locally Lipschitz \cite[Thm.\ 1.2(v)]{andersson98} and locally semi-convex, as it admits at every point smooth lower support functions with locally uniformly bounded Hessians \cite[Sect.\ 3]{andersson98}, \cite[Lemma 8.1]{minguzzi26f}. These are the functions $r+d(\gamma(r),\cdot)$, where $\gamma$ is a maximizing geodesic which realizes $\tau$ at the point and $r$ is small \cite[Prop.\ 3.1]{andersson98}, whose gradient is past directed, unit and timelike. At a differentiability point of $\tau$ the gradient of $\tau$ coincides with that of a lower support function, so $\nabla\tau$ is past directed with $g(\nabla\tau,\nabla\tau)=-1$ almost everywhere. Let $p\in M$ and let $(t,x)$ be Gaussian normal coordinates near $p$, with $\p_t$ future directed. Then the locally semi-concave function $u:=-\tau$ satisfies $\p_tu+\mathsf{H}(t,x,D_xu)=0$ at its differentiability points, with $\mathsf{H}(t,x,p)=(1+h^{ij}(t,x)p_ip_j)^{1/2}$, and no reversal of time is needed. Steps 1--3 of the proof of Theorem \ref{thmsbv} apply verbatim and show that $\nabla\tau$ is locally a special function of bounded variation. The gradient of a locally semi-convex function is continuous on the set of its differentiability points, by the upper semi-continuity of the superdifferential of $-\tau$ \cite[Thm.\ 2.5(iii)]{bianchini12}, so its approximate discontinuity set, on which the jump part is concentrated \cite[Sect.\ 3.9]{ambrosio00}, is contained in the set of non-differentiability points.
\end{proof}

\section{Volume monotonicity and rigidity} \label{secmono}

\subsection{Monotonicity}

Under the TCC, $\Box b\le 0$ at the points of twice differentiability of $b$ \cite[Cor.\ 6.11]{minguzzi26e}, so $\theta^b\ge 0$ almost everywhere by Lemma \ref{lemdict}(c). Equivalently, $\theta_H\ge 0$ by \cite[Thm.\ 15]{minguzzi14d}, since $H$ is achronal with future complete generators \cite[Thm.\ 5.12(i)]{minguzzi26e} and the null convergence condition for $\tilde g=g+\dd z^2$ is the TCC for $g$: a null vector of $\tilde M$ has the form $(X,a)$ with $X$ causal, and $\widetilde{Ric}((X,a),(X,a))=Ric(X,X)$. The dominant energy condition for $g$ does not imply it.

\begin{theorem}[Volume monotonicity theorem] \label{corvol}
Assume the TCC and that $(M,g)$ is future timelike complete, and let $v:=-\nabla b$, the almost everywhere defined field of initial unit tangents of the $b$-calibrated rays.
\begin{itemize}
\item[(a)] Let $\Omega$ be a calibrated tube, and let $\varphi$ be a non-negative locally Lipschitz function on a neighborhood of $\overline\Omega$ such that $\dd\varphi(v)\ge 0$ almost everywhere on $\Omega$. Then
\begin{equation} \label{volmon}
\begin{split}
\textrm{Vol}_b(B_2,\varphi)&= \textrm{Vol}_b(B_1,\varphi)+\int_\Omega\big[\varphi\,\theta^b+\dd\varphi(v)\big]\,\mu_g+2\int_{S^1\cap\Omega}\varphi\,\vert\omega_+\vert\\
&\ge \textrm{Vol}_b(B_1,\varphi).
\end{split}
\end{equation}
\item[(b)] Let $f$ be a locally Lipschitz Cauchy time function, let $t_1<t_2$, and set $K:=f^{-1}([t_1,t_2])$, $B_i:=f^{-1}(t_i)\cap \textrm{dom}\, b$, $i=1,2$, and $\Omega:=f^{-1}((t_1,t_2))\cap \textrm{dom}\, b$. Let $\varphi$ be a non-negative locally Lipschitz function on an open subset of $\textrm{dom}\, b$ containing $K\cap \textrm{dom}\, b$, bounded on $K\cap \textrm{dom}\, b$, such that $\dd\varphi(v)\ge 0$ almost everywhere on $\Omega$. Then (\ref{volmon}) holds in $[0,+\infty]$.
\item[(c)] In the setting of (b), let $u$ be a locally Lipschitz function on an open set containing $K$ such that $\dd u(v)\ge 1$ almost everywhere on $\Omega$. Then for every $s\in \mathbb{R}$ the sets $A_t(s):=f^{-1}(t)\cap\{b-u\le s\}$, $t\in [t_1,t_2]$, are compact subsets of $\textrm{dom}\, b$, and
\begin{equation} \label{volentry}
\begin{split}
\textrm{Vol}_b(A_{t_2}(s))&= \textrm{Vol}_b(A_{t_1}(s))+\int_{\Omega_s}\theta^b\,\mu_g+2\int_{S^1\cap\Omega_s}\vert\omega_+\vert+\Phi(s)\\
&\ge \textrm{Vol}_b(A_{t_1}(s)),
\end{split}
\end{equation}
where $\Omega_s:=\Omega\cap\{b-u\le s\}$ and $\Phi(s)\in[0,+\infty)$, the {\em entry term}, is the right derivative at $s$ of the non-decreasing function
\[
E(s):=\int_{\Omega_s}\big(\dd u(v)-1\big)\,\mu_g ,
\]
which is locally absolutely continuous. For almost every $s$, $\Phi(s)$ is the flux of the calibrated rays through the level set $\Omega\cap\{b-u=s\}$, namely the integral of $\vert\omega\vert$ over this set (see (\ref{entryflux}) below). In particular, $t\mapsto \textrm{Vol}_b(A_t(s))$ is finite and non-decreasing on $[t_1,t_2]$.
\end{itemize}
In particular, for $\varphi=1$, in cases (a) and (b)
\[
\textrm{Vol}_b(B_2)= \textrm{Vol}_b(B_1)+\int_\Omega\theta^b\,\mu_g+2\,\textrm{Vol}_b(S^1\cap\Omega)\ge \textrm{Vol}_b(B_1),
\]
where $\textrm{Vol}_b(S^1\cap\Omega):=\int_{S^1\cap\Omega}\vert\omega_+\vert$, and, in case (b), the function $t\mapsto\textrm{Vol}_b(f^{-1}(t)\cap \textrm{dom}\, b)\in [0,+\infty]$ is non-decreasing.
\end{theorem}

The condition $\dd\varphi(v)\ge 0$ holds if $\varphi$ is non-decreasing along the $b$-calibrated rays, for instance if $\varphi=F\circ b$ with $F$ non-decreasing, since $b$ increases by the proper time along them. It also holds if $\varphi=F(b-u)$, with $u$ as in (c) and $F$ smooth and non-increasing, as then $\dd\varphi(v)=F'(b-u)\,(1-\dd u(v))\ge 0$; part (c) is the limiting case in which $F$ is the characteristic function of $(-\infty,s]$. Without a condition of this type monotonicity fails: in a static product $(\mathbb{R}\times N,-\dd t^2+h)$ with $N$ compact and $Ric_h\ge 0$, $b=t-t_0$, $\theta^b=0$, and for $\varphi=\eu^{-b}$ we get $\textrm{Vol}_b(B_2,\varphi)=\eu^{-(t_2-t_1)}\,\textrm{Vol}_b(B_1,\varphi)$ on the slices $B_i=\{t_i\}\times N$.

The equalities in (\ref{volmon}) and (\ref{volentry}) rest on the SBV regularity of $\nabla b$ (Theorem \ref{thmsbv}); without it they hold as inequalities, with $\ge$ in place of the equality sign and with $\Phi(s)$ dropped, which suffices for all the monotonicity statements of this paper. By (\ref{volentry}) the calibrated volume of the truncated slices has three sources: the expansion of the calibrated congruence, its creases, and the calibrated rays which enter the set $\{b-u\le s\}$ between the two slices, as $b-u$ is non-increasing along them. Only the first can be negative, and the TCC excludes it. The TCC enters the theorem only through the sign of $\theta^b$: the identity in (\ref{volentry}), with $\Phi(s)\ge0$, holds without it, and so does that in (\ref{volmon}) for the tubes of (a). Examples in which only the first, or only the third, source is present are given in Section \ref{secfurther}.

\begin{proof}
(a). The equality in (\ref{volmon}) is the balance law (\ref{volbal}). Under the TCC $\theta^b\ge 0$ almost everywhere, so the integrand $\varphi\,\theta^b+\dd\varphi(v)$ is non-negative, which gives the inequality.

(b). The set $K$ is compact, being a closed subset of $J^+(f^{-1}(t_1))\cap J^-(f^{-1}(t_2))$, where the level sets of $f$ are compact Cauchy hypersurfaces. Let $C:=\sup_{K\cap \textrm{dom}\, b}\varphi$. Fix a smooth non-increasing function $F:\mathbb{R}\to [0,1]$ such that $F=1$ on $(-\infty,0]$, $F>0$ on $(-\infty,1)$, $F=0$ on $[1,+\infty)$ and $\vert F'\vert\le 2$, and for $c\in \mathbb{R}$ let $\varphi_c:=F(b-c)\,\varphi$ and $\Omega_c:=\Omega\cap \{b<c+1\}$. As $b:M\to (-\infty,+\infty]$ is continuous \cite[Thm.\ 5.12(iv)]{minguzzi26e}, $\overline{\Omega_c}\subset K\cap \{b\le c+1\}$ is a compact subset of $\textrm{dom}\, b$, and $\varphi_c$ is non-negative and Lipschitz on a neighborhood of $\overline{\Omega_c}$. By (\ref{fcpb}), $\{b<c+1\}$ is an open past set, so its boundary, which is contained in $\{b=c+1\}$, is a locally Lipschitz hypersurface \cite[Prop.\ 6.3.1]{hawking73}. Thus the proof of Theorem \ref{thmvol} applies to $\Omega_c$ and to the weight $\varphi_c+\delta$, $\delta>0$, which is positive, the only difference being that, besides subsets of the bases, $\p\Omega_c$ contains a subset of $\{b=c+1\}$, of finite $\mathcal{H}^n$ measure, whose contribution to the area functional \cite[Eq.\ (26)]{minguzzi14d} is proportional to $\delta$, as $\varphi_c$ vanishes there. All terms being finite and affine in $\delta$, letting $\delta\to 0$ we get (\ref{volin2}) for $\Omega_c$ and $\varphi_c$, with equality by Theorem \ref{thmsbv}, as $\overline{\Omega_c}$ is a compact subset of $\textrm{dom}\, b$. Since $\varphi_c$ vanishes on $\textrm{dom}\, b\cap \{b\ge c+1\}$ and, almost everywhere, $\dd\varphi_c(v)=F(b-c)\,\dd\varphi(v)+F'(b-c)\,\varphi$, as $\dd b(v)=-g(\nabla b,\nabla b)=1$ \cite[Lemma 5.16]{minguzzi26e}, it reads
\begin{equation} \label{volc}
\begin{split}
\textrm{Vol}_b(B_2,\varphi_c)={}& \textrm{Vol}_b(B_1,\varphi_c)+\int_\Omega F(b-c)\big[\varphi\,\theta^b+\dd\varphi(v)\big]\,\mu_g\\
&+2\int_{S^1\cap\Omega}\varphi_c\,\vert\omega_+\vert+\int_\Omega F'(b-c)\,\varphi\,\mu_g ,
\end{split}
\end{equation}
all terms being finite. As $c\to +\infty$, $F(b-c)$ increases to $1$ on $\textrm{dom}\, b$, so, the integrands being non-negative, by monotone convergence each term but the last converges to the corresponding term of (\ref{volmon}). The last term is bounded in absolute value by $2C\,\mu_g(K\cap\{c<b<c+1\})$, which tends to zero, since $K\cap\{c<b<c+1\}\subset K\cap\{c<b<+\infty\}$, $\mu_g(K)<+\infty$, and the latter sets decrease to the empty set. This proves the equality in (\ref{volmon}), in $[0,+\infty]$, and the inequality follows as in (a).

(c). As $u$ is continuous on $K$ and $b:M\to (-\infty,+\infty]$ is continuous, $K\cap\{b-u\le s\}$ is a closed subset of $K$, hence compact, and it is contained in $\textrm{dom}\, b$. Let $G:\mathbb{R}\to[0,1]$ be Lipschitz and non-increasing, with $G=1$ on $(-\infty,0]$ and $G=0$ on $[1,+\infty)$, and for $\epsilon\in(0,1]$ let $\varphi_\epsilon:=G((b-u-s)/\epsilon)$. It is non-negative, bounded and locally Lipschitz on the intersection of $\textrm{dom}\, b$ with the domain of $u$, and $\dd\varphi_\epsilon(v)=\epsilon^{-1}G'((b-u-s)/\epsilon)\,(1-\dd u(v))\ge 0$ almost everywhere on $\Omega$, as $\dd b(v)=1$. Since $0\le\varphi_\epsilon\le \varphi_1$ and $\varphi_1$ vanishes outside the compact subset $K\cap\{b-u\le s+1\}$ of $\textrm{dom}\, b$, for $c\ge s+1+\max_Ku$ we have $F(b-c)\,\varphi_\epsilon=\varphi_\epsilon$ on $K\cap\textrm{dom}\, b$, with $F$ as in the proof of (b). Thus (\ref{volc}), which does not use the TCC nor the sign of $\dd\varphi(v)$, reads
\begin{equation} \label{voleps}
\textrm{Vol}_b(B_2,\varphi_\epsilon)= \textrm{Vol}_b(B_1,\varphi_\epsilon)+\int_\Omega\varphi_\epsilon\,\theta^b\,\mu_g+2\int_{S^1\cap\Omega}\varphi_\epsilon\,\vert\omega_+\vert+\int_\Omega\dd\varphi_\epsilon(v)\,\mu_g ,
\end{equation}
all terms being finite. As $\epsilon\to 0$, $\varphi_\epsilon$ decreases to the characteristic function of $\{b-u\le s\}$, so the first four terms converge to the corresponding terms of (\ref{volentry}), by monotone convergence and, for the integral of $\varphi_\epsilon\theta^b$, by dominated convergence, as $\theta^b$ is integrable on the compact subsets of $\textrm{dom}\, b$. Hence the limit
\[
\Phi(s):=\lim_{\epsilon\to0}\int_\Omega\dd\varphi_\epsilon(v)\,\mu_g\in[0,+\infty)
\]
exists, it does not depend on $G$, and the identity in (\ref{volentry}) holds. Its right-hand side is at least $\textrm{Vol}_b(A_{t_1}(s))$ since $\theta^b\ge0$ under the TCC. The last statement follows by applying this to $t<t'$ in $[t_1,t_2]$.

It remains to identify $\Phi(s)$. Let $k:=b-u$, so that $\dd u(v)-1=-\dd k(v)\ge0$ almost everywhere on $\Omega$. Choose $G(r)=1-r$ on $[0,1]$. As $\dd k=0$ almost everywhere on the level sets of the locally Lipschitz function $k$, $\dd\varphi_\epsilon(v)=\epsilon^{-1}(\dd u(v)-1)$ almost everywhere on $\Omega\cap\{s<k\le s+\epsilon\}$, and it vanishes almost everywhere on the complement, so that $\int_\Omega\dd\varphi_\epsilon(v)\,\mu_g=[E(s+\epsilon)-E(s)]/\epsilon$. Thus $\Phi(s)$ is the right derivative of $E$ at $s$; the function $E$ is finite, since $\Omega_s$ is relatively compact in $\textrm{dom}\, b$ and $\dd u(v)$ is locally bounded, and non-decreasing. Let $h$ be the auxiliary Riemannian metric of Section \ref{secdict}. At the points where $b$ and $k$ are differentiable and $\dd k\ne0$ let $\rho_k:=\vert\omega(e_1,\dots,e_n)\vert$, where $e_1,\dots,e_n$ is an $h$-orthonormal basis of $\ker\dd k$. If $\nu$ is an $h$-unit vector $h$-orthogonal to $\ker \dd k$, then $\vert\dd k(v)\vert=\vert\dd k\vert_h\,\vert h(v,\nu)\vert$ and $\rho_k\,\mu_h=\vert h(v,\nu)\vert\,\mu_g$, where $\mu_h$ is the volume form of $h$, so that $\vert\dd k(v)\vert\,\mu_g=\rho_k\,\vert\dd k\vert_h\,\mu_h$. By the coarea formula \cite[Sect.\ 3.4]{evans92}, for $s_1<s_2$,
\begin{equation} \label{entryflux}
E(s_2)-E(s_1)=\int_{\Omega\cap\{s_1<k\le s_2\}}\vert\dd k(v)\vert\,\mu_g=\int_{s_1}^{s_2}\Big(\int_{\Omega\cap\{k=r\}}\rho_k\,\dd\mathcal{H}^n\Big)\dd r ,
\end{equation}
where, again by the coarea formula, for almost every $r$ the function $\rho_k$ is defined $\mathcal{H}^n$-almost everywhere on the level set $\Omega\cap\{k=r\}$. Thus $E$ is locally absolutely continuous, and $\Phi(s)=\int_{\Omega\cap\{k=s\}}\rho_k\,\dd\mathcal{H}^n$ for almost every $s$. The function $\rho_k$ is the density, with respect to $\mathcal{H}^n$, of the measure induced by $\omega$ on the hyperplanes $\ker\dd k$, to be compared with $\rho_B$ in Section \ref{secdict}.
\end{proof}

Two choices of $u$ in Theorem \ref{corvol}(c) are natural. The first is the signed distance $d_\Sigma$, which is locally Lipschitz on $M\backslash \Sigma$, and on $M$ if $\Sigma$ is a $C^{1,1}$ spacelike hypersurface \cite[Prop.\ 2.10]{minguzzi26e}.

If $d_\Sigma$ is locally Lipschitz on a neighborhood of $K$, which is the case if $K\cap\Sigma=\emptyset$ or if $\Sigma$ is a $C^{1,1}$ spacelike hypersurface, then $u=d_\Sigma$ satisfies the assumptions of Theorem \ref{corvol}(c). Indeed, $d_\Sigma$ is rushing, $d_\Sigma(q)-d_\Sigma(p)\ge d(p,q)$ for $p\le q$ \cite[Prop.\ 2.4]{minguzzi19d}, so if $p\in \Omega$ is a differentiability point of $b$ and $d_\Sigma$ and $\alpha$ is the proper time parametrized calibrated ray starting from $p$, then $d_\Sigma(\alpha(t))-d_\Sigma(p)\ge t$, which gives $\dd d_\Sigma(v)\ge 1$. In this case $b-d_\Sigma\ge 0$ by (\ref{cmlr}), with equality along the future $\Sigma$-rays \cite[Prop.\ 3.5]{minguzzi26e}, and the compact sets $f^{-1}(t)\cap \{b-d_\Sigma\le s\}$ exhaust $f^{-1}(t)\cap \textrm{dom}\, b$ as $s\to +\infty$. The second choice is $u=f/m$, where $f$ is a $C^1$ Cauchy temporal function and $m>0$ is such that $-g(\nabla f,\nabla f)\ge m^2$ on $K$, namely $f/m$ is steep on $K$ \cite[Thm.\ 1.23, Def.\ 1.24]{minguzzi18b}: by the reverse Cauchy-Schwarz inequality, $\dd f(v)\ge \sqrt{-g(\nabla f,\nabla f)}\ge m$ on $K$. Every globally hyperbolic spacetime admits smooth steep Cauchy temporal functions \cite[Thm.\ 2.4]{minguzzi19d}, for which $m=1$ and $A_t(s)=f^{-1}(t)\cap\{b\le s+t\}$. As the rushing functions are precisely the isotone almost everywhere steep functions \cite[Thm.\ 1.28]{minguzzi18b}, the first choice is a non-smooth analog of the second, adapted to $\Sigma$.

\subsection{Weights which decrease along the calibrated rays}

The proof of Theorem \ref{corvol}(a),(b) uses the TCC and the condition $\dd\varphi(v)\ge0$ only through the inequality $\varphi\,\theta^b+\dd\varphi(v)\ge0$ almost everywhere on $\Omega$. For a positive weight $\varphi=\eu^\psi$ this inequality reads $\theta^b+\dd\psi(v)\ge0$, and it can hold for weights which decrease along the calibrated rays, provided the curvature compensates. By Corollary \ref{cordiv}, the current $\eu^\psi v$ has divergence $\eu^\psi(\theta^b+\dd\psi(v))\,\mu_g+2\eu^\psi\vert\omega_+\vert\llcorner S^1$, so the inequality states the non-negativity of the production of the weighted calibrated volume. The weight is a datum: a weighted curvature condition is not by itself a thermodynamic inequality, and a physical interpretation requires a model which supplies $\psi$. The relevant curvature is the weighted Ricci curvature of Bakry--\'Emery type, and the relevant completeness is the $\epsilon$-completeness of Lu, Minguzzi and Ohta \cite{lu21,lu22}, which we now recall in the Lorentzian case. We do not assume that $\psi$ is $C^2$, since the weights of Section \ref{secdefw}, which are functions of the difference between $b$ and the cosmological time, are not.

Let $\psi$ be a locally Lipschitz function on $\textrm{dom}\, b$. We say that $\psi$ is {\em semi-concave along the calibrated rays} if for every proper time parametrized $b$-calibrated ray $\alpha:[0,+\infty)\to\textrm{dom}\, b$ and every $T>0$ there is $C\ge0$ such that $t\mapsto\psi(\alpha(t))-Ct^2$ is concave on $[0,T]$. This is the case if $\psi$ is $C^2$ and, more generally, if $\psi$ is locally semi-concave, as $\alpha$ is a smooth curve. Let $\psi_\alpha:=\psi\circ\alpha$. Then $\psi_\alpha$ has a right derivative $\psi_\alpha'(t)$ at every $t\ge0$, and on $[0,T]$ the function $\psi_\alpha'-2Ct$ is the right derivative of a concave Lipschitz function, hence it is non-increasing and right-continuous. Thus $\psi_\alpha'$ is right-continuous and locally of bounded variation, and its distributional derivative on $(0,+\infty)$ is
\[
D\psi_\alpha'=\psi_\alpha''\,\dd t+\nu_\alpha,
\]
where $\psi_\alpha''$ is the almost everywhere defined derivative of $\psi_\alpha'$, and $\nu_\alpha\le0$ is a measure which is singular with respect to the Lebesgue measure \cite[Sect.\ 3.2]{ambrosio00}.

Let $N\in(-\infty,0]\cup[n,+\infty]$, $N\ne n$. Where $\psi$ is $C^2$, for a timelike vector $v$ let
\[
Ric_N^\psi(v):=Ric(v,v)-\textrm{Hess}\,\psi(v,v)-\frac{\dd\psi(v)^2}{N-n},
\]
$Ric_\infty^\psi(v):=Ric(v,v)-\textrm{Hess}\,\psi(v,v)$,
and $Ric_n^\psi(v):=\lim_{N\downarrow n}Ric_N^\psi(v)$, which equals $-\infty$ where $\dd\psi(v)\ne0$. This is the weighted Ricci curvature of \cite{lu21,lu22} for the weight function $-\psi$, namely for the weighted measure $\eu^{\psi}\mu_g$; with our convention $\dim M=n+1$, $N$ is the synthetic dimension of \cite{lu21,lu22}. If $\psi$ is only semi-concave along the calibrated rays, the weighted Ricci curvature is defined along them: for a proper time parametrized calibrated ray $\alpha$ and for almost every $t>0$,
\[
Ric_N^\psi(\dot\alpha)(t):=Ric(\dot\alpha,\dot\alpha)-\psi_\alpha''(t)-\frac{\psi_\alpha'(t)^2}{N-n},
\]
with the same conventions for $N=n,+\infty$. This agrees with the previous definition if $\psi$ is $C^2$, since $\textrm{Hess}\,\psi(\dot\alpha,\dot\alpha)=\psi_\alpha''$ along a geodesic. The singular part $\nu_\alpha$ of the second derivative does not enter the definition: being non-positive, it would contribute a non-negative term. Finally, if $\psi$ is locally semi-concave on $\textrm{dom}\, b$, it is twice differentiable almost everywhere, namely it has almost everywhere a second order expansion, whose quadratic term defines $\textrm{Hess}\,\psi$, and $Ric_N^\psi(v)$ is defined at these points by the first formula. The two definitions are consistent: for $v=-\nabla b$, and for the calibrated ray $\alpha$ starting from almost every point, $Ric_N^\psi(\dot\alpha)(t)$ is the value of $Ric_N^\psi(v)$ at $\alpha(t)$ for almost every $t$ (Corollary \ref{corpoint}). For $n<N<+\infty$ and $N'\le0$, $Ric_n^\psi\le Ric_N^\psi\le Ric_\infty^\psi\le Ric_{N'}^\psi\le Ric_0^\psi$. The $\epsilon$-range is \cite[Def.\ 5.3]{lu22}
\[
\epsilon=0\ \textrm{ for } N=0,\qquad \vert\epsilon\vert<\sqrt{\frac{N}{N-n}}\ \textrm{ for } N\ne0,n,\qquad \epsilon\in\mathbb{R}\ \textrm{ for } N=n,
\]
with the constant $c=c(N,\epsilon):=\frac1n\big(1-\epsilon^2\frac{N-n}{N}\big)>0$, $c(0,0):=1/n$. Note that $\epsilon=0$ is admissible for every $N$, while $\epsilon=1$ is admissible exactly for $N\in[n,+\infty)$. Along a proper time parametrized timelike geodesic $\alpha$, the $\epsilon$-proper time is
\[
\phi_\alpha(t):=\int_0^t\eu^{2(1-\epsilon)\psi(\alpha(t'))/n}\,\dd t' .
\]

\begin{lemma}[Weighted Raychaudhuri inequality \cite{lu21}] \label{lemwray}
Let $X$ be a $C^1$ unit timelike geodesic vector field without rotation on a neighborhood of a segment of an integral curve $\alpha$, with expansion $\theta=\nabla\cdot X$, and let $\psi$ be $C^2$ on this neighborhood. Let $\Theta:=\theta+(\psi\circ\alpha)'$ and $\theta_\epsilon:=\eu^{-2(1-\epsilon)\psi/n}\,\Theta$ along $\alpha$. For $N\in(-\infty,0]\cup[n,+\infty]$ and $\epsilon$ in the $\epsilon$-range,
\[
\frac{\dd\theta_\epsilon}{\dd\phi_\alpha}\le-\eu^{-4(1-\epsilon)\psi/n}\,Ric_N^\psi(\dot\alpha)-c\,\theta_\epsilon^2 .
\]
\end{lemma}

\begin{proof}
This is \cite[Prop.\ 5.7]{lu21} (see also \cite[Thm.\ 5.6]{lu22}); we give the short computation in the Lorentzian case. Write $\psi'$ for $(\psi\circ\alpha)'$ and $E:=\eu^{-2(1-\epsilon)\psi/n}$, so that $\dd\phi_\alpha/\dd t=E^{-1}$. The Raychaudhuri equation gives $\theta'\le-\theta^2/n-Ric(\dot\alpha,\dot\alpha)$, and
\[
\frac{\dd\theta_\epsilon}{\dd\phi_\alpha}=E^2\Big[\Theta'-\frac{2(1-\epsilon)}n\psi'\Theta\Big]\le E^2\Big[-\frac{\Theta^2}n+\frac{2\epsilon}n\psi'\Theta-\frac{\psi'^2}n+\psi''-Ric(\dot\alpha,\dot\alpha)\Big]
\]
after substituting $\theta=\Theta-\psi'$. For $N\ne0,n$, using $\psi''-Ric(\dot\alpha,\dot\alpha)=-Ric_N^\psi(\dot\alpha)-\psi'^2/(N-n)$, the bracket equals
\[
-Ric_N^\psi(\dot\alpha)-c\,\Theta^2-\frac1n\Big(\epsilon\sqrt{\tfrac{N-n}{N}}\,\Theta-\sqrt{\tfrac{N}{N-n}}\,\psi'\Big)^2\le-Ric_N^\psi(\dot\alpha)-c\,\Theta^2 ,
\]
where $(N-n)/N>0$. For $N=0$ and $\epsilon=0$ the bracket is $-Ric_0^\psi(\dot\alpha)-\Theta^2/n$, and the case $N=n$ follows by the limit $N\downarrow n$. Multiplying by $E^2$ gives the claim.
\end{proof}

The computation is pointwise. Hence, if $\psi$ is only semi-concave along the calibrated rays and $\alpha$ is a calibrated ray, the inequality between the two brackets, and the bound of the second one by $-Ric_N^\psi(\dot\alpha)-c\,\Theta^2$, hold at every $t$ at which the right derivative $\psi_\alpha'$ is differentiable, with $\psi'=\psi_\alpha'(t)$, $\psi''=\psi_\alpha''(t)$ and $\Theta'=\theta'+\psi_\alpha''(t)$.

\begin{proposition} \label{propBE}
Assume that $(M,g)$ is future timelike complete, let $\psi$ be a locally Lipschitz function on $\textrm{dom}\, b$ which is semi-concave along the calibrated rays, for instance $\psi\in C^2(\textrm{dom}\, b)$, and let $N\in(-\infty,0]\cup[n,+\infty]$ and $\epsilon$ in the $\epsilon$-range. Assume that for almost every $p\in\textrm{dom}\, b$ the proper time parametrized $b$-calibrated ray $\alpha$ starting from $p$ satisfies
\begin{itemize}
\item[(i)] $Ric_N^\psi(\dot\alpha)(t)\ge0$ for almost every $t>0$, and
\item[(ii)] $\alpha$ is future $\epsilon$-complete: $\phi_\alpha(t)\to+\infty$ as $t\to+\infty$.
\end{itemize}
Then $\theta^b+\dd\psi(v)\ge0$ almost everywhere on $\textrm{dom}\, b$.
\end{proposition}

\begin{proof}
Let $\Box_\psi:=\Box+g(\nabla\psi,\nabla\,\cdot\,)$, so that $\Box_\psi b=-\theta^b-\dd\psi(v)$ at the points at which $b$ is twice differentiable and $\psi$ is differentiable, by Lemma \ref{lemdict}(c). Let $p$ be such a point, and assume that the calibrated ray $\alpha$ starting from $p$ satisfies (i) and (ii); by Rademacher's theorem almost every point of $\textrm{dom}\, b$ has these properties. Let $s>0$, $r:=d(\cdot,\alpha(s))$ and $b_{p,s}=b(p)+s-r$, the smooth upper support function of $b$ at $p$ of Section \ref{secprel}. Both $b$ and $b_{p,s}$ have gradient $-\dot\alpha(0)$ at $p$, and $\textrm{Hess}\,b(p)\le\textrm{Hess}\,b_{p,s}(p)$ as quadratic forms. The difference $D$ of the two Hessians vanishes on $\dot\alpha(0)$, because $b\circ\alpha$ and $b_{p,s}\circ\alpha$ are affine in $t$ and $D$ is positive semi-definite, so its trace with respect to $g$ is its trace on $\dot\alpha(0)^\perp$, which is non-negative. Hence $\Box_\psi b(p)\le\Box_\psi b_{p,s}(p)$, the first order terms being equal.

Along $\alpha\vert_{[0,s)}$, $\nabla r=\dot\alpha$, and $X:=\nabla r$ is the unit, geodesic, irrotational field of the timelike geodesics which end at $\alpha(s)$, with expansion $\theta=\Box r$, which is smooth on $[0,s)$. Let $\psi_\alpha'$ be the right derivative of $\psi_\alpha=\psi\circ\alpha$, and let
\[
\Theta:=\theta+\psi_\alpha',\qquad E:=\eu^{-2(1-\epsilon)\psi_\alpha/n},\qquad y:=E\,\Theta ,
\]
so that $y$ is right-continuous and locally of bounded variation on $[0,s)$. Since $\psi$ is differentiable at $p$, $\psi_\alpha'(0)=\dd\psi(v)(p)$, hence $\Box_\psi b_{p,s}(p)=-\Theta(0)=-y(0)/E(0)$. As $E$ is locally Lipschitz, with $E'=-\frac{2(1-\epsilon)}{n}\psi_\alpha'E$ almost everywhere, the Leibniz rule for the product of a locally Lipschitz function and a function of bounded variation \cite[Sect.\ 3.1]{ambrosio00} gives, on $(0,s)$,
\[
Dy=E\Big[\theta'+\psi_\alpha''-\frac{2(1-\epsilon)}n\psi_\alpha'\Theta\Big]\dd t+E\,\nu_\alpha .
\]
By the observation which follows Lemma \ref{lemwray}, the bracket is bounded almost everywhere by $-Ric_N^\psi(\dot\alpha)-c\,\Theta^2$, so, by (i) and $\nu_\alpha\le0$, $Dy\le-c\,E\,\Theta^2\,\dd t=-c\,E^{-1}y^2\,\dd t$ as measures. Since $y$ is right-continuous, for every $t\in[0,s)$
\[
y(t)\le Y(t):=y(0)-c\int_0^tE^{-1}y^2\,\dd t' .
\]
Suppose that $y(0)<0$. Then $y\le Y\le y(0)<0$ on $[0,s)$, so $y^2\ge Y^2$, and the locally absolutely continuous function $Y$ satisfies $Y'=-c\,E^{-1}y^2\le-c\,E^{-1}Y^2$ almost everywhere. Thus $(1/Y)'\ge c\,E^{-1}$, and, as $\phi_\alpha(t)=\int_0^tE^{-1}\dd t'$,
\[
0>\frac1{Y(t)}\ge\frac1{y(0)}+c\,\phi_\alpha(t),\qquad t\in[0,s) .
\]
Letting $t\to s$ we get $y(0)\ge-1/(c\,\phi_\alpha(s))$, which holds also if $y(0)\ge0$. Therefore
\[
\Box_\psi b(p)\le\Box_\psi b_{p,s}(p)=-\frac{y(0)}{E(0)}\le\frac{\eu^{2(1-\epsilon)\psi(p)/n}}{c\,\phi_\alpha(s)} ,
\]
which tends to $0$ as $s\to+\infty$ by (ii). This proves $\theta^b+\dd\psi(v)\ge0$ at $p$.
\end{proof}

\begin{corollary}[Weighted volume monotonicity] \label{corBE}
Assume that $(M,g)$ is future timelike complete, and let $\psi$ be a locally Lipschitz function on $\textrm{dom}\, b$ such that $\theta^b+\dd\psi(v)\ge0$ almost everywhere on $\textrm{dom}\, b$, for instance under the assumptions of Proposition \ref{propBE}. The following statements hold without the TCC.
\begin{itemize}
\item[(a)] Theorem \ref{corvol}(a),(b) hold for the weights $\varphi=\chi\,\eu^\psi$, where $\chi$ is a non-negative locally Lipschitz function, with the domain required there for $\varphi$, such that $\dd\chi(v)\ge0$ almost everywhere on $\Omega$, provided $\varphi$ is bounded on $K\cap\textrm{dom}\, b$ in case (b). In particular, they hold for $\varphi=\eu^\psi$.
\item[(b)] In the setting of Theorem \ref{corvol}(c), for every $s\in\mathbb{R}$
\begin{align*}
\textrm{Vol}_b(A_{t_2}(s),\eu^\psi)\ge{}& \textrm{Vol}_b(A_{t_1}(s),\eu^\psi)+\int_{\Omega_s}\eu^\psi\big[\theta^b+\dd\psi(v)\big]\mu_g\\
&+2\int_{S^1\cap\Omega_s}\eu^\psi\vert\omega_+\vert\ \ge\ \textrm{Vol}_b(A_{t_1}(s),\eu^\psi),
\end{align*}
all terms being finite. In particular, $t\mapsto\textrm{Vol}_b(A_t(s),\eu^\psi)$ is finite and non-decreasing on $[t_1,t_2]$.
\end{itemize}
\end{corollary}

\begin{proof}
(a). Almost everywhere on $\Omega$, $\varphi\,\theta^b+\dd\varphi(v)=\chi\,\eu^\psi\,[\theta^b+\dd\psi(v)]+\eu^\psi\,\dd\chi(v)\ge0$, and, as observed at the beginning of this subsection, this is the only property of the integrand used in the proofs of Theorem \ref{corvol}(a),(b).

(b). Let $\varphi_\epsilon$ be the functions of the proof of Theorem \ref{corvol}(c), for which $\dd\varphi_\epsilon(v)\ge0$ almost everywhere on $\Omega$. The weight $\varphi_\epsilon\,\eu^\psi$ is bounded on $K\cap\textrm{dom}\, b$, since $\varphi_\epsilon$ vanishes outside the compact subset $K\cap\{b-u\le s+1\}$ of $\textrm{dom}\, b$. By (a), (\ref{volmon}) holds for it, and dropping the non-negative term $\int_\Omega\eu^\psi\,\dd\varphi_\epsilon(v)\,\mu_g$,
\[
\textrm{Vol}_b(B_2,\varphi_\epsilon\eu^\psi)\ge \textrm{Vol}_b(B_1,\varphi_\epsilon\eu^\psi)+\int_\Omega\varphi_\epsilon\,\eu^\psi\big[\theta^b+\dd\psi(v)\big]\mu_g+2\int_{S^1\cap\Omega}\varphi_\epsilon\,\eu^\psi\vert\omega_+\vert .
\]
All the integrands are non-negative, and the claim follows for $\epsilon\to0$ by monotone convergence, as in the proof of Theorem \ref{corvol}(c).
\end{proof}

The monotone quantity, the $\eu^\psi$-weighted calibrated volume, does not depend on $N$ and $\epsilon$, which only determine the sufficient conditions, and the hypotheses can be traded against each other along the $\epsilon$-range.
\begin{itemize}
\item For $N\in(n,+\infty)$ the value $\epsilon=1$ is admissible, $\phi_\alpha(t)=t$, and (ii) is the future completeness of the calibrated rays, which always holds: the only assumption is $Ric_N^\psi(\dot\alpha)=Ric(\dot\alpha,\dot\alpha)-\psi_\alpha''-(\psi_\alpha')^2/(N-n)\ge0$.
\item For $N=+\infty$, $\vert\epsilon\vert<1$, the curvature condition is $Ric(\dot\alpha,\dot\alpha)\ge\psi_\alpha''$, and (ii) holds with $\epsilon<1$ if $\psi$ is bounded from below along the calibrated rays, the setting of \cite{case10} and of the $\psi$-completeness ($\epsilon=0$) of \cite{woolgar16,woolgar18}.
\item For $N\le0$ the curvature condition is weaker, and it is weakest for $N=0$, $Ric_0^\psi(v)=Ric(v,v)-\textrm{Hess}\,\psi(v,v)+\frac1n\dd\psi(v)^2$, but then $\epsilon=0$ is forced and (ii) reads $\int_0^{+\infty}\eu^{2\psi(\alpha(t))/n}\dd t=+\infty$.
\item For $\psi=0$ all the conditions reduce to the TCC in the calibrated directions, and Proposition \ref{propBE} and Corollary \ref{corBE} contain the case $\varphi=1$ of Theorem \ref{corvol}.
\item If $\psi$ is non-increasing along the calibrated rays, $\dd\psi(v)\le0$, the conclusion $\theta^b\ge-\dd\psi(v)\ge0$ of Proposition \ref{propBE} is stronger than the inequality $\theta^b\ge0$ which follows from the TCC. In this case Theorem \ref{corvol}, whose proof uses the TCC only through the latter inequality, holds as it stands, and Corollary \ref{corBE} improves it: the calibrated volume increases at least as much as is needed to compensate the decrease of the weight.
\item The second derivative of $\psi$ along the rays enters only through its absolutely continuous part. A concave kink of $\psi_\alpha$, at which $\psi_\alpha'$ jumps downwards, makes the weighted expansion $\Theta$ jump downwards, which is compatible with the Riccati comparison. A convex kink is not: this is the reason for the semi-concavity assumption.
\end{itemize}
The hypotheses cannot be dropped: in the static product considered after Theorem \ref{corvol}, with $\psi=-b$, $Ric_N^\psi(\dot\alpha)=1/(n-N)$ is negative for $N\in[n,+\infty)$, while for $N\le0$ and $N=+\infty$ it is non-negative but (ii) fails, as $\epsilon<1$ in these cases; and indeed the weighted volume decreases.

For $(N,\epsilon)=(0,0)$ and $\psi$ of class $C^2$ Proposition \ref{propBE} has a transparent meaning in terms of the product trick. Since $\psi$ does not depend on $z$, the conformal metric $\hat g:=\eu^{2\psi/n}(g+\dd z^2)$ on $\tilde M$ has the same null geodesics as $\tilde g$, up to parametrization, so $H$ is a horizon for $\hat g$ with the same generators, and the $n$-dimensional areas of its sections are multiplied by $\eu^\psi$: by the dictionary of Section \ref{secdict}, the $\hat g$-area of a section of $H$ is the $\eu^\psi$-weighted calibrated volume of its projection. For the generator $k=(\dot\alpha,1)$, the transformation law of the Ricci tensor under a conformal change in dimension $n+2$ gives $\widehat{Ric}(k,k)=Ric_0^\psi(\dot\alpha)$, and $\phi_\alpha$ is an affine parameter of the generators for $\hat g$. Thus (i) is the null convergence condition of $\hat g$ along the generators of $H$ and (ii) is their future completeness for $\hat g$, and Proposition \ref{propBE}, with Corollary \ref{corBE}, is the area theorem for $H$ in the metric $\hat g$. The other values of $N$ and $\epsilon$ have no such interpretation. If $\psi$ is only locally Lipschitz and semi-concave along the calibrated rays, the metric $\hat g$ is only locally Lipschitz, and the area theorem for $\hat g$ is not available in this form; Proposition \ref{propBE} replaces it.

In spacetimes with an initial singularity, the weights $F(w)$, $w=b^{\textnormal{\textsc{m}}}-\tau$, of Section \ref{seccanvol} with $F$ increasing decrease along the calibrated rays, as $w$ does. The function $w$ is not $C^2$, but these weights are semi-concave along the calibrated rays, and Proposition \ref{propBE} applies to them with $(N,\epsilon)=(0,0)$ (Theorem \ref{thmdefw}).

\subsection{Whole slices and rigidity}

Once the slices leave $\textrm{dom}\, b$ their calibrated volume becomes infinite, as we now show.

\begin{proposition} \label{propinfvol}
Assume that $(M,g)$ is future timelike complete, and let $f$ be a smooth Cauchy temporal function such that $f^{-1}(t_0)\not\subset \textrm{dom}\, b$ for some $t_0\in\mathbb{R}$. Then for every $t>t_0$ there are $c>0$ and $s_0\in\mathbb{R}$ such that $\textrm{Vol}_b(f^{-1}(t)\cap \{b\le s\})\ge c\,(s-s_0)$ for $s\ge s_0$. In particular, $\textrm{Vol}_b(f^{-1}(t)\cap \textrm{dom}\, b)=+\infty$.
\end{proposition}

\begin{proof}
Let $t>t_0$, let $Y:=f^{-1}(t)$, a smooth compact spacelike hypersurface, connected since $M$ is homeomorphic to $\mathbb{R}\times Y$, with induced Riemannian metric $h_Y$, volume measure $\mu_Y$ and future unit normal $N$, and let $B:=Y\cap \textrm{dom}\, b$. Let $p\in f^{-1}(t_0)\backslash \textrm{dom}\, b$. As $\textrm{dom}\, b$ is an open past set, $I^+(p)\cap \textrm{dom}\, b=\emptyset$, so $Y\backslash B$ contains the non-empty open set $I^+(p)\cap Y$ and $\lambda_1:=\mu_Y(Y\backslash B)>0$. Moreover $B\ne\emptyset$, since the past set $\textrm{dom}\, b$ contains the future $\Sigma$-rays \cite[Prop.\ 3.5]{minguzzi26e}, which meet $Y$ or lie in its future. For $\mathcal{H}^n$-almost every $x\in B$ we have $x\notin S$, as $Y$ is achronal \cite[Thm.\ 5.21(c)]{minguzzi26e}, and at such points the calibrated density is $-g(v,N)=:\gamma$, while the $h_Y$-gradient $\nabla^Y b$ of $b\vert_B$ is the projection $-v+\gamma N$ of $\nabla b=-v$ on $T_xY$, so that $\vert\nabla^Y b\vert^2=\gamma^2-1$. Thus
\[
\textrm{Vol}_b(B)=\int_B\gamma\,\mu_Y\ge \int_B\vert\nabla^Y b\vert\,\mu_Y .
\]
Let $m_0:=\min_Y b$, which is finite as $b:M\to(-\infty,+\infty]$ is continuous and $B\ne \emptyset$, and let $\lambda_0:=\mu_Y(Y\cap\{b<m_0+1\})>0$. For an integer $k\ge 1$ let $w_k:=\min\{\max\{b-m_0-k,0\},1\}$ on $B$ and $w_k:=1$ on $Y\backslash B$. As $Y\cap\{b>m_0+k+1\}$ is an open neighborhood of $Y\backslash B$ in $Y$, $w_k$ is Lipschitz on $Y$, and $\vert\nabla^Yw_k\vert\le \vert \nabla^Yb\vert$ almost everywhere on $B_k:=B\cap\{m_0+k\le b\le m_0+k+1\}$, while $\nabla^Yw_k=0$ almost everywhere on $Y\backslash B_k$. Since $w_k=0$ on $Y\cap\{b<m_0+1\}$ and $w_k=1$ on $Y\backslash B$, for every constant $c$ we have $\int_Y\vert w_k-c\vert\,\mu_Y\ge \lambda_0\vert c\vert+\lambda_1\vert 1-c\vert\ge \min\{\lambda_0,\lambda_1\}$. By the Poincar\'e inequality on the compact connected Riemannian manifold $(Y,h_Y)$, there is $C>0$, independent of $k$, such that $\int_Y\vert w_k-\bar w_k\vert\,\mu_Y\le C\int_Y\vert\nabla^Yw_k\vert\,\mu_Y$, where $\bar w_k$ is the mean value of $w_k$. Therefore $\int_{B_k}\vert \nabla^Yb\vert\,\mu_Y\ge \min\{\lambda_0,\lambda_1\}/C$ for every $k\ge 1$. Since every point of $B$ belongs to at most two sets $B_k$, for every integer $K\ge 1$
\[
\textrm{Vol}_b(Y\cap\{b\le m_0+K+1\})\ge \int_{Y\cap\{b\le m_0+K+1\}}\vert\nabla^Y b\vert\,\mu_Y\ge \frac{K}{2C}\min\{\lambda_0,\lambda_1\},
\]
which gives the claim with $c:=\min\{\lambda_0,\lambda_1\}/(2C)$ and $s_0:=m_0+2$.
\end{proof}

The divergence can be compensated by weights which vanish at $\p\,\textrm{dom}\, b$, provided they are non-decreasing along the calibrated rays. With $u$ as in Theorem \ref{corvol}(c), the weights $F(b-u)$, with $F$ non-increasing and $F(s)\to 0$ for $s\to+\infty$, have both properties, since $b\to+\infty$ at $\p\,\textrm{dom}\, b$ while $u$ is locally bounded.

\begin{corollary}[Weighted calibrated volume of the slices] \label{corweight}
Assume the TCC and that $(M,g)$ is future timelike complete. Let $f$ be a locally Lipschitz Cauchy time function and let $u$ be a locally Lipschitz function on $M$ such that $\dd u(v)\ge 1$ almost everywhere on $\textrm{dom}\, b$. For a smooth non-increasing function $F:\mathbb{R}\to [0,+\infty)$ such that $F(s)\to 0$ for $s\to +\infty$, and for $t\in\mathbb{R}$, let
\[
V_F(t):=\textrm{Vol}_b\big(f^{-1}(t)\cap\textrm{dom}\, b,\,F(b-u)\big)\in [0,+\infty].
\]
\begin{itemize}
\item[(a)] $V_F(t)=\int_{\mathbb{R}}(-F'(s))\,\textrm{Vol}_b(A_t(s))\,\dd s$, where $A_t(s):=f^{-1}(t)\cap\{b-u\le s\}$, and $V_F$ is non-decreasing.
\item[(b)] There is a positive such function $F$ for which $V_F(t)<+\infty$ for every $t$.
\item[(c)] If $f$ is a smooth Cauchy temporal function and $f^{-1}(t_0)\not\subset \textrm{dom}\, b$ for some $t_0\in \mathbb{R}$, then $V_F(t)<+\infty$ for some $t>t_0$ only if $\int_0^{+\infty}F(s)\,\dd s<+\infty$.
\end{itemize}
\end{corollary}

\begin{proof}
(a). As $F(s)\to 0$ for $s\to+\infty$, $F(w)=\int_w^{+\infty}(-F'(s))\,\dd s$ for every $w\in\mathbb{R}$, so that $F(b-u)=\int_{\mathbb{R}}(-F'(s))\,\chi_{\{b-u\le s\}}\,\dd s$ on $\textrm{dom}\, b$, and the identity follows from Tonelli's theorem. By Theorem \ref{corvol}(c), applied to $f^{-1}([t,t'])$ for $t<t'$, the function $t\mapsto \textrm{Vol}_b(A_t(s))$ is non-decreasing for every $s$, hence so is $V_F$.

(b). For every integer $k$ let $\tau_k:=k+1$, and let $\epsilon_k:=2^{-\vert k\vert}/(1+\textrm{Vol}_b(A_{\tau_k}(k+1)))>0$, the volume being finite by Theorem \ref{corvol}(c). Let $\psi$ be a smooth positive function such that $\psi\le \epsilon_k$ on $[k,k+1]$ for every $k$, and let $F(s):=\int_s^{+\infty}\psi(r)\,\dd r$, which is positive, smooth, decreasing and tends to zero at $+\infty$. Let $t\in\mathbb{R}$ and let $k_0$ be the smallest integer such that $k_0\ge t$. For $k\ge k_0$ and $s\in [k,k+1]$ we have $t\le \tau_k$, so that, by Theorem \ref{corvol}(c) and the inclusion $A_t(s)\subset A_t(k+1)$, $\textrm{Vol}_b(A_t(s))\le \textrm{Vol}_b(A_{\tau_k}(k+1))$, and hence $\int_k^{k+1}\psi(s)\,\textrm{Vol}_b(A_t(s))\,\dd s\le 2^{-k}$. For $s\le k_0$ we have $\textrm{Vol}_b(A_t(s))\le \textrm{Vol}_b(A_t(k_0))<+\infty$, while $\int_{\mathbb{R}}\psi\,\dd s\le \sum_k\epsilon_k<+\infty$. By (a), $V_F(t)<+\infty$.

(c). Let $t>t_0$, $Y:=f^{-1}(t)$ and $c_u:=\min_Y u$. Since $A_t(s)\supset Y\cap\{b\le s+c_u\}$, by Proposition \ref{propinfvol} there are $c>0$ and $s_1\in \mathbb{R}$ such that $\textrm{Vol}_b(A_t(s))\ge c\,(s-s_1)$ for $s\ge s_1$. By (a) and Tonelli's theorem,
\begin{align*}
V_F(t)&\ge c\int_{s_1}^{+\infty}(-F'(s))(s-s_1)\,\dd s\\
&=c\int_{s_1}^{+\infty}\!\!\int_r^{+\infty}(-F'(s))\,\dd s\,\dd r=c\int_{s_1}^{+\infty}F(r)\,\dd r ,
\end{align*}
and $F$ is bounded on $[0,+\infty)$.
\end{proof}

Thus the decay of $F$ needed for the finiteness of $V_F$ is dictated by the growth of $s\mapsto \textrm{Vol}_b(A_t(s))$, which is at least linear. This growth is governed by the areas of the sections of the horospheres by the slices, rather than by the rate at which $b$ diverges at $\p\,\textrm{dom}\, b$. Indeed, let $f$ be a smooth steep Cauchy temporal function and $u=f$, let $t\in\mathbb{R}$, $Y:=f^{-1}(t)$, and let $L_t(r):=\mathcal{H}^{n-1}(Y\cap\{b=r\})$, computed with the induced metric $h_Y$. As in the proof of Proposition \ref{propinfvol}, the calibrated density with respect to $\mu_Y$ lies between $\vert\nabla^Yb\vert$ and $1+\vert\nabla^Y b\vert$, so that, by the coarea formula \cite[Sect.\ 3.4]{evans92},
\[
\int_{\mathbb{R}}F(r-t)\,L_t(r)\,\dd r\le V_F(t)\le F\big(\min_Y b-t\big)\,\mu_Y(Y)+\int_{\mathbb{R}}F(r-t)\,L_t(r)\,\dd r .
\]
In particular, if the areas $L_t(r)$ are bounded for large $r$, then $V_F(t)$ is finite for every $F$ integrable at $+\infty$, for instance $F(s)=\eu^{-s}$. We do not know whether such a bound holds in general. In Section \ref{secdiv} we show that their essential lower limit for $r\to+\infty$ is at least the area of the section of the future cosmological horizon by the slice, to which they are expected to converge, at least in the Ces\`aro sense. An eventual upper bound on $L_t(r)$ would already give the finiteness of $V_F(t)$ for every $F$ integrable at $+\infty$, without identifying the slope.

\begin{remark}[Spacetimes with initial singularity] \label{remcosmvol}
Let $(M,g)$ be future timelike complete with regular cosmological time $\tau$, as in Section \ref{secinit}, and let $b=b^{\textnormal{\textsc{m}}}$, which is allowed as explained in Section \ref{secinit}. Then $u=\tau$ satisfies the assumptions of Theorem \ref{corvol}(c) and Corollary \ref{corweight} on the whole of $M$, with no regularity requirement on the reference hypersurfaces. Indeed, $\tau$ is locally Lipschitz \cite[Thm.\ 1.2(v)]{andersson98} and rushing, so that $\dd\tau(v)\ge 1$ almost everywhere on $\textrm{dom}\, b^{\textnormal{\textsc{m}}}$, by the argument given above for $d_\Sigma$. Here $\tau$ plays the role of $d_\Sigma$, since $d^+_{T_a}=\tau-a$ on $J^+(T_a)$ \cite[Prop.\ 5.23(a)]{minguzzi26e}: the analog of \cite[Prop.\ 2.10]{minguzzi26e} is \cite[Thm.\ 1.2(v)]{andersson98}.

The slices are the level sets $T_a=\tau^{-1}(a)$, $a>0$, which are compact Cauchy hypersurfaces \cite[Prop.\ 5.23(a)]{minguzzi26e}. As $\tau$ tends to $0$ in the past direction and to $+\infty$ in the future direction along every inextendible causal curve, $f:=\log\tau$ is a locally Lipschitz Cauchy time function with $f^{-1}(t)=T_{\eu^t}$. Fix $F$ as in Corollary \ref{corweight}. The weight $F(b^{\textnormal{\textsc{m}}}-\tau)$ is a function on $M$ which does not depend on the slice, and on $T_a$ it reduces to $F(b^{\textnormal{\textsc{m}}}-a)$, since $\tau=a$ there. If moreover the TCC holds, by Theorem \ref{corvol}(c) and Corollary \ref{corweight}, for every $s\in\mathbb{R}$ the functions
\begin{align*}
&a\mapsto \textrm{Vol}_b\big(T_a\cap\{b^{\textnormal{\textsc{m}}}\le a+s\}\big),\\
&a\mapsto \textrm{Vol}_b\big(T_a\cap\textrm{dom}\, b^{\textnormal{\textsc{m}}},\,F(b^{\textnormal{\textsc{m}}}-\tau)\big)=\int_{T_a\cap\textrm{dom}\, b^{\textnormal{\textsc{m}}}}F(b^{\textnormal{\textsc{m}}}-a)\,\rho_{T_a}\,\dd\mathcal{H}^n
\end{align*}
are non-decreasing in $a>0$; the former is finite, and the latter is finite for every $a$ for a suitable positive $F$. These quantities depend only on $(M,g)$ and on $s$ or $F$. Moreover, $b^{\textnormal{\textsc{m}}}-\tau\ge 0$ by \cite[Prop.\ 5.23(b)]{minguzzi26e}, so only the values of $F$ on $[0,+\infty)$ matter.
\end{remark}

Let $f$ be a smooth Cauchy temporal function. Since $\textrm{dom}\, b$ is an open past set, the set of $t$ such that $f^{-1}(t)\subset \textrm{dom}\, b$ is an interval $(-\infty,t_*)$, $t_*\in[-\infty,+\infty]$. For $t<t_*$ the slices are compact subsets of $\textrm{dom}\, b$, with finite calibrated volume, while for $t>t_*$ their calibrated volume is infinite by Proposition \ref{propinfvol}. Thus, for $\varphi=1$, Theorem \ref{corvol}(b) adds nothing to the case of tubes without side, except possibly at $t=t_*$, and for $t>t_*$ the monotonicity of the calibrated volume is expressed by Theorem \ref{corvol}(c), or by the weighted volumes $V_F$ of Corollary \ref{corweight}, which are finite for a suitable $F$.

\begin{theorem}[Rigidity] \label{thmvolrig}
Assume the TCC and that $(M,g)$ is future timelike complete, and let $\Omega$ be a calibrated tube such that $\textrm{Vol}_b(B_1)=\textrm{Vol}_b(B_2)$. Then $b$ is smooth on $\Omega$, $\Box b=0$, and $v=-\nabla b$ is a parallel unit timelike vector field on $\Omega$, so that $\Omega$ is locally isometric to a static product $(\mathbb{R}\times N,-\dd t^2+h)$ with $v=\p_t$. In particular no creases are present: $S\cap\Omega=\emptyset$. This applies to $f^{-1}((t_1,t_2))$ if $f$ is a locally Lipschitz Cauchy time function, $t_1<t_2$, $f^{-1}([t_1,t_2])\subset \textrm{dom}\, b$ and $\textrm{Vol}_b(f^{-1}(t_1))=\textrm{Vol}_b(f^{-1}(t_2))$, as $f^{-1}((t_1,t_2))$ is then a calibrated tube without side.
\end{theorem}

By the balance law (\ref{volbal}), under the TCC the assumption $\textrm{Vol}_b(B_1)=\textrm{Vol}_b(B_2)$ is equivalent to $\theta^b=0$ almost everywhere on $\Omega$ and $\mathcal{H}^n(S\cap\Omega)=0$. The proof which follows does not use Theorem \ref{thmsbv}. The analogous statement for the canonical truncated volumes, where also the entry term of Theorem \ref{corvol}(c) has to vanish, is Theorem \ref{thmcanbal}.

\begin{proof}
By Theorem \ref{corvol}(a) with $\varphi=1$ and (\ref{volin}), $\int_\Omega\theta^b\mu_g=0$, so $\theta^b=0$ almost everywhere on $\Omega$, and equality holds in (\ref{volin}), so, by \cite[Thm.\ 13]{minguzzi14d} and Lemma \ref{lemreg}(a), the Hessian measures of the local graphing functions of $H$ have no singular part on $\Psi(\Omega)$. By \cite[Thm.\ 17]{minguzzi14d}, $H$ is smooth on $\Psi(\Omega)$. Since $\p_z$ is transverse to the null hyperplanes orthogonal to the semitangents $(u,1)$, $u\in U_p$, which are the graphs $\{(X,-g(u,X))\}$ over $T_pM$, the implicit function theorem shows that $b$ is smooth on $\Omega$, and $-\Box b=\theta^b=0$ by Lemma \ref{lemdict}(c). The field $v=-\nabla b$ is a smooth unit timelike geodesic field, irrotational as a gradient, with vanishing expansion, so the Raychaudhuri equation gives $\vert\textrm{shear}\vert^2+Ric(v,v)=0$, and by the TCC the shear vanishes. Thus $\nabla v=0$: $v$ is a Killing field orthogonal to the level sets of $b$, and in coordinates $(b,x)$ adapted to the flow of $v$ the metric reads $-\dd b^2+h$, with $h$ independent of $b$.
\end{proof}

\begin{remark}
Theorem \ref{thmvolrig} is a local splitting statement which requires only future timelike completeness. Under timelike completeness and the TCC, a Cauchy hypersurface contained in $\textrm{dom}\, b^+$ already forces the global splitting \cite[Cor.\ 7.10]{minguzzi26e}, and under future timelike completeness alone it still forces a $b^+$-calibrated timelike line, possibly past incomplete \cite[Rem.\ 7.11]{minguzzi26e}, so in a counterexample to Bartnik's conjecture no level set of a Cauchy time function $f$ is contained in $\textrm{dom}\, b^+$, namely $t_*=-\infty$. If $f$ is a smooth Cauchy temporal function, the slices $f^{-1}(t)\cap \textrm{dom}\, b^+$ then have infinite calibrated volume for every $t$ (Proposition \ref{propinfvol}), and the monotonicity of the calibrated volume is expressed by the finite weighted volumes of Corollary \ref{corweight}, or by Theorem \ref{corvol}(c), for instance for the compact sets $f^{-1}(t)\cap\{b^+\le s+t\}$ if $f$ is steep, or $f^{-1}(t)\cap\{b^+-d_\Sigma\le s\}$ if $\Sigma$ is a $C^{1,1}$ spacelike hypersurface. Theorems \ref{corvol} and \ref{thmvolrig} are also of interest in spacetimes with an initial singularity, for instance for the canonical function $b^{\textnormal{\textsc{m}}}$, to which all the results of this paper apply (Section \ref{secinit}). The balance law (\ref{volbal}) shows that the creases of $b$ always act as sources of calibrated volume, that a negative expansion of the calibrated congruence is the only possible sink, and that there are no other sources. Under the TCC there are no sinks, as the calibrating semi-flow is non-contracting, $\theta^b\ge 0$ (\cite[Thm.\ 6.9(a)]{minguzzi26e} and Lemma \ref{lemdict}(c)), so the calibrated volume is non-decreasing (Theorem \ref{corvol}). Without energy conditions sinks may occur: in the warped product of \cite[Ex.\ 5.18]{minguzzi26e} with $a=1+\eu^{-t}$ over the round $3$-sphere, which satisfies the dominant energy condition, $b=t-t_0$ and $\theta^b=3\dot a/a<0$.
\end{remark}

\subsection{Elliptic regularity of the calibrating semi-flow} \label{secellreg}

The smoothness part of Theorem \ref{thmvolrig} rests on the elliptic regularity of horizons \cite[Thm.\ 17]{minguzzi14d}. In terms of $b$ the ellipticity has a transparent origin. At the points where $b$ is twice differentiable, $\nabla^2b(v,\cdot)=0$ (proof of Lemma \ref{lemdict}(c)), so that $-\Box b$ is the trace of $-\nabla^2 b$ on $v^\perp$, namely the future mean curvature of the horospheres, whose future unit normal is $v$. Thus a bound on $\Box b$ is a weak mean curvature bound for the horospheres, and $\Box b=0$ states that they are maximal, a condition governed by the elliptic maximal hypersurface equation, although $\Box$ is hyperbolic. No energy condition is needed in the following result.

\begin{proposition} \label{propellreg}
Assume that $(M,g)$ is future timelike complete. Let $O\subset \textrm{dom}\, b$ be open and assume that $b\in W^{2,1}_{loc}(O)$, namely that the distributional Hessian of $b$ has no singular part on $O$.
\begin{itemize}
\item[(a)] If $\Box b\in L^\infty_{loc}(O)$, equivalently $\theta^b\in L^\infty_{loc}(O)$, then $b\in C^{1,1}_{loc}(O)$.
\item[(b)] If $\Box b=0$ almost everywhere on $O$, equivalently $\theta^b=0$, then $b$ has on $O$ the same regularity as the metric: it is smooth, and analytic if $(M,g)$ is analytic; for metrics of class $C^l$, $l\ge 3$, it is of class $C^{l,\alpha}$ for some $\alpha\in(0,1]$. Moreover, the horospheres $B_c\cap O$ are smooth maximal spacelike hypersurfaces, orthogonal to the calibrated rays, which are the integral curves of the smooth geodesic field $v=-\nabla b$.
\end{itemize}
\end{proposition}

\begin{proof}
(a). Let $K\subset O$ be compact and contained in a coordinate chart. As $b$ is locally semi-concave \cite[Thm.\ 5.12(vi)]{minguzzi26e} and locally Lipschitz, there is $C>0$ such that, at the points of $K$ where $b$ is twice differentiable, the covariant Hessian satisfies $\nabla^2b(X,X)\le C\vert X\vert^2$, where $\vert\cdot\vert$ is the coordinate Euclidean norm; indeed, the coordinate Hessian is bounded from above and it differs from the covariant one by the term $-\Gamma^k_{ij}\p_k b$, which is bounded. At these points, which form a set of full measure, $p\notin S$ and $\nabla^2b(p)(v,\cdot)=0$. Let $e_1,\dots,e_n$ be a $g$-orthonormal basis of $v^\perp$, and let $\lambda_1,\dots,\lambda_n$ be the eigenvalues of $\nabla^2b(p)$ restricted to $v^\perp$ with respect to $g\vert_{v^\perp}$. Since $v$ ranges in a compact set of timelike vectors over $K$ \cite[Thm.\ 5.12(v)]{minguzzi26e}, the vectors $e_i$ are bounded in coordinates, uniformly on $K$, so $\lambda_i\le C'$ for some constant $C'$. Moreover $\sum_i\lambda_i=\Box b(p)\ge -\Vert \Box b\Vert_{L^\infty(K)}$, since $\Box b=-\nabla^2b(v,v)+\sum_i\nabla^2b(e_i,e_i)$. Thus $\lambda_i\ge -\Vert \Box b\Vert_{L^\infty(K)}-(n-1)C'$, so $\nabla^2b$, which vanishes on $v$, is bounded almost everywhere on $K$. As the distributional Hessian of $b$ has no singular part, it is the function given almost everywhere by the Hessian of $b$ \cite[Sect.\ 6.3, 6.4]{evans92}, which is locally bounded. Hence $\nabla b\in W^{1,\infty}_{loc}(O)$, namely $\nabla b$ is locally Lipschitz \cite[Sect.\ 4.2]{evans92}, and $b\in C^{1,1}_{loc}(O)$. Alternatively, (a) follows from the second part of \cite[Thm.\ 17]{minguzzi14d}, through Lemmas \ref{lemdict}(c) and \ref{lemreg}(a).

(b). By Lemma \ref{lemdict}(c), $\theta_H=\theta^b\circ \Psi^{-1}=0$ almost everywhere on $\Psi(O)$, a condition which does not depend on the normalization of the semitangents, and by Lemma \ref{lemreg}(a) the Hessian measures of the local graphing functions of $H$ have no singular part on $\Psi(O)$. By \cite[Thm.\ 17]{minguzzi14d} and its proof, $H$ has on $\Psi(O)$ the same regularity as the metric $\tilde g=g+\dd z^2$, which is that of $g$: it is smooth, analytic if $g$ is analytic, and of class $C^{l,\alpha}$ if $g$ is of class $C^l$, $l\ge 3$, the coordinates of the cylinders of \cite[Sect.\ 2.2]{minguzzi14d} being smooth (analytic in the analytic case). Since $\p_z$ is transverse to the null hyperplanes orthogonal to the semitangents $(u,1)$, $u\in U_p$, which are the graphs $\{(X,-g(u,X))\}$ over $T_pM$, $H$ is locally the graph of $b$ over $O$ along $\p_z$, and the implicit function theorem, which holds in each of these regularity classes, shows that $b$ has the same regularity on $O$. Then $v=-\nabla b$ is a smooth unit timelike vector field. Its integral curves are the calibrated rays: if $\alpha$ is the calibrated ray starting from $p\in O$, then, for every $t$ such that $\alpha(t)\in O$, $\alpha\vert_{[t,+\infty)}$ is the unique calibrated ray starting from $\alpha(t)$, so $\dot\alpha(t)=v(\alpha(t))$ by \cite[Lemma 5.16]{minguzzi26e}. As the calibrated rays are timelike geodesics \cite[Thm.\ 5.12(i)]{minguzzi26e}, $v$ is a geodesic vector field, $\nabla_vv=0$. Finally, the level sets of $b$ are smooth spacelike hypersurfaces orthogonal to $v$, with future mean curvature $-\Box b=0$.
\end{proof}

Under the TCC, Proposition \ref{propellreg}(b) applies to the tubes of Theorem \ref{thmvolrig}, and the Raychaudhuri equation then upgrades the maximality of the horospheres to the vanishing of their second fundamental form, namely to the local static structure.

\section{The transport form of the volume monotonicity theorem} \label{sectransp}

Theorem \ref{corvol} compares whole slices, or the bases of tubes whose sides are ruled by calibrated rays. In this section we localize it to arbitrary Borel subsets of the slices, by means of the past semi-flow of the generalized characteristics of $b$ introduced in \cite{minguzzi26f}. The result is applied to spacetimes with an initial singularity in Section \ref{secinitsing}.

\subsection{The past projection}

Assume that $(M,g)$ is future timelike complete. A {\em past characteristic} of $b$ is a locally absolutely continuous curve $x$ in $\textrm{dom}\, b$ whose velocity is, almost everywhere, a negative multiple of a vector of the convex hull $\textrm{co}\, U_x$. Fix a smooth Cauchy temporal function $\mathfrak{t}$. By \cite[Thm.\ 1.3]{minguzzi26f} there is a unique map $\Gamma:[0,+\infty)\times \textrm{dom}\, b\to\textrm{dom}\, b$, the {\em past semi-flow}, such that for every $p$ the curve $s\mapsto\Gamma_s(p)$ is a past characteristic with $\Gamma_0(p)=p$ and $\mathfrak{t}(\Gamma_s(p))=\mathfrak{t}(p)-s$; it is jointly locally Lipschitz, $\Gamma_{s+s'}=\Gamma_s\circ\Gamma_{s'}$, and two past characteristics with the same starting point have the same oriented image as long as both are defined. We shall use three consequences.
\begin{itemize}
\item[(i)] The trajectory $s\mapsto \Gamma_s(p)$ is a past inextendible causal curve, as $\mathfrak{t}\to-\infty$ along it. It is locally uniformly timelike \cite[Sect.\ 1.2]{minguzzi26f}: for every compact set $C\subset\textrm{dom}\, b$ and $s_0>0$ there is $c_0>0$ such that the Lorentzian length of the trajectory of $p\in C$ between the parameters $s<s'$ in $[0,s_0]$ is at least $c_0(s'-s)$.
\item[(ii)] If $\alpha:[0,+\infty)\to\textrm{dom}\, b$ is a $b$-calibrated ray, then for every $t>0$ the curve $\alpha\vert_{[0,t]}$, traversed backwards, is a past characteristic, since $\dot\alpha(r)\in U_{\alpha(r)}$. By uniqueness, the trajectory of $\alpha(t)$ runs backwards along $\alpha$ up to $\alpha(0)$, and then coincides with the trajectory of $\alpha(0)$.
\item[(iii)] If $y\in \textrm{dom}\, b\backslash S$, every curve starting from $y$ which, traversed backwards, is a past characteristic, is contained in the calibrated ray starting from $y$ \cite[Thm.\ 7.1(2)]{minguzzi26f}.
\end{itemize}

\begin{definition}
A locally Lipschitz Cauchy time function $f$ is {\em locally steep} if for every compact set $K\subset M$ there is $c>0$ such that $f(q)-f(p)\ge c\, d(p,q)$ for every $p,q\in K$ with $p\le q$.
\end{definition}

Every $C^1$ Cauchy temporal function $f$ is locally steep. Indeed, $K':=J^+(K)\cap J^-(K)$ is compact, so $-g(\nabla f,\nabla f)\ge m^2$ on $K'$ for some $m>0$, and by the reverse Cauchy-Schwarz inequality $f$ increases at least by $m$ times the Lorentzian length along the causal curves contained in $K'$, among which there is a maximizing geodesic from $p$ to $q$. If $(M,g)$ is future timelike complete with regular cosmological time $\tau$, then $\log\tau$ is locally steep, as $\log\tau(q)-\log\tau(p)\ge (\tau(q)-\tau(p))/\tau(q)\ge d(p,q)/\max_K\tau$. If $f$ is locally steep, $K=f^{-1}([t_1,t_2])$ and $c$ is the constant of $K$, then $u:=f/c$ satisfies the assumption of Theorem \ref{corvol}(c): at a differentiability point $p\in f^{-1}((t_1,t_2))$ of $b$ and $f$, if $\alpha$ is the calibrated ray starting from $p$, $f(\alpha(t))-f(p)\ge c\,t$ for small $t$, hence $\dd f(v)\ge c$.

\begin{lemma} \label{lempast}
Assume that $(M,g)$ is future timelike complete, let $f$ be a locally steep Cauchy time function, $t_1\in\mathbb{R}$ and $Y:=f^{-1}(t_1)$. For $p\in \textrm{dom}\, b$ with $f(p)\ge t_1$ let $\varsigma(p)\ge 0$ be the unique parameter such that $\Gamma_{\varsigma(p)}(p)\in Y$, and let $\varsigma(p):=0$ if $f(p)<t_1$. Then $\hat\pi(p):=\Gamma_{\varsigma(p)}(p)$ defines a locally Lipschitz map $\hat\pi:\textrm{dom}\, b\to\textrm{dom}\, b$ which is the identity on $\{f\le t_1\}$ and which maps $\{f\ge t_1\}\cap \textrm{dom}\, b$ onto $Y\cap\textrm{dom}\, b$. It does not depend on $\mathfrak{t}$, and it is constant along every $b$-calibrated curve contained in $\{f\ge t_1\}$. Moreover, if $x\in Y\cap\textrm{dom}\, b\backslash S$ and $t_2>t_1$, the set $\hat\pi^{-1}(x)\cap f^{-1}(t_2)$ consists of the point at which the calibrated ray starting from $x$ meets $f^{-1}(t_2)$.
\end{lemma}

We call $\pi:=\hat\pi\vert_{\{f\ge t_1\}}$ the {\em past projection} on $Y$: $\pi(p)$ is the point at which the past characteristic starting from $p$ meets $Y$.

\begin{proof}
By (i) the trajectory of $p$ is a past inextendible causal curve, so if $f(p)\ge t_1$ it meets the Cauchy hypersurface $Y$ exactly once, and $\varsigma$ is well defined. Moreover $\varsigma(p)=\mathfrak{t}(p)-\mathfrak{t}(\hat\pi(p))\le \mathfrak{t}(p)-\min_Y\mathfrak{t}$, as $Y$ is compact, so $\varsigma$ is locally bounded. Let $C\subset\textrm{dom}\, b$ be compact and let $s_0>0$ be a bound of $\varsigma$ on $C$. The function $G(s,p):=f(\Gamma_s(p))$ is Lipschitz on $[0,s_0]\times C$, with some constant $L$ with respect to an auxiliary distance on $M$, and for $p\in C$ and $s<s'$ in $[0,s_0]$, by the local steepness of $f$ on the compact set $\Gamma([0,s_0]\times C)$ and by (i),
\[
G(s,p)-G(s',p)\ge c\, d(\Gamma_{s'}(p),\Gamma_s(p))\ge \kappa\,(s'-s),\qquad \kappa:=c\,c_0>0 .
\]
Let $p,p'\in C$. If $f(p),f(p')\ge t_1$, then $G(\varsigma(p),p)=t_1=G(\varsigma(p'),p')$, so $\kappa\vert\varsigma(p)-\varsigma(p')\vert\le \vert G(\varsigma(p),p')-G(\varsigma(p'),p')\vert=\vert G(\varsigma(p),p')-G(\varsigma(p),p)\vert$, which is bounded by $L$ times the distance of $p$ and $p'$. If $f(p')<t_1\le f(p)$, then $\kappa\,\varsigma(p)\le G(0,p)-G(\varsigma(p),p)=f(p)-t_1\le f(p)-f(p')$. Thus $\varsigma$ is locally Lipschitz, and so is $\hat\pi=\Gamma(\varsigma(\cdot),\cdot)$. The independence of $\mathfrak{t}$ follows from the fact that the oriented images of the past characteristics starting from $p$ coincide. Let $\alpha$ be a calibrated curve contained in $\{f\ge t_1\}$, which we can assume to be a portion of a calibrated ray starting from $\alpha(0)$. By (ii) the trajectory of $\alpha(t)$ reaches $\alpha(0)$ without leaving $\{f\ge t_1\}$ before, as $f$ decreases along it, so $\hat\pi(\alpha(t))=\hat\pi(\alpha(0))$. Finally, let $x\in Y\cap\textrm{dom}\, b\backslash S$. The point at which the calibrated ray from $x$ meets $f^{-1}(t_2)$ is projected to $\hat\pi(x)=x$. Conversely, if $q\in f^{-1}(t_2)$ and $\hat\pi(q)=x$, the portion of the trajectory of $q$ up to $x$, traversed backwards, is a curve as in (iii), so $q$ belongs to the calibrated ray starting from $x$.
\end{proof}

The calibrated rays starting from almost every point spend almost no time in a given null set. This is used in Section \ref{secdefw} to pass from conditions which hold almost everywhere on $\textrm{dom}\, b$ to conditions which hold almost everywhere along the rays.

\begin{lemma} \label{lemnull}
Assume that $(M,g)$ is future timelike complete, and let $Z\subset\textrm{dom}\, b$ be a $\mu_g$-null set. Then for almost every $p\in\textrm{dom}\, b$ the proper time parametrized calibrated ray $\alpha$ starting from $p$ satisfies $\alpha(t)\notin Z$ for almost every $t>0$.
\end{lemma}

\begin{proof}
The function $\beta(s,q):=b(\Gamma_s(q))$ is locally Lipschitz on $[0,+\infty)\times\textrm{dom}\, b$. By (\ref{fcpb}) and (i), for every compact set $C\subset\textrm{dom}\, b$ and $s_0>0$ there is $c_0>0$ such that, for $q\in C$ and $s<s'$ in $[0,s_0]$,
\[
\beta(s,q)-\beta(s',q)\ge d(\Gamma_{s'}(q),\Gamma_s(q))\ge c_0\,(s'-s) .
\]
Let $\mathcal{D}$ be the set of the pairs $(q,t)\in\textrm{dom}\, b\times[0,+\infty)$ for which there is $\sigma=\sigma(q,t)\ge0$, necessarily unique, such that $\beta(\sigma,q)=b(q)-t$; thus $\Gamma_{\sigma(q,t)}(q)$ is the point of the trajectory of $q$ at which $b$ has decreased by $t$. If $(q_0,t_0)\in\mathcal{D}$ and $s_1>\sigma(q_0,t_0)$, then $\beta(s_1,q)<b(q)-t\le\beta(0,q)$ for $(q,t)$ near $(q_0,t_0)$, by continuity, so $(q,t)\in\mathcal{D}$ and $\sigma(q,t)<s_1$. Hence $\mathcal{D}$ is relatively open and $\sigma$ is locally bounded. Moreover, for $(q,t)$ and $(q',t')$ in a compact subset of $\mathcal{D}$, with $c_0$ the constant of a suitable compact set and of a bound $s_0$ of $\sigma$,
\begin{align*}
c_0\,\vert\sigma(q,t)-\sigma(q',t')\vert&\le\vert\beta(\sigma(q,t),q')-\beta(\sigma(q',t'),q')\vert\\
&=\vert\beta(\sigma(q,t),q')-\beta(\sigma(q,t),q)+b(q)-b(q')+t'-t\vert ,
\end{align*}
which is bounded by a constant times the distance of $(q,t)$ and $(q',t')$. Thus $\sigma$ is locally Lipschitz on $\mathcal{D}$, and so is the map $\Xi:\mathcal{D}\to\textrm{dom}\, b\times[0,+\infty)$, $\Xi(q,t):=(\Gamma_{\sigma(q,t)}(q),t)$.

Let $p\in\textrm{dom}\, b\backslash S$, let $\alpha$ be the calibrated ray starting from $p$, and let $t>0$. By (ii) the trajectory of $\alpha(t)$ runs backwards along $\alpha$ up to $p$, and $b(\alpha(t))-b(p)=t$, so $(\alpha(t),t)\in\mathcal{D}$ and $\Xi(\alpha(t),t)=(p,t)$. Therefore the set of the pairs $(p,t)$, $p\in\textrm{dom}\, b\backslash S$, $t>0$, such that the calibrated ray starting from $p$ is in $Z$ at the proper time $t$, is contained in $\Xi(\mathcal{D}\cap(Z\times(0,+\infty)))$. The set $Z\times(0,+\infty)$ is null for the product measure $\mu_g\otimes\dd t$, and locally Lipschitz maps between manifolds of the same dimension map null sets to null sets \cite[Sect.\ 2.4]{evans92}. Hence the above set of pairs is contained in a Borel null set, and the claim follows from Fubini's theorem, as $S$ is $\mu_g$-null.
\end{proof}

\begin{corollary}[Pointwise form of the weighted curvature condition] \label{corpoint}
Assume that $(M,g)$ is future timelike complete, and let $\psi$ be a locally Lipschitz and locally semi-concave function on $\textrm{dom}\, b$. For almost every $p\in\textrm{dom}\, b$ the proper time parametrized calibrated ray $\alpha$ starting from $p$ satisfies, for almost every $t>0$,
\[
\psi_\alpha'(t)=\dd\psi(v),\qquad \psi_\alpha''(t)=\textrm{Hess}\,\psi(v,v)\qquad \textrm{at } \alpha(t),
\]
where $v=-\nabla b$, so that $Ric_N^\psi(\dot\alpha)(t)$ is the value of $Ric_N^\psi(v)$ at $\alpha(t)$. In particular, if
\[
Ric_N^\psi(v)\ge0 \qquad \textrm{almost everywhere on } \textrm{dom}\, b ,
\]
then assumption (i) of Proposition \ref{propBE} holds.
\end{corollary}

\begin{proof}
Let $Z$ be the set of the points of $\textrm{dom}\, b$ at which $b$ is not differentiable or $\psi$ is not twice differentiable, which is $\mu_g$-null. By Lemma \ref{lemnull}, for almost every $p$ the calibrated ray $\alpha$ starting from $p$ satisfies $\alpha(t)\notin Z$ for almost every $t>0$. Let $t>0$ be such that $\alpha(t)\notin Z$ and the right derivative $\psi_\alpha'$ is differentiable at $t$. Then $\dot\alpha(t)=v(\alpha(t))$, and $\psi_\alpha(t+\eta)=\psi_\alpha(t)+\psi_\alpha'(t)\,\eta+\frac12\psi_\alpha''(t)\,\eta^2+o(\eta^2)$, as $\psi_\alpha$ is the integral of $\psi_\alpha'$. On the other hand, the second order expansion of $\psi$ at $\alpha(t)$, evaluated along the geodesic $\alpha$, gives an expansion of the same form with coefficients $\dd\psi(v)$ and $\textrm{Hess}\,\psi(v,v)$. Hence the coefficients coincide. For the last statement it suffices to add to $Z$ the null set on which $Ric_N^\psi(v)<0$.
\end{proof}

\subsection{The transport inequality}

\begin{theorem}[Volume monotonicity theorem, transport form] \label{thmtransp}
Assume the TCC and that $(M,g)$ is future timelike complete. Let $f$ be a locally steep Cauchy time function, $t_1<t_2$, $B_i:=f^{-1}(t_i)\cap\textrm{dom}\, b$, $i=1,2$, and let $\pi:B_2\to B_1$ be the past projection. Let $u$ be as in Theorem \ref{corvol}(c). Then for every Borel set $E\subset B_1$ and every $s\in\mathbb{R}$
\begin{equation} \label{transp}
\begin{split}
\textrm{Vol}_b\big(\pi^{-1}(E)\cap\{b-u\le s\}\big)&\ge \textrm{Vol}_b\big(E\cap\{b-u\le s\}\big),\\
\textrm{Vol}_b\big(\pi^{-1}(E)\big)&\ge \textrm{Vol}_b(E) .
\end{split}
\end{equation}
Moreover, $\pi^{-1}(E\backslash S)$ is the set $P(E)$ of the points at which the calibrated rays starting from the points of $E\backslash S$ meet $f^{-1}(t_2)$, and $\textrm{Vol}_b(E\cap S)=0$, so that
\begin{equation} \label{transp2}
\begin{split}
\textrm{Vol}_b\big(P(E)\cap\{b-u\le s\}\big)&\ge \textrm{Vol}_b\big(E\cap\{b-u\le s\}\big),\\
\textrm{Vol}_b\big(P(E)\big)&\ge \textrm{Vol}_b(E) .
\end{split}
\end{equation}
\end{theorem}

Thus the calibrated volume of every portion of a slice does not decrease when the portion is transported to a later slice by the calibrated observers. The first inequality in (\ref{transp}) is stronger, as $\pi^{-1}(E)$ also contains the points of the later slice whose past characteristic reaches $E$ at a crease; the fibres $\pi^{-1}(x)$, $x\in S$, are studied in \cite[Sect.\ 7]{minguzzi26f}. No tube with ruled sides is needed, and the sets $\pi^{-1}(E)$, $P(E)$ do not depend on any choice.

\begin{proof}
Let $K:=f^{-1}([t_1,t_2])$, $\Omega:=f^{-1}((t_1,t_2))\cap\textrm{dom}\, b$, and let $\psi$ be a non-negative bounded Lipschitz function on $M$. By Lemma \ref{lempast}, $\varphi_1:=\psi\circ\hat\pi$ is a non-negative, bounded, locally Lipschitz function on $\textrm{dom}\, b$, which coincides with $\psi$ on $B_1$ and with $\psi\circ\pi$ on $B_2$. If $p\in \Omega$ is a differentiability point of $b$ and $\varphi_1$, and $\alpha$ is the calibrated ray starting from $p$, which is contained in $\{f\ge t_1\}$, then $\varphi_1\circ\alpha$ is constant by Lemma \ref{lempast}, hence $\dd\varphi_1(v)=0$. Let $\varphi_\epsilon=G((b-u-s)/\epsilon)$ be as in the proof of Theorem \ref{corvol}(c). The function $\varphi_1\varphi_\epsilon$ is non-negative, bounded and locally Lipschitz on the intersection of $\textrm{dom}\, b$ with the domain of $u$, and $\dd(\varphi_1\varphi_\epsilon)(v)=\varphi_1\,\dd\varphi_\epsilon(v)\ge 0$ almost everywhere on $\Omega$. By Theorem \ref{corvol}(b), $\textrm{Vol}_b(B_2,\varphi_1\varphi_\epsilon)\ge\textrm{Vol}_b(B_1,\varphi_1\varphi_\epsilon)$, and letting $\epsilon\to 0$, by dominated convergence as in the proof of Theorem \ref{corvol}(c),
\[
\int_{B_2\cap\{b-u\le s\}}(\psi\circ\pi)\,\rho_{B_2}\,\dd\mathcal{H}^n\ge \int_{B_1\cap\{b-u\le s\}}\psi\,\rho_{B_1}\,\dd\mathcal{H}^n .
\]
Let $\mu_i$ be the restriction of the measure $\rho_{B_i}\mathcal{H}^n$ to $B_i\cap\{b-u\le s\}$, a finite Borel measure by Theorem \ref{corvol}(c). The inequality reads $\int\psi\,\dd(\pi_*\mu_2)\ge\int\psi\,\dd\mu_1$. If $C\subset B_1$ is closed in $M$, the functions $\psi_k:=\max\{0,1-k\,\textrm{dist}(\cdot,C)\}$ decrease to the characteristic function of $C$, so, the measures being finite, $\pi_*\mu_2(C)\ge\mu_1(C)$. As a finite Borel measure on a metric space is inner regular with respect to the closed sets, and the closed subsets of the compact set $B_1\cap\{b-u\le s\}$ are closed in $M$, $\pi_*\mu_2(E)\ge\mu_1(E)$ for every Borel set $E\subset B_1$, which is the first inequality in (\ref{transp}). The second one follows by letting $s\to+\infty$, by monotone convergence; here we can take $u=f/c$, as observed above, so that the assumption on $u$ is not restrictive. The identity $\pi^{-1}(E\backslash S)=P(E)$ is the last statement of Lemma \ref{lempast}, and $\textrm{Vol}_b(E\cap S)=0$ since $\mathcal{H}^n(B_1\cap S)=0$, so (\ref{transp2}) follows from (\ref{transp}) applied to $E\backslash S$.
\end{proof}

\begin{remark}[Weighted transport] \label{remwtransp}
The proof of Theorem \ref{thmtransp} uses the TCC only through Theorem \ref{corvol}(b),(c). Hence, if $\psi$ is a locally Lipschitz function on $\textrm{dom}\, b$ such that $\theta^b+\dd\psi(v)\ge0$ almost everywhere, for instance under the assumptions of Proposition \ref{propBE}, then (\ref{transp}) and (\ref{transp2}) hold, without the TCC, for the weighted calibrated volumes $\textrm{Vol}_b(\,\cdot\,,\eu^\psi)$ in place of $\textrm{Vol}_b$. Indeed, the weights $\varphi_1\varphi_\epsilon\,\eu^\psi$ are of the form considered in Corollary \ref{corBE}(a), with $\chi=\varphi_1\varphi_\epsilon$ and $\dd\chi(v)=\varphi_1\,\dd\varphi_\epsilon(v)\ge0$, they are bounded, since $\varphi_\epsilon$ vanishes outside a compact subset of $\textrm{dom}\, b$, and the measures $\eu^\psi\rho_{B_i}\mathcal{H}^n$ restricted to $B_i\cap\{b-u\le s\}$ are finite by Corollary \ref{corBE}(b). The rest of the proof is unchanged.
\end{remark}

\section{Canonical volumes in spacetimes with an initial singularity} \label{secinitsing}

In this section we apply the transport form of the volume monotonicity theorem to spacetimes with an initial singularity, where it gives monotone quantities and an invariant which depend only on $(M,g)$. Only the first part of Section \ref{secdiv} concerns the Cauchy Busemann function of an arbitrary compact Cauchy hypersurface.

\subsection{The deficit and the asymptotic calibrated volume} \label{seccanvol}

In this subsection $(M,g)$ is future timelike complete with regular cosmological time $\tau$, $b=b^{\textnormal{\textsc{m}}}$ (Section \ref{secinit}), and $T_a=\tau^{-1}(a)$. The function
\[
w:=b^{\textnormal{\textsc{m}}}-\tau:\ M\to[0,+\infty]
\]
is finite exactly on $\textrm{dom}\, b^{\textnormal{\textsc{m}}}$, and it depends only on $(M,g)$. By \cite[Prop.\ 5.23(d)]{minguzzi26e}, for every future inextendible causal curve $x$ starting from $p$ the limit $L^{\textnormal{\textsc{m}}}(x)=\lim_{t\to+\infty}[\tau(x(t))-\ell(x\vert_{[0,t]})]$ exists, and $b^{\textnormal{\textsc{m}}}(p)$ is the infimum of $L^{\textnormal{\textsc{m}}}(x)$ over these curves, attained on the calibrated ones. Thus $w(p)$ is the least asymptotic delay, with respect to the cosmological time, accumulated after $p$ by an observer leaving $p$. Let $\alpha$ be a proper time parametrized $b^{\textnormal{\textsc{m}}}$-calibrated ray. Since $b^{\textnormal{\textsc{m}}}(\alpha(t))=b^{\textnormal{\textsc{m}}}(\alpha(0))+t$, while $\tau(\alpha(t'))-\tau(\alpha(t))\ge t'-t$ for $t<t'$ by the rushing property,
\begin{equation} \label{wlyap}
\begin{split}
&t\mapsto w(\alpha(t)) \textrm{ is non-increasing, and}\\
&w(\alpha(t))=b^{\textnormal{\textsc{m}}}(\alpha(0))-[\tau(\alpha(t))-t]\to 0 \quad \textrm{for } t\to+\infty .
\end{split}
\end{equation}
Thus $w$ is a Lyapunov function of the canonical calibrating semi-flow of \cite[Sect.\ 5.1]{minguzzi26e}: the sets $\{w\le s\}$, $s>0$, are invariant under it, and every calibrated ray eventually enters each of them. Moreover, as $\tau\circ\alpha$ is locally Lipschitz, hence locally absolutely continuous, (\ref{wlyap}) gives
\begin{equation} \label{wint}
w(\alpha(t))=\int_t^{+\infty}\Big[\frac{\dd}{\dd r}\tau(\alpha(r))-1\Big]\dd r ,
\end{equation}
where the integrand is non-negative by the rushing property: the deficit at an event of a calibrated ray is the excess of the cosmological age over the proper time which the observer will still accumulate. The integrand has a kinematical meaning. Let $p$ be a differentiability point of both $\tau$ and $b^{\textnormal{\textsc{m}}}$. Then $u_\tau:=-\nabla\tau(p)$ is the unit tangent at $p$ of the maximizing geodesic which realizes $\tau(p)$, as the smooth lower support functions of $\tau$ at $p$ used in the proof of Theorem \ref{thmasym}(c) show, and
\[
\gamma:=\dd\tau(v)=-g(u_\tau,v)\ge1
\]
is the Lorentz factor of the calibrated observer with respect to the observer comoving with the cosmological time, so that
\[
\dd w(v)=1-\gamma\le0 ,
\]
and the integrand of (\ref{wint}) is $\gamma-1$ at the points of the ray at which $\tau$ is differentiable. The same factor relates the calibrated volume to the volume induced by $g$ on the slices of the cosmological time. Indeed, at $\mathcal{H}^n$-almost every point $x$ of $T_a$ there is a tangent hyperplane, which is spacelike and whose future unit normal $u_\tau$ is the final tangent of a maximizing geodesic which realizes $\tau(x)$ (proof of Theorem \ref{thmasym}(c)). On it $\omega=i_v\mu_g$ restricts to $\gamma=-g(u_\tau,v)$ times the induced volume form (Section \ref{secdict}), so, denoting by $\mathcal{A}_\tau$ the area measure induced by $g$ on $T_a$, for every Borel set $E\subset T_a\cap\textrm{dom}\, b^{\textnormal{\textsc{m}}}$
\begin{equation} \label{volgamma}
\textrm{Vol}_b(E)=\int_E\gamma\,\dd\mathcal{A}_\tau .
\end{equation}
Thus the calibrated volume is not the induced volume, unless the two families of observers coincide, and the relative boost which enlarges the calibrated volume is the same which makes the deficit decrease. The induced area of the compact Lipschitz hypersurface $T_a$ is finite, so the divergence of the calibrated volume at the future cosmological horizon (Proposition \ref{propinfvol}, Section \ref{secdiv}) is due to the divergence of $\gamma$, namely to the fact that the calibrated observers which start near the horizon are increasingly boosted with respect to the cosmological time.

For $a>0$ and $s\ge 0$ let
\[
N(a,s):=\textrm{Vol}_b\big(T_a\cap\{w\le s\}\big),
\]
the calibrated volume of the compact set of the events of cosmological time $a$ whose least asymptotic delay is at most $s$. It is finite, and under the TCC it is non-decreasing in $a$ (Remark \ref{remcosmvol}); it is non-decreasing in $s$. Equivalently, let $\nu_a$ be the push-forward under $w$ of the calibrated volume measure $\rho_{T_a}\mathcal{H}^n$ of $T_a\cap\textrm{dom}\, b^{\textnormal{\textsc{m}}}$, a Borel measure on $[0,+\infty)$ which is finite on the bounded sets and which depends only on $(M,g)$ and $a$; we call it the {\em deficit spectrum} at the cosmological time $a$. Then $N(a,s)=\nu_a([0,s])$ is its cumulative distribution function, and $\textrm{Vol}_b(T_a\cap\textrm{dom}\, b^{\textnormal{\textsc{m}}},F(w))=\int F\,\dd\nu_a$ for every non-negative Borel function $F$. The family $\{\nu_a\}_{a>0}$ is the object selected by the geometry, and the results of this section describe its dependence on $a$.

\begin{theorem}[Asymptotic calibrated volume] \label{thmasym}
Let $(M,g)$ be future timelike complete with regular cosmological time, and assume the TCC.
\begin{itemize}
\item[(a)] The cumulative distributions of the deficit spectra are ordered: for $0<a_1<a_2$ and every $s\ge0$,
\[
\nu_{a_1}([0,s])=N(a_1,s)\le N(a_2,s)=\nu_{a_2}([0,s]) .
\]
The limit
\[
V_\infty:=\lim_{a\to+\infty}N(a,s)\in(0,+\infty]
\]
does not depend on $s>0$, and $N(a,s)\le V_\infty$ for every $a>0$ and $s\ge 0$.
\item[(b)] For every locally absolutely continuous non-increasing function $F:[0,+\infty)\to[0,+\infty)$ with $F(s)\to 0$ for $s\to+\infty$, the weighted volume $V_F(a)=\textrm{Vol}_b(T_a\cap \textrm{dom}\, b^{\textnormal{\textsc{m}}},F(w))$ is non-decreasing in $a$ and
\[
\lim_{a\to+\infty}V_F(a)=F(0)\,V_\infty ,
\]
with the convention $0\cdot(+\infty)=0$.
\item[(c)] If $V_\infty<+\infty$, then $\textrm{dom}\, b^{\textnormal{\textsc{m}}}=M$. If $\textrm{dom}\, b^{\textnormal{\textsc{m}}}=M$, then $V_\infty=\lim_{a\to+\infty}\textrm{Vol}_b(T_a)$.
\end{itemize}
\end{theorem}

We call $V_\infty$ the {\em asymptotic calibrated volume} of $(M,g)$. It depends only on $(M,g)$, with no auxiliary choice. By (b), all the weighted volumes of Corollary \ref{corweight} have the same limit up to the normalization $F(0)$: the weight only affects the time at which the volume is counted. By (c), if the future cosmological horizon $\p\,\textrm{dom}\, b^{\textnormal{\textsc{m}}}$ is non-empty, then $V_\infty=+\infty$, although each $N(a,s)$ is finite. The converse does not hold: $V_\infty=+\infty$ also in spacetimes without horizon, and $V_\infty<+\infty$ does not imply that the calibrated congruence is eventually non-expanding (Example \ref{exwarp2}). Thus a finite $V_\infty$ is an obstruction to the existence of the horizon, while $V_\infty$ does not distinguish the spacetimes which have one; the information is in the family $\nu_a$. The ordering in (a) is an ordering of cumulative distributions: it does not state that $\nu_{a_2}-\nu_{a_1}$ is a positive measure, and indeed the calibrated rays which enter the set $\{w\le s\}$ leave the set $\{w>s\}$ (Theorem \ref{thmcanbal}). In (a) the condition $s>0$ cannot be removed: $N(a,0)$, the calibrated volume of the events at which the calibrated observers are comoving with the cosmological time forever, can vanish for every $a$ while $V_\infty>0$ (Example \ref{exflat}).

\begin{proof}
(a). The ordering is Theorem \ref{corvol}(c) for $f=\log\tau$ and $u=\tau$ (Remark \ref{remcosmvol}). Let $0<s'\le s$ and $0<a<a'$, and apply Theorem \ref{thmtransp} to $f=\log\tau$, which is locally steep, to $t_1=\log a$, $t_2=\log a'$, so that $B_1=T_a\cap \textrm{dom}\, b^{\textnormal{\textsc{m}}}$, $B_2=T_{a'}\cap \textrm{dom}\, b^{\textnormal{\textsc{m}}}$, and to $u=\tau$, which is admitted by Remark \ref{remcosmvol}. Let $E:=T_a\cap\{w\le s\}\backslash S$, and for $x\in E$ let $P_{a'}(x)$ be the point at which the calibrated ray starting from $x$ meets $T_{a'}$. The map $P_{a'}$ is continuous on $E$, as the initial tangent $v$ of the ray is continuous on $\textrm{dom}\, b^{\textnormal{\textsc{m}}}\backslash S$ \cite[Thm.\ 5.21(a)]{minguzzi26e}, so
\[
E_{a'}:=\{x\in E:\ w(P_{a'}(x))\le s'\}
\]
is a Borel set. By (\ref{wlyap}) the sets $E_{a'}$ increase with $a'$ and their union is $E$, and $P_{a'}(E_{a'})\subset T_{a'}\cap\{w\le s'\}$. Since $E_{a'}\subset\{w\le s\}$, (\ref{transp2}) gives
\[
\textrm{Vol}_b(E_{a'})\le \textrm{Vol}_b\big(P_{a'}(E_{a'})\cap\{w\le s\}\big)\le N(a',s')\le V_\infty(s'):=\sup_{a''>0}N(a'',s') .
\]
Letting $a'\to+\infty$, $N(a,s)=\textrm{Vol}_b(E)\le V_\infty(s')$, as $\mathcal{H}^n(T_a\cap S)=0$. Thus $V_\infty(s)\le V_\infty(s')$, while the opposite inequality follows from $N(a,s')\le N(a,s)$. Since $N(\cdot,s)$ is non-decreasing, $V_\infty(s)$ is its limit, which is therefore independent of $s>0$, and $N(a,0)\le N(a,s)\le V_\infty$. Finally, $V_\infty>0$: by (\ref{wlyap}), $T_a\cap\{w<s\}$ is a non-empty open subset of $T_a$ for large $a$, and the density $\rho_{T_a}$ is positive almost everywhere, since the tangent hyperplanes of the acausal hypersurface $T_a$ do not contain timelike vectors, so that $\omega=i_v\mu_g$ does not vanish on them.

(b). As in Corollary \ref{corweight}(a), whose proof does not use the smoothness of $F$, and since $w\ge 0$, $V_F(a)=\int_0^{+\infty}(-F'(s))\,N(a,s)\,\dd s$. By (a) the integrand is non-decreasing in $a$ and converges to $(-F'(s))\,V_\infty$ for every $s>0$, so the claim follows from the monotone convergence theorem and $\int_0^{+\infty}(-F')\,\dd s=F(0)$.

(c). Let $V_\infty<+\infty$. Letting $s\to+\infty$ in $N(a,s)\le V_\infty$ we get $\textrm{Vol}_b(T_a\cap \textrm{dom}\, b^{\textnormal{\textsc{m}}})\le V_\infty$ for every $a>0$. Suppose that $\textrm{dom}\, b^{\textnormal{\textsc{m}}}\ne M$, so that $T_{a_0}\not\subset \textrm{dom}\, b^{\textnormal{\textsc{m}}}$ for some $a_0>0$, and let $a>a_0$. We show that $\textrm{Vol}_b(T_a\cap \textrm{dom}\, b^{\textnormal{\textsc{m}}})=+\infty$, a contradiction, by the argument of Proposition \ref{propinfvol} applied to $Y:=T_a$, which is only a Lipschitz hypersurface. Two points have to be checked. First, $Y$ is the image of a smooth compact connected Cauchy hypersurface under a bi-Lipschitz map, the projection along the flow of a smooth timelike vector field, so the Poincar\'e inequality $\inf_c\int_Y\vert \phi-c\vert\,\dd\mathcal{H}^n\le C\int_Y\vert\nabla^Y\phi\vert\,\dd\mathcal{H}^n$ holds for the Lipschitz functions $\phi$ on $Y$, being invariant under bi-Lipschitz maps up to the value of the constant, where $\nabla^Y$ is the tangential gradient with respect to the auxiliary Riemannian metric $h$. Second, there is $c_1>0$ such that $\rho_Y\ge c_1\vert\nabla^Y b\vert$ at $\mathcal{H}^n$-almost every point of $Y\cap\textrm{dom}\, b^{\textnormal{\textsc{m}}}$. Indeed, let $x\in Y$ be a point at which $Y$ has a tangent hyperplane $T_xY$, and let $\gamma:(0,a]\to M$ be a unit speed maximizing timelike geodesic ending at $x$ with $\tau(\gamma(r))=r$, with final tangent $u_x$ \cite[Prop.\ 2.3]{andersson98}. For $r\in(0,a)$ the function $r+d(\gamma(r),\cdot)$ is smooth near $x$ \cite[Prop.\ 3.1]{andersson98}, it is bounded from above by $\tau$, by the rushing property, and it coincides with $\tau$ at $x$, so its level set through $x$, a smooth spacelike hypersurface orthogonal to $u_x$, lies in $\{\tau\ge a\}$ and touches $Y$ at $x$. Representing both hypersurfaces as graphs in a chart, the difference of the graphing functions has a minimum at $x$, where both are differentiable, hence $T_xY=u_x^\perp$. The vectors $u_x$, $x\in Y$, belong to a compact set of timelike vectors \cite[Prop.\ 2.4]{andersson98}, so on the hyperplanes $T_xY$ the metric induced by $g$ is uniformly equivalent to that induced by $h$. Thus, at the points where also $b$ is differentiable, with $\gamma_x:=-g(v,u_x)\ge 1$, $\rho_Y$ is bounded from below by a positive constant times $\gamma_x$, while $\vert\nabla^Yb\vert$ is bounded from above by a constant times the norm $(\gamma_x^2-1)^{1/2}$ of the $g$-orthogonal projection of $\nabla b=-v$ on $u_x^\perp$. With these two facts the proof of Proposition \ref{propinfvol} applies, with $\mu_Y$ replaced by $\mathcal{H}^n$ and the inequality $\textrm{Vol}_b(B)\ge\int_B\vert\nabla^Yb\vert\,\mu_Y$ by $\textrm{Vol}_b(B)\ge c_1\int_B\vert\nabla^Yb\vert\,\dd\mathcal{H}^n$. Finally, if $\textrm{dom}\, b^{\textnormal{\textsc{m}}}=M$ then $N(a,s)\le \textrm{Vol}_b(T_a)$, and $\textrm{Vol}_b(T_a)=\lim_{s\to+\infty}N(a,s)\le V_\infty$, so $\textrm{Vol}_b(T_a)\to V_\infty$ by (a).
\end{proof}

\begin{remark}[The deficit spectrum, and a weight which depends only on the spacetime] \label{remspec}
Since $V_F(a)=\int F\,\dd\nu_a$, the weights $F$ of Corollary \ref{corweight} are test functions for the family of the deficit spectra $\nu_a$. Theorem \ref{thmasym} states that, under the TCC, $\nu_a([0,s])$ is non-decreasing in $a$ for every $s$, and that it converges to $V_\infty$ for every $s>0$: as $a\to+\infty$ the calibrated volume concentrates on the events of vanishing asymptotic delay, and if $V_\infty<+\infty$ then $\nu_a\to V_\infty\,\delta_0$ weakly.

The existence of a weight $F$ with finite $V_F$ (Corollary \ref{corweight}(b)) can also be made independent of choices. Since $N(a,s)\le N(s,s)$ for $a\le s$, the function
\[
F^{\textnormal{\textsc{m}}}(s):=\int_s^{+\infty}\frac{\eu^{-r}}{1+N(r,r)}\,\dd r ,
\]
which depends only on $(M,g)$, is positive, decreasing and locally absolutely continuous, and
\[
V_{F^{\textnormal{\textsc{m}}}}(a)=\int_0^{+\infty}\frac{\eu^{-s}\,N(a,s)}{1+N(s,s)}\,\dd s\le N(a,a)+\eu^{-a}<+\infty
\]
for every $a>0$. Whether a universal weight such as $F(s)=\eu^{-s}$ has finite weighted volume in every spacetime depends on the growth of $N(a,s)$ in $s$, on which we have no upper bound. The function $F^{\textnormal{\textsc{m}}}$ is determined by $(M,g)$ once the formula is fixed, but the formula is a convention, and it requires a choice of the units of time and volume, since $s$ is a time and $N$ a volume: under the rescaling $g\mapsto\lambda^2g$, $\lambda>0$, we have $\tau\mapsto\lambda\tau$, $w\mapsto\lambda w$ and $N(a,s)\mapsto\lambda^nN(a/\lambda,s/\lambda)$. For this reason we regard the family $\nu_a$, rather than any weighted volume, as the canonical object. Weighted volumes which are finite in every spacetime and which do not involve units are obtained by truncating at a fixed value of the dimensionless ratio $w/\tau$ (Corollary \ref{correl}).
\end{remark}

On the horospheres the calibrated observers are at rest, so the calibrated volume is the induced area, with no Lorentz factor. Let $B_c:=(b^{\textnormal{\textsc{m}}})^{-1}(c)$, an acausal edgeless set \cite[Prop.\ 6.1]{minguzzi26e}. On $B_c$ we have $w=c-\tau$, so that $B_c\cap\{w\le s\}=B_c\cap J^+(T_{c-s})$ is the portion of the horosphere which lies above the slice of cosmological time $c-s$.

\begin{corollary}[Areas of the horospheres] \label{corhoro}
Let $(M,g)$ be future timelike complete with regular cosmological time, and assume the TCC. For every $s\ge 0$ the function
\[
c\mapsto \textrm{Vol}_b\big(B_c\cap J^+(T_{c-s})\big), \qquad c>s,
\]
is finite and non-decreasing, and it is the area of $B_c\cap J^+(T_{c-s})$, namely the integral over this set of the density, with respect to $\mathcal{H}^n$, of the volume induced by $g$ on the tangent hyperplanes of $B_c$.
\end{corollary}

\begin{proof}
Let $0<c_1<c_2$ and $0<\epsilon<c_1/(c_2+s)$. The function $m:=\min\{b^{\textnormal{\textsc{m}}}-s,\tau\}$ is finite, as $\tau$ is, and locally Lipschitz: this is clear on $\textrm{dom}\, b^{\textnormal{\textsc{m}}}$, while every point outside it has a neighborhood on which $b^{\textnormal{\textsc{m}}}-s>\tau$, by the continuity of $b^{\textnormal{\textsc{m}}}:M\to(-\infty,+\infty]$. Both $b^{\textnormal{\textsc{m}}}$ and $\tau$ are strictly increasing along the causal curves, so the same holds for $m$ and for $m_\epsilon:=\max\{m,\epsilon\tau\}$. Since $\epsilon\tau\le m_\epsilon\le\tau$, $f:=\log m_\epsilon$ is a locally Lipschitz Cauchy time function. For $c\in[c_1,c_2]$,
\[
f^{-1}(\log c)=\{m=c,\ \epsilon\tau\le c\}\cup\{\epsilon\tau=c,\ m\le c\} .
\]
On the second set $\tau=c/\epsilon>c$, so $m\le c$ would give $b^{\textnormal{\textsc{m}}}\le c+s$, which contradicts $b^{\textnormal{\textsc{m}}}\ge\tau=c/\epsilon>c_2+s$; thus it is empty. On $\{m=c\}$ we have $\tau\le b^{\textnormal{\textsc{m}}}\le c+s<c/\epsilon$ if $b^{\textnormal{\textsc{m}}}-s=c$, and $\tau=c$ otherwise, so the condition $\epsilon\tau\le c$ is satisfied. Therefore
\[
Y_c:=f^{-1}(\log c)=\big(B_{c+s}\cap\{\tau\ge c\}\big)\cup\big(T_c\cap\{b^{\textnormal{\textsc{m}}}\ge c+s\}\big),
\]
a compact Cauchy hypersurface obtained by gluing a portion of a horosphere with a portion of a slice of the cosmological time, and $Y_c\cap\{w\le s\}=B_{c+s}\cap\{\tau\ge c\}=B_{c+s}\cap J^+(T_c)$, since on $T_c\cap\{b^{\textnormal{\textsc{m}}}\ge c+s\}$ the condition $w\le s$ gives $b^{\textnormal{\textsc{m}}}=c+s$. By Theorem \ref{corvol}(c), applied to $f$, to $[t_1,t_2]=[\log c_1,\log c_2]$ and to $u=\tau$ (Remark \ref{remcosmvol}), $c\mapsto\textrm{Vol}_b(B_{c+s}\cap J^+(T_c))$ is finite and non-decreasing on $[c_1,c_2]$, hence on $(0,+\infty)$, which is the claim up to the replacement of $c+s$ with $c$. Finally, at $\mathcal{H}^n$-almost every point of $A:=B_{c+s}\cap J^+(T_c)$ the Lipschitz hypersurfaces $Y_c$ and $B_{c+s}$, which both contain $A$, have the same tangent hyperplane, namely at the points of density one of $A$ at which both are differentiable, and $b^{\textnormal{\textsc{m}}}$ is differentiable there, with $v$ orthogonal to the tangent hyperplane of $B_{c+s}$, as $b^{\textnormal{\textsc{m}}}$ is constant on $B_{c+s}$. There $\omega=i_v\mu_g$ restricts to the volume form induced by $g$ (Section \ref{secdict}).
\end{proof}

Thus, under the TCC, the area of the portion of the horosphere $B_c$ which lies in the future of the slice $T_{c-s}$ is non-decreasing in $c$. Both families of hypersurfaces, and the parameter $s$, a bound on the asymptotic delay, have an intrinsic meaning.

\subsection{Weights increasing with the deficit} \label{secdefw}

We keep the assumptions and the notation of Section \ref{seccanvol}. The weights $F(w)$ of Theorem \ref{thmasym} are non-increasing functions of the deficit $w$, hence, by (\ref{wlyap}), they are non-decreasing along the calibrated rays, and the TCC suffices for the monotonicity of their weighted volumes. We now consider weights $F(w)$ with $F$ positive and non-decreasing, which do not increase along the calibrated rays. Proposition \ref{propBE} applies to them, and the condition of Bakry--\'Emery type can be expressed through the Hessian of the cosmological time.

The cosmological time $\tau$ is locally semi-convex, since it admits at every point smooth lower support functions whose Hessians are locally uniformly bounded \cite[Sect.\ 3]{andersson98} (see also \cite[Lemma 8.1]{minguzzi26f}), so it is twice differentiable almost everywhere \cite[Thm.\ 1.2(v)]{andersson98}, namely it has almost everywhere a second order expansion, whose quadratic term defines the Hessian $\textrm{Hess}\,\tau$. Hence at almost every point of $\textrm{dom}\, b^{\textnormal{\textsc{m}}}$ the vector $v=-\nabla b^{\textnormal{\textsc{m}}}$ and the quantities $\dd\tau(v)$ and $\textrm{Hess}\,\tau(v,v)$ are defined, and $\dd\tau(v)\ge1$ (Remark \ref{remcosmvol}). Moreover, $w$ is locally semi-concave on $\textrm{dom}\, b^{\textnormal{\textsc{m}}}$, as $b^{\textnormal{\textsc{m}}}$ is, and so is $\psi=\ln F(w)$ for $F$ of class $C^2$, positive and non-decreasing. Thus the weighted Ricci curvature $Ric_N^\psi(v)$ is defined almost everywhere (Section \ref{secmono}), and the natural assumption is the non-negativity of the largest one, $Ric_0^\psi(v)$. Let
\[
N_F(a,s):=\textrm{Vol}_b\big(T_a\cap\{w\le s\},F(w)\big),\qquad a>0,\ s\ge0 .
\]

\begin{theorem}[Weights increasing with the deficit] \label{thmdefw}
Let $(M,g)$ be future timelike complete with regular cosmological time, let $b=b^{\textnormal{\textsc{m}}}$ and $w=b^{\textnormal{\textsc{m}}}-\tau$. Let $F\in C^2([0,+\infty))$ be positive and non-decreasing, let $\psi:=\ln F(w)$ and $h:=F^{-1/n}$, and let $v=-\nabla b^{\textnormal{\textsc{m}}}$. At almost every point of $\textrm{dom}\, b^{\textnormal{\textsc{m}}}$ the weighted Ricci curvature of synthetic dimension $N=0$ is
\begin{equation} \label{ric0w}
Ric_0^\psi(v)=Ric(v,v)+(\ln F)'(w)\,\textrm{Hess}\,\tau(v,v)+n\,\frac{h''(w)}{h(w)}\,\big(\dd\tau(v)-1\big)^2 .
\end{equation}
Assume that
\begin{equation} \label{defwcond}
Ric_0^\psi(v)\ge0 \qquad \textrm{almost everywhere on } \textrm{dom}\, b^{\textnormal{\textsc{m}}} .
\end{equation}
Then the following statements hold, without the TCC.
\begin{itemize}
\item[(a)] The function $\psi$ is locally Lipschitz and locally semi-concave on $\textrm{dom}\, b^{\textnormal{\textsc{m}}}$, hence semi-concave along the calibrated rays, and $\theta^b+\dd\psi(v)\ge0$ almost everywhere, namely
\[
\theta^b\ge(\ln F)'(w)\,\big(\dd\tau(v)-1\big)\ge0 \qquad \textrm{almost everywhere on } \textrm{dom}\, b^{\textnormal{\textsc{m}}} .
\]
In particular, the conclusions of Theorems \ref{corvol}, \ref{thmtransp} and \ref{thmasym} and of Corollary \ref{corhoro} hold, and Corollary \ref{corBE} and Remark \ref{remwtransp} apply to the weight $F(w)$.
\item[(b)] For every $s\ge0$ the function $a\mapsto N_F(a,s)$ is finite and non-decreasing on $(0,+\infty)$. More precisely, for $0<a<a'$, with $\Omega_s:=\{a<\tau<a'\}\cap\{w\le s\}$,
\begin{align*}
N_F(a',s)\ge{}& N_F(a,s)+\int_{\Omega_s}F(w)\big[\theta^b-(\ln F)'(w)(\dd\tau(v)-1)\big]\mu_g\\
&+2\int_{S^1\cap\Omega_s}F(w)\,\vert\omega_+\vert .
\end{align*}
\item[(c)] For every $s>0$, $\lim_{a\to+\infty}N_F(a,s)=F(0)\,V_\infty$, where $V_\infty$ is the asymptotic calibrated volume of Theorem \ref{thmasym}. In particular, if $V_\infty<+\infty$, for every $a>0$ and $s\ge0$
\[
\textrm{Vol}_b\big(T_a\cap\{w\le s\},F(w)-F(0)\big)\le F(0)\,\big[V_\infty-N(a,s)\big] .
\]
\end{itemize}
If $h$ is convex, for instance if $F$ is concave, or if $F(s)=\eu^{\lambda s}$ with $\lambda\ge0$, the last term of (\ref{ric0w}) is non-negative, and (\ref{defwcond}) follows from
\begin{equation} \label{defwcond2}
Ric(v,v)+(\ln F)'(w)\,\textrm{Hess}\,\tau(v,v)\ge0 .
\end{equation}
\end{theorem}

For $F=1$ the assumption is $Ric(v,v)\ge0$, the TCC in the calibrated directions, which is all that the proofs of Theorems \ref{corvol}, \ref{thmtransp} and \ref{thmasym} use of the TCC. For $F$ increasing, the calibrating semi-flow expands at least at the rate $(\ln F)'(w)(\dd\tau(v)-1)$, which is positive where the calibrated observers move with respect to those who have aged maximally since the initial singularity, and the volume weighted by the increasing function $F(w)$ of the deficit is non-decreasing, although the deficit decreases along the calibrated rays. The last inequality of (c) bounds the excess of the weighted volume in terms of the calibrated volume which is still to be created; for instance, for $F(s)=1+\lambda s$ it bounds the integral of $\lambda w$ over $T_a\cap\{w\le s\}$ with respect to the calibrated volume by $V_\infty-N(a,s)$.

\begin{proof}
(a). The function $w$ is locally Lipschitz and locally semi-concave on $\textrm{dom}\, b^{\textnormal{\textsc{m}}}$, since $b^{\textnormal{\textsc{m}}}$ is locally Lipschitz and locally semi-concave there (Section \ref{secprel}) and $\tau$ is locally Lipschitz and locally semi-convex. Let $G:=\ln F\in C^2([0,+\infty))$, so that $G'\ge0$. A non-decreasing function of class $C^2$ of a locally Lipschitz and locally semi-concave function is locally Lipschitz and locally semi-concave, so $\psi=G(w)$ has these properties, and it is semi-concave along the calibrated rays (Section \ref{secmono}). Let $x$ be a point at which both $b^{\textnormal{\textsc{m}}}$ and $\tau$ are twice differentiable, as is the case almost everywhere. Then $w$ and $\psi$ are twice differentiable at $x$, $\dd w(v)=1-\dd\tau(v)$, and $\textrm{Hess}\,w(v,v)=-\textrm{Hess}\,\tau(v,v)$, as $\nabla^2b^{\textnormal{\textsc{m}}}(v,\cdot)=0$ (proof of Lemma \ref{lemdict}(c)). Hence
\begin{align*}
\dd\psi(v)&=-G'(w)\,\big(\dd\tau(v)-1\big),\\
\textrm{Hess}\,\psi(v,v)&=G''(w)\,\big(\dd\tau(v)-1\big)^2-G'(w)\,\textrm{Hess}\,\tau(v,v),
\end{align*}
and $Ric_0^\psi(v)=Ric(v,v)-\textrm{Hess}\,\psi(v,v)+\frac1n\dd\psi(v)^2$ is given by (\ref{ric0w}), since $n\,h''/h=G'^2/n-G''$. By (\ref{defwcond}) and Corollary \ref{corpoint}, assumption (i) of Proposition \ref{propBE} holds with $N=0$. Assumption (ii) holds with $\epsilon=0$: along a calibrated ray $\alpha$ the function $w\circ\alpha$ takes values in $[0,w(\alpha(0))]$ by (\ref{wlyap}), so $F(w\circ\alpha)\ge F(0)$ and $\phi_\alpha(t)=\int_0^tF(w(\alpha(t')))^{2/n}\dd t'\ge F(0)^{2/n}\,t$. By Proposition \ref{propBE}, $\theta^b+\dd\psi(v)\ge0$ almost everywhere.

Hence $\theta^b\ge0$ almost everywhere, which is the only consequence of the TCC used in the proofs of Theorems \ref{corvol}, \ref{thmtransp} and \ref{thmasym} and of Corollary \ref{corhoro}.

(b). This is Corollary \ref{corBE}(b) for $f=\log\tau$ and $u=\tau$, which are admitted by Remark \ref{remcosmvol}, since $A_t(s)=T_a\cap\{w\le s\}$ for $a=\eu^t$.

(c). The proof of Theorem \ref{thmasym}(a) applies to the weighted volumes $\textrm{Vol}_b(\,\cdot\,,F(w))$, the weighted transport inequality of Remark \ref{remwtransp} replacing (\ref{transp2}) and (b) replacing the monotonicity of $N(\cdot,s)$. It shows that the limit $V_\infty^F:=\lim_{a\to+\infty}N_F(a,s)=\sup_{a>0}N_F(a,s)$ does not depend on $s>0$. As $F$ is non-decreasing, $F(0)\le F(w)\le F(s)$ on $\{w\le s\}$, hence $F(0)\,N(a,s)\le N_F(a,s)\le F(s)\,N(a,s)$. Since $N(a,s)\to V_\infty$ for $a\to+\infty$ by (a) and Theorem \ref{thmasym}(a), we get $F(0)\,V_\infty\le V_\infty^F\le F(s)\,V_\infty$ for every $s>0$, and letting $s\to0$, $V_\infty^F=F(0)\,V_\infty$. The last inequality follows from $N_F(a,s)\le V_\infty^F=F(0)\,V_\infty$ and $\textrm{Vol}_b(T_a\cap\{w\le s\},F(0))=F(0)\,N(a,s)$, for $s>0$, and for $s=0$ its left-hand side vanishes.

Finally, $h''=\frac1n\big(\frac1n+1\big)F^{-1/n-2}F'^2-\frac1nF^{-1/n-1}F''$, which is non-negative if $F$ is concave, and $h=\eu^{-\lambda s/n}$ is convex for $F=\eu^{\lambda s}$.
\end{proof}

\begin{remark}[On the condition $Ric_0^\psi(v)\ge0$] \label{remdefw}
(1) Let $u:=-\nabla\tau$, defined at the differentiability points of $\tau$, where it is the final unit tangent of the maximizing geodesic which realizes $\tau$; it is the frame of the observers which have aged maximally since the initial singularity, while $v$ is the frame of those which age maximally into the far future. At the points at which $\tau$ is twice differentiable, $\textrm{Hess}\,\tau(u,\cdot)=0$, by the argument given for $b$ in the proof of Lemma \ref{lemdict}(c), applied to the smooth lower support functions of $\tau$. Hence
\[
\textrm{Hess}\,\tau(v,v)=\textrm{Hess}\,\tau(v^\perp,v^\perp),\quad v^\perp:=v+g(u,v)\,u,\quad g(v^\perp,v^\perp)=\dd\tau(v)^2-1 ,
\]
where $v^\perp$ is the component of $v$ orthogonal to $u$, which points in the direction of the motion of the second frame relative to the first, and $\dd\tau(v)=-g(u,v)$ is their relative Lorentz factor. Where $\tau$ is $C^2$, $-\textrm{Hess}\,\tau$ restricted to $u^\perp$ is the second fundamental form $K(X,Y)=g(\nabla_Xu,Y)$ of the slices of the cosmological time, so (\ref{ric0w}) and (\ref{defwcond2}) read
\begin{align*}
Ric_0^\psi(v)&=Ric(v,v)-(\ln F)'(w)\,K(v^\perp,v^\perp)+n\,\frac{h''(w)}{h(w)}\,\big(\dd\tau(v)-1\big)^2,\\
Ric(v,v)&\ge(\ln F)'(w)\,K(v^\perp,v^\perp) .
\end{align*}
The second is the timelike convergence condition in the calibrated directions, strengthened by a term proportional to the expansion of the slices of the cosmological time in the direction of the relative velocity of the two frames. This term is of second order in the relative velocity, while the last term of $Ric_0^\psi(v)$ is of fourth order. All the terms but the first vanish where the two frames coincide.

(2) Under the TCC, and for $F$ non-decreasing with $F^{-1/n}$ convex, $Ric_0^\psi(v)\ge0$ at the points at which $\textrm{Hess}\,\tau(v,v)\ge0$, namely where the slices of the cosmological time do not expand in the direction of $v^\perp$, and it requires a strict form of the TCC where they do. The latter points cannot be avoided: if $\textrm{Hess}\,\tau(v,v)\ge0$ almost everywhere on $\textrm{dom}\, b^{\textnormal{\textsc{m}}}$, then $b^{\textnormal{\textsc{m}}}=\tau$ there. Indeed, by Corollary \ref{corpoint} applied to the locally semi-concave function $-\tau$, for almost every $p$ the function $\tau_\alpha:=\tau\circ\alpha$ along the calibrated ray $\alpha$ starting from $p$ satisfies $\tau_\alpha''=\textrm{Hess}\,\tau(v,v)\ge0$ almost everywhere, so it is convex, the singular part of its second derivative being non-negative by the semi-convexity of $\tau$ (only its sign is used here; by Proposition \ref{proptausbv} the singular part of the Hessian of $\tau$ is concentrated on the non-differentiability set of $\tau$), so $t\mapsto\tau_\alpha(t)-t$ is convex, non-decreasing and bounded by (\ref{wlyap}), hence constant; thus $w$ is constant along $\alpha$, and $w(p)=0$ by (\ref{wlyap}). Therefore, if $w$ does not vanish identically and $F'>0$, (\ref{defwcond2}) requires $Ric(v,v)>0$ on a set of positive measure, and it does not hold in vacuum. The singular part of the Hessian of $\tau$, which is concentrated on the set of non-differentiability points of $\tau$ and is non-negative, does not enter the condition. For $F$ non-increasing the signs are reversed: the weight is not semi-concave along the calibrated rays, but then it is non-decreasing along them, and the TCC suffices (Theorem \ref{thmasym}).

(3) The pair $(N,\epsilon)=(0,0)$ used in the proof is the optimal one for these weights. Indeed, $Ric_0^\psi$ is the largest of the weighted Ricci curvatures, and the $\epsilon$-completeness of the calibrated rays is automatic for every $\epsilon$, since $F(w)$ is bounded and bounded away from zero along each of them, so nothing is gained from the other values. In terms of the product trick, Theorem \ref{thmdefw} is an area theorem for the horizon $H$ in the metric $F(w)^{2/n}(g+\dd z^2)$ of $\tilde M$, which is only locally Lipschitz.

(4) The bound of Theorem \ref{thmdefw}(a) is consistent with (2). If the TCC holds in the calibrated directions and $\textrm{Hess}\,\tau(v,v)\ge0$ almost everywhere, then (\ref{defwcond2}) holds for $F(s)=\eu^{\lambda s}$ for every $\lambda\ge0$, and Theorem \ref{thmdefw}(a) gives $\theta^b\ge\lambda\,(\dd\tau(v)-1)$ for every $\lambda$, hence $\dd\tau(v)=1$ almost everywhere, namely $v=u$ and $\dd w(v)=0$. By Corollary \ref{corpoint}, $w$ is then constant along the calibrated ray starting from almost every point, which again gives $b^{\textnormal{\textsc{m}}}=\tau$ on $\textrm{dom}\, b^{\textnormal{\textsc{m}}}$.
\end{remark}

\subsection{The divergence at the future cosmological horizon} \label{secdiv}

If the future cosmological horizon $\p\,\textrm{dom}\, b$ is non-empty, the calibrated volume of the slices which meet it is infinite (Proposition \ref{propinfvol}), and finite quantities are obtained by truncation (Theorem \ref{corvol}(c)) or by weights which vanish at the horizon (Corollary \ref{corweight}). In this subsection we describe the structure of the divergence. Once the slices have left the future development, the truncated volumes diverge at least linearly in the truncation parameter. Their lower asymptotic slope, which may be infinite, is non-decreasing in time and, for smooth slices and almost every time, it is bounded from below by the area of the section of the horizon. Under an additional convergence assumption on the areas of the sections of the horospheres by the slices, the slope equals that area. Thus the horizon contributes to the calibrated volume through its area, and the weights are not needed to obtain quantities which depend only on the spacetime.

In the first part of this subsection $b$ is, as in Section \ref{secdict}, the Cauchy Busemann function of a compact Cauchy hypersurface. Let $Y$ be a smooth compact spacelike Cauchy hypersurface, with induced Riemannian metric $h_Y$, volume measure $\mu_Y$, Hausdorff measure $\mathcal{H}^{n-1}$ and tangential gradient $\nabla^Y$, and let $B:=Y\cap\textrm{dom}\, b$, an open subset of $Y$. For $s,r\in\mathbb{R}$ let
\[
N_Y(s):=\textrm{Vol}_b\big(Y\cap\{b\le s\}\big),\qquad L_Y(r):=\mathcal{H}^{n-1}\big(Y\cap\{b=r\}\big),
\]
and let $P_Y\in[0,+\infty]$ be the perimeter of $B$ in $(Y,h_Y)$, namely the total variation of the characteristic function of $B$ \cite[Sect.\ 5.1]{evans92}. The volume $N_Y(s)$ is finite, as $Y\cap\{b\le s\}$ is a compact subset of $\textrm{dom}\, b$, and it is non-decreasing in $s$.

\begin{proposition} \label{propslope}
Assume that $(M,g)$ is future timelike complete, and let $Y$ be a smooth compact spacelike Cauchy hypersurface.
\begin{itemize}
\item[(a)] For every Borel set $E\subset B$,
\[
\textrm{Vol}_b(E)=\int_E\sqrt{1+\vert\nabla^Yb\vert^2}\ \mu_Y ,
\]
namely the calibrated volume of $E$ is the area of the graph of $b\vert_E$ in the Riemannian product $(Y\times\mathbb{R},h_Y+\dd z^2)$.
\item[(b)] For every $s\in\mathbb{R}$,
\[
\int_{-\infty}^sL_Y(r)\,\dd r\le N_Y(s)\le\int_{-\infty}^sL_Y(r)\,\dd r+\mu_Y\big(Y\cap\{b\le s\}\big) .
\]
\item[(c)] The areas of the level sets of $b$ on $Y$ satisfy
\[
\operatorname*{ess\,lim\,inf}_{r\to+\infty}L_Y(r)\ge P_Y,\qquad \textrm{hence}\qquad \liminf_{s\to+\infty}\frac{N_Y(s)}{s}\ge P_Y .
\]
\item[(d)] If $\mathcal{H}^{n-1}(Y\cap\p\,\textrm{dom}\, b)<+\infty$, then $P_Y\le\mathcal{H}^{n-1}(Y\cap\p\,\textrm{dom}\, b)$, with equality if the horizon $\p\,\textrm{dom}\, b$ is differentiable at $\mathcal{H}^{n-1}$-almost every point of $Y\cap\p\,\textrm{dom}\, b$. If $f$ is a smooth Cauchy temporal function, both conditions hold for $Y=f^{-1}(t)$ for almost every $t$.
\end{itemize}
\end{proposition}

By (a) the graph of $b$ over $B$ is a hypersurface of $Y\times\mathbb{R}$ which rises to infinity over $Y\cap\p\,\textrm{dom}\, b$, and by (b) the calibrated volume below the height $s$ is, up to a bounded term, the integral of the areas of its horizontal sections. By (c) and (d), for every $\epsilon>0$ these areas are, for almost every sufficiently large height, at least the area of the section of the horizon minus $\epsilon$, so that the graph has an end which is asymptotically not thinner than the vertical cylinder over $Y\cap\p\,\textrm{dom}\, b$. Proposition \ref{propinfvol} gives a linear lower bound for $N_Y(s)$ whose constant comes from a Poincar\'e inequality; (c) identifies the geometric constant.

\begin{proof}
(a). As recalled in the proof of Proposition \ref{propinfvol}, at $\mu_Y$-almost every point of $B$ the calibrated density with respect to $\mu_Y$ is $\gamma=-g(v,\nu)$, where $\nu$ is the future unit normal of $Y$, and $\vert\nabla^Yb\vert^2=\gamma^2-1$. The hypersurface $Y\times\mathbb{R}$ of $\tilde M$ has induced metric $h_Y+\dd z^2$, and $\Psi(E)$ is the graph of $b\vert_E$ in it, whose area is given by the stated integral; this is consistent with Lemma \ref{lemdict}(b).

(b). As $0\le\gamma-\vert\nabla^Yb\vert\le1$, $N_Y(s)$ differs from $\int_{Y\cap\{b\le s\}}\vert\nabla^Yb\vert\,\mu_Y$ by a non-negative term bounded by $\mu_Y(Y\cap\{b\le s\})$. The function $b_s$ which equals $\min\{b,s\}$ on $B$ and $s$ on $Y\backslash B$ is Lipschitz on $Y$, since $Y\cap\{b>s\}$ is an open neighborhood of $Y\backslash B$ on which it is constant, and its gradient vanishes almost everywhere on $\{b_s=s\}$. By the coarea formula \cite[Sect.\ 3.4]{evans92},
\[
\int_{Y\cap\{b\le s\}}\vert\nabla^Yb\vert\,\mu_Y=\int_Y\vert\nabla^Yb_s\vert\,\mu_Y=\int_{-\infty}^sL_Y(r)\,\dd r .
\]

(c). We use some facts on the sets of finite perimeter, which hold on Riemannian manifolds as on $\mathbb{R}^n$, being local or consequences of the Euclidean statements through charts. For $r\in\mathbb{R}$ let $E_r:=Y\cap\{b<r\}$, an open set whose boundary in $Y$ is contained in $Y\cap\{b=r\}$, as $b:M\to(-\infty,+\infty]$ is continuous. By (b), $L_Y(r)<+\infty$ for almost every $r$. For such $r$ the set $E_r$ has finite perimeter \cite[Sect.\ 5.11]{evans92}, and its perimeter is the $\mathcal{H}^{n-1}$-measure of its measure theoretic boundary \cite[Sect.\ 5.7, 5.8]{evans92}, which is contained in its topological boundary, so that $P(E_r)\le L_Y(r)$. As $r\to+\infty$ the sets $E_r$ increase to $B$, so their characteristic functions converge to that of $B$ in $L^1(Y)$, and by the lower semi-continuity of the perimeter \cite[Sect.\ 5.2]{evans92}, $P_Y\le\liminf_kP(E_{r_k})\le\liminf_kL_Y(r_k)$ for every sequence $r_k\to+\infty$ of such values. This proves the first inequality. If $\lambda<P_Y$, then $L_Y(r)\ge\lambda$ for almost every $r\ge r_\lambda$, for some $r_\lambda$, and by (b) $N_Y(s)\ge\lambda\,(s-r_\lambda)$ for $s\ge r_\lambda$, which gives the second one.

(d). The boundary of $B$ in $Y$ is contained in $Y\cap\p\,\textrm{dom}\, b$, so, if the latter has finite $\mathcal{H}^{n-1}$-measure, $B$ has finite perimeter and, as in (c), $P_Y$ is the $\mathcal{H}^{n-1}$-measure of the measure theoretic boundary of $B$, namely of the set of the points of $Y$ at which neither $B$ nor $Y\backslash B$ has density zero, which is at most $\mathcal{H}^{n-1}(Y\cap\p\,\textrm{dom}\, b)$. Let $q\in Y\cap\p\,\textrm{dom}\, b$ be a point at which $\p\,\textrm{dom}\, b$ is differentiable. Its tangent hyperplane at $q$ contains no timelike vector, as $\p\,\textrm{dom}\, b$ is achronal, and it contains the tangent of a lightlike geodesic contained in $\p\,\textrm{dom}\, b$, so it is a null hyperplane. Let $(x^0,x)$ be coordinates near $q$ such that $\p_0$ is timelike and $Y=\{x^0=0\}$. As $\p\,\textrm{dom}\, b$ is the boundary of the open past set $\textrm{dom}\, b$, there is a Lipschitz function $\eta$ such that, locally, $\textrm{dom}\, b=\{x^0<\eta(x)\}$ \cite[Prop.\ 6.3.1]{hawking73}, so that $B=\{\eta>0\}$ and $Y\cap\p\,\textrm{dom}\, b=\{\eta=0\}$. The function $\eta$ is differentiable at the point $x_q$ which corresponds to $q$, and its differential does not vanish there, since otherwise the tangent hyperplane of $\p\,\textrm{dom}\, b$ at $q$ would be that of $Y$, which is spacelike. Hence both $\{\eta>0\}$ and $\{\eta\le0\}$ have density $\frac12$ at $x_q$, and $q$ belongs to the measure theoretic boundary of $B$. This proves the statement on the equality.

Finally, let $f$ be a smooth Cauchy temporal function and $t_1<t_2$. The compact set $\p\,\textrm{dom}\, b\cap f^{-1}([t_1,t_2])$ is covered by the images of finitely many bi-Lipschitz maps $\Phi:U\to\p\,\textrm{dom}\, b$, $U\subset\mathbb{R}^n$ open and bounded, and by Rademacher's theorem the set $Z_\Phi\subset U$ of the points which are mapped to non-differentiability points of $\p\,\textrm{dom}\, b$ is Lebesgue null. By the coarea formula for the Lipschitz function $f\circ\Phi$ \cite[Sect.\ 3.4]{evans92}, for almost every $t$ the level set $(f\circ\Phi)^{-1}(t)$ has finite $\mathcal{H}^{n-1}$-measure and meets $Z_\Phi$ in an $\mathcal{H}^{n-1}$-null set. As $\Phi$ is bi-Lipschitz, and the Hausdorff measures of $h_Y$ and of an auxiliary Riemannian metric of $M$ are comparable on $Y=f^{-1}(t)$, both conditions hold for almost every $t\in[t_1,t_2]$.
\end{proof}

For a smooth Cauchy temporal function $f$ let $Y_t:=f^{-1}(t)$, and let
\[
\sigma_f(t):=\liminf_{s\to+\infty}\frac{N_{Y_t}(s)}{s}\in[0,+\infty],\qquad A(t):=\mathcal{H}^{n-1}\big(Y_t\cap\p\,\textrm{dom}\, b\big) .
\]
We call $\sigma_f(t)$ the {\em slope} of the calibrated volume of $Y_t$ at the horizon; it depends on $f$ only through the hypersurface $Y_t$. For almost every $t$, $A(t)$ is the area of the section of the horizon considered in Theorem \ref{thmhorarea}, since at the differentiability points of the horizon the area form restricts to the induced area form of its spacelike sections (Section \ref{secdict}). In spacetimes with regular cosmological time, with the notation of Section \ref{seccanvol}, let
\[
\sigma(a):=\liminf_{s\to+\infty}\frac{N(a,s)}{s}\in[0,+\infty],\qquad a>0 ,
\]
which depends only on $(M,g)$.

\begin{theorem}[The slope at the horizon] \label{thmslope}
Assume the TCC and that $(M,g)$ is future timelike complete.
\begin{itemize}
\item[(a)] Let $f$ be a smooth Cauchy temporal function. The function $\sigma_f$ is non-decreasing, $\sigma_f(t)=0$ if $Y_t\subset\textrm{dom}\, b$, $\sigma_f(t)>0$ if $Y_{t_0}\not\subset\textrm{dom}\, b$ for some $t_0<t$, and
\[
\sigma_f(t)\ge P_{Y_t} \ \textrm{ for every } t,\qquad \sigma_f(t)\ge A(t) \ \textrm{ for almost every } t .
\]
\item[(b)] If the cosmological time is regular, the function $\sigma$ is non-decreasing, $\sigma(a)=0$ if $T_a\subset\textrm{dom}\, b^{\textnormal{\textsc{m}}}$, and $\sigma(a)>0$ if $T_{a_0}\not\subset\textrm{dom}\, b^{\textnormal{\textsc{m}}}$ for some $a_0<a$.
\end{itemize}
\end{theorem}

\begin{proof}
(a). Let $t_1<t_2$, $K:=f^{-1}([t_1,t_2])$, and let $m>0$ be such that $-g(\nabla f,\nabla f)\ge m^2$ on $K$, so that $u=f/m$ satisfies the assumption of Theorem \ref{corvol}(c). Since $A_t(s)=Y_t\cap\{b\le s+t/m\}$, that theorem gives $N_{Y_{t_1}}(s+t_1/m)\le N_{Y_{t_2}}(s+t_2/m)$ for every $s$. Dividing by $s$ and taking the lower limit for $s\to+\infty$ we get $\sigma_f(t_1)\le\sigma_f(t_2)$. If $Y_t\subset\textrm{dom}\, b$ then $b$ is bounded on $Y_t$ and $N_{Y_t}$ is eventually constant. The positivity follows from Proposition \ref{propinfvol}, and the last two inequalities from Proposition \ref{propslope}(c),(d).

(b). By Remark \ref{remcosmvol}, $N(a,s)$ is non-decreasing in $a$ for every $s$, so $\sigma$ is non-decreasing. If $T_a\subset\textrm{dom}\, b^{\textnormal{\textsc{m}}}$ then $w$ is bounded on $T_a$ and $N(a,\cdot)$ is eventually constant. If $T_{a_0}\not\subset\textrm{dom}\, b^{\textnormal{\textsc{m}}}$ and $a>a_0$, the argument of Proposition \ref{propinfvol}, which applies to $T_a$ as shown in the proof of Theorem \ref{thmasym}(c), gives $c>0$ and $s_0$ such that $N(a,s)=\textrm{Vol}_b(T_a\cap\{b^{\textnormal{\textsc{m}}}\le a+s\})\ge c\,(s-s_0)$ for large $s$.
\end{proof}

Thus the monotonicity of the truncated volumes for every value of the truncation parameter implies the monotonicity of their lower asymptotic rate of growth, and for almost every time this rate is at least the area of the section of the horizon. A monotone slope bounded from below by $A(t)$ does not, by itself, imply the monotonicity of $A(t)$. By Proposition \ref{propslope}(b), $N_{Y_t}(s)=\int_{r_0}^sL_{Y_t}(r)\,\dd r+O(1)$ for every $r_0$, so the slope is identified by a Ces\`aro limit. Suppose that
\begin{equation} \label{areaconv}
\lim_{s\to+\infty}\frac1s\int_{r_0}^sL_{Y_t}(r)\,\dd r=A(t)<+\infty ,
\end{equation}
which holds, for instance, if $\frac1s\int_{r_0}^s\vert L_{Y_t}(r)-A(t)\vert\,\dd r\to0$, and in particular if $L_{Y_t}(r)\to A(t)$ essentially; this allows exceptional heights. Then
\[
N_{Y_t}(s)=A(t)\,s+o(s),\qquad \sigma_f(t)=A(t),
\]
and if (\ref{areaconv}) holds at $t_1<t_2$, Theorem \ref{thmslope}(a) gives $A(t_1)\le A(t_2)$, which is the content of Theorem \ref{thmhorarea} under the TCC: the area theorem for the horizon is the leading term, for large values of the truncation parameter, of the volume monotonicity theorem. Moreover, under (\ref{areaconv}) with $A(t)>0$, by the inequalities which follow Corollary \ref{corweight}, the weighted volume $V_F(t)$ of that corollary, for $u=f$ steep, is finite if and only if $F$ is integrable at $+\infty$. We do not know whether (\ref{areaconv}) holds in general; Proposition \ref{propslope}(c),(d) gives the lower bound, and the upper bound requires a control of the horospheres near the horizon which is uniform in $r$, at least on average.

\begin{remark}[The finite part] \label{remfinite}
If moreover $\int^{+\infty}\vert L_{Y_t}(r)-A(t)\vert\,\dd r<+\infty$, then the limit
\[
R:=\lim_{s\to+\infty}\big[N_{Y_t}(s)-A(t)\,s\big]
\]
exists and is finite. Indeed, by the proof of Proposition \ref{propslope}(b), $N_{Y_t}(s)-\int_{-\infty}^sL_{Y_t}\,\dd r=\int_{Y_t\cap\{b\le s\}}(\gamma-\vert\nabla^{Y_t}b\vert)\,\mu_{Y_t}$, which is non-decreasing in $s$ and bounded by $\mu_{Y_t}(Y_t)$. This renormalized calibrated volume depends on the function used for the truncation: replacing $\{b\le s\}$ with $\{b\le s+c\}$ changes it by $c\,A(t)$. In spacetimes with regular cosmological time the truncation $\{w\le s\}$ is canonical, and so is the corresponding finite part of $N(a,s)$, whenever it exists. We do not know whether it is monotone.
\end{remark}

\begin{remark}[Truncations and weights] \label{remtrunc}
Let the cosmological time be regular, and assume the TCC. The function $N(a,s)$ is non-decreasing in both variables. Hence $a\mapsto N(a,s(a))$ is non-decreasing for every non-decreasing function $s(a)$: besides the sets $\{w\le s\}$, this covers the sets $\{w\le\varsigma\,\tau\}$, $\varsigma>0$, which are invariant under the rescalings of the metric, and the sets $\{b^{\textnormal{\textsc{m}}}-c\,\tau\le s\}$, $c\ge1$. More generally, every function of $b^{\textnormal{\textsc{m}}}$ and $\tau$ which is increasing in $b^{\textnormal{\textsc{m}}}$ has on each slice $T_a$ the sublevel sets of $w$, so all these truncations select the same family of subsets of each slice, and differ only in the way in which the bound depends on $a$. Among these functions, those of $w$ are the ones which are stationary along the calibrated rays where these are tangent to the maximizing geodesics which realize $\tau$, since $\dd w(v)=1-\dd\tau(v)$, which vanishes exactly there. The weighted volumes are superpositions of the truncated ones, $V_F(a)=\int_0^{+\infty}(-F')\,N(a,s)\,\dd s$, so the function $N$, equivalently the family of the deficit spectra $\nu_a$ of Section \ref{seccanvol}, contains all the information, and no choice of a weight is needed.

No weight $F(w)$ with $F$ as in Theorem \ref{thmasym}(b) and $F(0)>0$ gives a finite limit for $a\to+\infty$ when the horizon is non-empty, since $V_F(a)\to F(0)\,V_\infty=+\infty$ (Theorem \ref{thmasym}): the infinite calibrated volume which lies near the horizon on one slice is transported, on the later slices, to the set $\{w\le s\}$, for every $s>0$. The weights $F(w)$ of Theorem \ref{thmdefw}, with $F$ non-decreasing, are bounded away from zero at the horizon, so for them only the truncated volumes $N_F(a,s)$, or the volumes with the weights $G(w)F(w)$, where $G$ is non-increasing and vanishes sufficiently fast at infinity, can be finite; under the assumptions of that theorem the latter are non-decreasing by Corollary \ref{corBE}(a), with $\chi=G(w)$.
\end{remark}

The horospheres offer a way of avoiding the horizon without truncations or weights, since on them $w$ is bounded.

\begin{corollary}[Total area of the horospheres] \label{corhoro2}
Let $(M,g)$ be future timelike complete with regular cosmological time, and let $B_c=(b^{\textnormal{\textsc{m}}})^{-1}(c)$, $c>0$. Then $w<c$ on $B_c$, and for every $\epsilon>0$ the set $B_c\cap\{\tau\ge\epsilon\}$ is a compact subset of $\textrm{dom}\, b^{\textnormal{\textsc{m}}}$. If the TCC holds, the total area of the horosphere,
\[
\textrm{Area}(B_c):=\lim_{s\uparrow c}\textrm{Vol}_b\big(B_c\cap J^+(T_{c-s})\big)\in[0,+\infty],
\]
is non-decreasing in $c$.
\end{corollary}

\begin{proof}
On $B_c$, $w=c-\tau<c$ and $\tau\le b^{\textnormal{\textsc{m}}}=c$. The set $B_c$ is closed, as $b^{\textnormal{\textsc{m}}}:M\to(-\infty,+\infty]$ is continuous, and $\tau^{-1}([\epsilon,c])$ is compact, being a closed subset of $J^+(T_\epsilon)\cap J^-(T_c)$, so $B_c\cap\{\tau\ge\epsilon\}$ is compact. For $0\le s<c$ let $A(c,s):=\textrm{Vol}_b(B_c\cap J^+(T_{c-s}))$, which is non-decreasing in $s$, as the sets $B_c\cap J^+(T_{c-s})=B_c\cap\{\tau\ge c-s\}$ increase with $s$, and non-decreasing in $c>s$ by Corollary \ref{corhoro}. Their union over $s<c$ is $B_c$, so the limit exists, and for $c<c'$, $\textrm{Area}(B_c)=\sup_{s<c}A(c,s)\le\sup_{s<c}A(c',s)\le\textrm{Area}(B_{c'})$.
\end{proof}

Thus the horospheres do not approach the horizon: where the slices of the cosmological time meet it, the horospheres bend towards the initial singularity, and the finiteness of their area is decided there, where the volumes are expected to be small, rather than at the horizon. For instance, suppose that $\tau$ is $C^2$ on $\{\tau<c'\}$ for some $c'>c$, with $\Box\tau\le0$ there, namely that up to the time $c'$ the maximizing geodesics which realize $\tau$ form a smooth expanding congruence, with unit tangent field $u=-\nabla\tau$. Then
\[
\textrm{Area}(B_c)\le\textrm{Vol}_g(T_c)<+\infty,
\]
where $\textrm{Vol}_g(T_c)$ is the volume of the compact hypersurface $T_c$ for the metric induced by $g$. Indeed, as $B_c$ is acausal, the flow of $u$ maps $B_c$ injectively into $T_c$; the flux form $i_u\mu_g$ satisfies $\dd(i_u\mu_g)=-\Box\tau\,\mu_g\ge0$, so the flux of $u$ through a portion of $B_c$ does not exceed that through its image, which is the induced volume of the image, as $u$ is normal to $T_c$; and at the differentiability points of $b^{\textnormal{\textsc{m}}}$ on $B_c$ the density of the flux of $u$ with respect to the area is $-g(u,v)\ge1$.

\subsection{Balance law, rigidity, relative delay and examples} \label{secfurther}

We keep the assumptions and the notation of Section \ref{seccanvol}: $(M,g)$ is future timelike complete with regular cosmological time, $b=b^{\textnormal{\textsc{m}}}$, $w=b^{\textnormal{\textsc{m}}}-\tau$, and $\gamma=\dd\tau(v)\ge1$ is the almost everywhere defined Lorentz factor of the calibrated observers with respect to the cosmological time. The exact form of Theorem \ref{corvol}(c) separates the three mechanisms by which the canonical volumes $N(a,s)$ change, and it gives a rigidity statement for them.

We begin with the case in which the future cosmological horizon is empty, where no truncation is needed.

\begin{corollary}[The horizon-free case] \label{corhf}
Let $(M,g)$ be future timelike complete with regular cosmological time, and assume that $\textrm{dom}\, b^{\textnormal{\textsc{m}}}=M$. Then the calibrated volume
\[
\mathcal{V}(a):=\textrm{Vol}_b(T_a)=\int_{T_a}\gamma\,\dd\mathcal{A}_\tau
\]
is finite for every $a>0$, and for $0<a_1<a_2$, with $Z:=\{a_1<\tau<a_2\}$,
\begin{equation} \label{hfbal}
\mathcal{V}(a_2)-\mathcal{V}(a_1)=\int_Z\theta^b\,\mu_g+2\,\textrm{Vol}_b(S^1\cap Z) .
\end{equation}
Assume moreover the TCC. Then $\mathcal{V}$ is non-decreasing, $\mathcal{V}(a)\to V_\infty$ for $a\to+\infty$, and if $\mathcal{V}(a_1)=\mathcal{V}(a_2)$ then $b^{\textnormal{\textsc{m}}}$ is smooth on $Z$, $v$ is parallel there, and $Z$ is locally isometric to a static product $(\mathbb{R}\times N,-\dd t^2+h)$ with $v=\p_t$.
\end{corollary}

\begin{proof}
The function $f=\log\tau$ is a locally Lipschitz Cauchy time function (Remark \ref{remcosmvol}), and $\tau^{-1}([a_1,a_2])$ is a compact subset of $M=\textrm{dom}\, b^{\textnormal{\textsc{m}}}$, so $Z=f^{-1}((\log a_1,\log a_2))$ is a calibrated tube without side, whose bases $T_{a_i}$ are compact subsets of $\textrm{dom}\, b^{\textnormal{\textsc{m}}}$, with finite calibrated volume. The integral expression of $\mathcal{V}$ is (\ref{volgamma}). The identity (\ref{hfbal}) is the balance law (\ref{volbal}) with $\varphi=1$, which does not require the TCC. Under the TCC $\theta^b\ge0$, so $\mathcal{V}$ is non-decreasing; its limit is $V_\infty$ by Theorem \ref{thmasym}(c), and the rigidity statement is Theorem \ref{thmvolrig}.
\end{proof}

Since $w$ is bounded on the compact set $\tau^{-1}([a_1,a_2])$, for $s\ge\max w$ we have $N(a_i,s)=\mathcal{V}(a_i)$, and the entry term of Theorem \ref{thmcanbal} below vanishes: in the horizon-free case the threshold on the delay plays no role. If the horizon is non-empty, $\mathcal{V}(a)=+\infty$ for large $a$ (Theorem \ref{thmasym}(c)), and the following result replaces Corollary \ref{corhf}.

\begin{theorem}[Balance law and rigidity for the canonical volumes] \label{thmcanbal}
Let $(M,g)$ be future timelike complete with regular cosmological time, let $0<a_1<a_2$, $Z:=\{a_1<\tau<a_2\}$, and for $s\ge0$ let $\Omega_s:=Z\cap\{w\le s\}$.
\begin{itemize}
\item[(a)] For every $s\ge0$,
\begin{equation} \label{canbal}
N(a_2,s)-N(a_1,s)=\int_{\Omega_s}\theta^b\,\mu_g+2\,\textrm{Vol}_b(S^1\cap\Omega_s)+\Phi(s),
\end{equation}
where $\Phi(s)\in[0,+\infty)$ is the right derivative at $s$ of the non-decreasing, locally absolutely continuous function $E(s):=\int_{\Omega_s}(\gamma-1)\,\mu_g$. For almost every $s$, $\Phi(s)$ is the flux of the calibrated rays through the level set $Z\cap\{w=s\}$.
\item[(b)] Assume the TCC, and let $N(a_1,s)=N(a_2,s)$ for some $s>0$. Then on the open set $U:=Z\cap\{w<s\}$ the function $b^{\textnormal{\textsc{m}}}$ is smooth, $\Box b^{\textnormal{\textsc{m}}}=0$, and $v=-\nabla b^{\textnormal{\textsc{m}}}$ is a parallel unit timelike vector field, so that $U$ is locally isometric to a static product $(\mathbb{R}\times N,-\dd t^2+h)$ with $v=\p_t$; in particular $S\cap U=\emptyset$. Moreover $\Phi(s)=0$.
\item[(c)] Assume the TCC, and let $N(a_1,s)=N(a_2,s)$ for every $s$ in an open interval $J\subset(0,+\infty)$. Then either $w$ takes no value in $J$ on $Z$, or $w$ is constant on $Z$, with value $c\in J$. In the latter case $Z\subset\textrm{dom}\, b^{\textnormal{\textsc{m}}}$, the cosmological time $\tau=b^{\textnormal{\textsc{m}}}-c$ is smooth on $Z$, $-\nabla\tau=v$ is parallel, and $Z$ is locally isometric to a static product whose static slices are the level sets of $\tau$.
\end{itemize}
\end{theorem}

Statement (a) does not require the TCC. The three terms on the right-hand side of (\ref{canbal}) are the contributions of the expansion of the calibrated congruence, of its creases, and of the calibrated observers whose delay falls below $s$ between the two cosmological times. The last term does not correspond to an expansion of a region comoving with the calibrated observers, and it can be the only non-vanishing one (Example \ref{exflat}). By (b), under the TCC the equality of the canonical volumes forces the first two terms to vanish, which gives the local static structure on the set of the events with delay less than $s$, and it forces the entry term to vanish; by (c), if the entry term vanishes for an interval of values of the bound, then no event of $Z$ has delay in that interval, unless the two families of observers coincide on $Z$.

\begin{proof}
(a). By Remark \ref{remcosmvol}, $f=\log\tau$ and $u=\tau$ satisfy the assumptions of Theorem \ref{corvol}(c), with $t_i=\log a_i$, $\Omega=Z\cap\textrm{dom}\, b^{\textnormal{\textsc{m}}}$, $b-u=w$, $A_{t_i}(s)=T_{a_i}\cap\{w\le s\}$ and $\dd u(v)=\gamma$. As observed after that theorem, the identity in (\ref{volentry}) does not use the TCC.

(b). Under the TCC $\theta^b\ge0$ almost everywhere, so the three terms on the right-hand side of (\ref{canbal}) are non-negative, and they vanish. Thus $\Phi(s)=0$, $\theta^b=0$ almost everywhere on $U\subset\Omega_s$, and $\int_{S^1\cap U}\vert\omega_+\vert=0$, which gives $\mathcal{H}^n(S^1\cap U)=0$, as the density of $\vert\omega_+\vert$ with respect to $\mathcal{H}^n$ on $S^1$ is positive almost everywhere (proof of Theorem \ref{thmvol}), hence $\mathcal{H}^n(S\cap U)=0$. The set $U$ is open, since $w:M\to[0,+\infty]$ is continuous, and it is contained in $\textrm{dom}\, b^{\textnormal{\textsc{m}}}$. By Theorem \ref{thmsbv}, $b^{\textnormal{\textsc{m}}}\in W^{2,1}_{loc}(U)$, and by Proposition \ref{propellreg}(b), whose proof does not depend on the results of this section, $b^{\textnormal{\textsc{m}}}$ is smooth on $U$, with $\Box b^{\textnormal{\textsc{m}}}=-\theta^b=0$, and $v$ is a smooth unit timelike geodesic field. The rest follows from the Raychaudhuri equation as in the proof of Theorem \ref{thmvolrig}.

(c). By (a) and the TCC, $\Phi=0$ on $J$. As $E$ is locally absolutely continuous and its derivative coincides almost everywhere with $\Phi$, $E$ is constant on $J$, namely $\gamma=1$ almost everywhere on the open set $W:=Z\cap w^{-1}(J)$. At the differentiability points of $\tau$ and $b^{\textnormal{\textsc{m}}}$, $\gamma=-g(u_\tau,v)$ with $u_\tau=-\nabla\tau$ a future directed unit timelike vector (Section \ref{seccanvol}), so, by the equality case of the reverse Cauchy--Schwarz inequality, $\gamma=1$ if and only if $v=u_\tau$, namely $\nabla w=0$. Thus the locally Lipschitz function $w$ has vanishing gradient almost everywhere on $W$, and hence it is locally constant on $W$. Suppose that $W\ne\emptyset$, let $C$ be a connected component of $W$, which is open, and let $c\in J$ be the value of $w$ on $C$. If $x\in Z$ belongs to the closure of $C$, then $w(x)=c$ by continuity, so $x\in W$, and the connected component of $W$ which contains $x$ meets $C$, hence $x\in C$. Thus $C$ is open and closed in $Z$, which is connected, being homeomorphic to $(a_1,a_2)\times T_{a_1}$, where $T_{a_1}$ is connected as $M$ is. Therefore $w=c$ on $Z$ and, by continuity, on $T_{a_1}\cup T_{a_2}$. In particular $\tau^{-1}([a_1,a_2])\subset\textrm{dom}\, b^{\textnormal{\textsc{m}}}$ and $N(a_i,s)=\textrm{Vol}_b(T_{a_i})$ for $s\ge c$. Choosing $s\in J$, $s\ge c$, we get $\textrm{Vol}_b(T_{a_1})=\textrm{Vol}_b(T_{a_2})$, and Theorem \ref{thmvolrig}, applied to $f=\log\tau$, gives that $b^{\textnormal{\textsc{m}}}$ is smooth on $Z$ with $v$ parallel. The remaining statements follow from $\tau=b^{\textnormal{\textsc{m}}}-c$.
\end{proof}

The truncations considered so far compare the delay with a fixed duration $s$. One can instead compare it with the cosmological age. Let
\[
\varrho:=\frac{w}{\tau}=\frac{b^{\textnormal{\textsc{m}}}}{\tau}-1:\ M\to[0,+\infty],
\]
the {\em relative delay}, which is invariant under the rescalings $g\mapsto\lambda^2g$, $\lambda>0$, of the metric, and which satisfies $\dd\varrho(v)=(\tau-\gamma\, b^{\textnormal{\textsc{m}}})/\tau^2\le0$ almost everywhere on $\textrm{dom}\, b^{\textnormal{\textsc{m}}}$, as $b^{\textnormal{\textsc{m}}}\ge\tau$ and $\gamma\ge1$.

\begin{corollary}[Volumes at bounded relative delay] \label{correl}
Let $(M,g)$ be future timelike complete with regular cosmological time, and assume the TCC.
\begin{itemize}
\item[(a)] For every $\kappa>0$ the function
\[
Q_\kappa(a):=\textrm{Vol}_b\big(T_a\cap\{\varrho\le\kappa\}\big)=N(a,\kappa a),\qquad a>0,
\]
is finite and non-decreasing, and $Q_\kappa(a)\to V_\infty$ for $a\to+\infty$.
\item[(b)] Let $\psi:[0,+\infty)\to[0,+\infty)$ be locally absolutely continuous and non-increasing, with $\psi(r)\to0$ for $r\to+\infty$. Then
\[
W_\psi(a):=\textrm{Vol}_b\big(T_a\cap\textrm{dom}\, b^{\textnormal{\textsc{m}}},\psi(\varrho)\big)=\int_0^{+\infty}(-\psi'(\kappa))\,Q_\kappa(a)\,\dd\kappa
\]
is non-decreasing in $a$, and $W_\psi(a)\to\psi(0)\,V_\infty$ for $a\to+\infty$, with the convention $0\cdot(+\infty)=0$. If $\psi$ vanishes on $[R,+\infty)$, then $W_\psi(a)\le\psi(0)\,N(a,Ra)<+\infty$ for every $a>0$.
\end{itemize}
\end{corollary}

\begin{proof}
(a). On $T_a$ the condition $\varrho\le\kappa$ reads $w\le\kappa a$, so $Q_\kappa(a)=N(a,\kappa a)$, which is finite, and non-decreasing in $a$ because $N$ is non-decreasing in both variables (Theorem \ref{thmasym}(a)). For every $s>0$ and $a\ge s/\kappa$ we have $N(a,s)\le Q_\kappa(a)\le V_\infty$, and the claim on the limit follows from Theorem \ref{thmasym}(a).

(b). As in the proof of Corollary \ref{corweight}(a), $\psi(\varrho)=\int_0^{+\infty}(-\psi'(\kappa))\,\chi_{\{\varrho\le\kappa\}}\,\dd\kappa$ on $\textrm{dom}\, b^{\textnormal{\textsc{m}}}$, and the integral expression of $W_\psi$ follows from Tonelli's theorem. By (a) the integrand is non-decreasing in $a$ and converges to $(-\psi'(\kappa))V_\infty$ for every $\kappa>0$, so the monotonicity and the limit follow from the monotone convergence theorem and $\int_0^{+\infty}(-\psi')\,\dd\kappa=\psi(0)$. The last statement follows from $\psi(\varrho)\le\psi(0)\,\chi_{\{\varrho\le R\}}$.
\end{proof}

A simple choice is $\psi(r)=\max\{1-r,0\}$. These volumes involve no unit of time: under the rescaling $g\mapsto\lambda^2g$ they change as $Q_\kappa(a)\mapsto\lambda^nQ_\kappa(a/\lambda)$, with the same $\kappa$. They are finite in every spacetime which satisfies the assumptions, also in the presence of the future cosmological horizon, while for the weights $F(w)$ with $F$ positive the finiteness for a given $F$ is open (Remark \ref{remspec}). They still depend on a dimensionless convention, the constant $\kappa$ or the profile $\psi$, so they are not uniquely selected; what the geometry selects is the family $\nu_a$.

We end this section with two examples, in which the calibrated volume changes only by expansion and only by entry, respectively.

\begin{example}[Warped products: expansion only] \label{exwarp2}
Let $(S,h)$ be a compact Riemannian manifold of dimension $n$, let $A:(0,+\infty)\to(0,+\infty)$ be smooth, and let $M=(0,+\infty)\times S$, $g=-\dd t^2+A(t)^2h$, be future timelike complete, namely $\int^{+\infty}\min\{A,1\}\,\dd t=+\infty$ \cite[Ex.\ 5.18]{minguzzi26e}; here the scale factor is denoted $A$, as $a$ denotes the values of the cosmological time. As shown there, $d(p,q)\le t(q)-t(p)$, with equality on the integral curves of $\p_t$, and $t\to0$ along the past inextendible causal curves. Hence $\tau=t$, which is regular, $T_a=\{a\}\times S$, $b^{\textnormal{\textsc{m}}}=t$, and
\[
w=0,\qquad \gamma=1,\qquad S=\emptyset,\qquad \theta^b=n\frac{\dot A}{A},\qquad \nu_a=A(a)^n\,\textrm{vol}_h(S)\,\delta_0 ,
\]
so that $N(a,s)=A(a)^n\,\textrm{vol}_h(S)$ for every $s\ge0$, $\textrm{dom}\, b^{\textnormal{\textsc{m}}}=M$, and there is no horizon. In (\ref{canbal}) only the expansion term is present. By the expression of the Ricci tensor recalled in \cite[Ex.\ 5.18]{minguzzi26e}, $Ric(\p_t,\p_t)=-n\ddot A/A$, $Ric(\p_t,V)=0$ and $Ric(V,V)=Ric_h(V,V)+(A\ddot A+(n-1)\dot A^2)\,h(V,V)$ for $V$ tangent to $S$.

(i) Let $A(t)=t$ and $Ric_h\ge-(n-1)h$, for instance let $(S,h)$ be a round sphere, a flat torus or a compact hyperbolic manifold of curvature $-1$. Then $Ric(\p_t,\p_t)=0$ and $Ric(V,V)\ge0$, so the TCC holds, and $V_\infty=\lim_{a\to+\infty}a^n\,\textrm{vol}_h(S)=+\infty$. Thus $V_\infty=+\infty$ does not imply that the future cosmological horizon is non-empty. If $(S,h)$ is a compact hyperbolic manifold of curvature $-1$, then $Ric_h=-(n-1)h$ and $Ric=0$; in fact $g$ is flat, being locally the Milne universe. Thus $\mathcal{V}(a)=a^n\,\textrm{vol}_h(S)$ grows strictly in a flat spacetime, consistently with Corollary \ref{corhf}, as $v=\p_t$ is not parallel.

(ii) Let $n\ge2$, let $(S,h)$ be the round unit sphere, and let $A(t)=L(1-\eu^{-t/L})$, $L>0$. With $x:=\eu^{-t/L}\in(0,1)$ we have $\dot A=x$, $\ddot A=-x/L$,
\[
Ric(\p_t,\p_t)=\frac{n\,x}{L^2(1-x)}>0,\qquad Ric(V,V)=\big[(n-1)-x+n\,x^2\big]\,h(V,V)>0 ,
\]
so the TCC holds, and the spacetime is future timelike complete, as $A\to L$. Here $V_\infty=L^n\,\textrm{vol}(S^n)<+\infty$, although $\theta^b>0$ at every time. Thus a finite asymptotic calibrated volume does not imply that the spacetime is eventually static.

(iii) Let $n=3$, let $(S,h)$ be a flat torus, and let $A(t)=C\,t^{1/2}$, $C>0$, the scale factor of the spatially flat Friedmann--Lema\^itre--Robertson--Walker model filled with radiation. Then $Ric(\p_t,\p_t)=3/(4t^2)$ and $Ric(V,V)=(A\ddot A+2\dot A^2)\,h(V,V)=\frac{C^2}{4t}\,h(V,V)$, so the TCC holds, the spacetime is future timelike complete, and $\mathcal{V}(a)=C^3a^{3/2}\,\textrm{vol}_h(S)$, with $\theta^b=3/(2t)$.
\end{example}

\begin{example}[A flat spacetime: entry only] \label{exflat}
Let $(T^n,h)$ be a flat torus, with distance $d_h$ and diameter $D$, let $\phi:T^n\to[0,+\infty)$ be a smooth non-constant function with $\min\phi=0$ and $\vert D\phi\vert_h<1$, and let
\[
M:=\{(t,x)\in\mathbb{R}\times T^n:\ t>\phi(x)\},\qquad g=-\dd t^2+h .
\]
Thus $M$ is the chronological future of the spacelike graph of $\phi$ in the static spacetime $\mathbb{R}\times T^n$. It is globally hyperbolic, with compact Cauchy hypersurfaces $\{t=c\}$, $c>\max\phi$, it is future timelike complete and flat, so the TCC holds, and, as $M$ is a future set, its Lorentzian distance is the restriction of that of $\mathbb{R}\times T^n$, namely $d((t',z),(t,x))=[(t-t')^2-d_h(z,x)^2]^{1/2}$ if $t-t'\ge d_h(z,x)$. Therefore
\[
\tau(t,x)=\max\big\{[(t-\phi(z))^2-d_h(z,x)^2]^{1/2}:\ z\in T^n,\ t-\phi(z)\ge d_h(z,x)\big\}\le t ,
\]
the distance from the graph of $\phi$, which is finite and tends to zero along the past inextendible causal curves, as these approach the graph. Hence the cosmological time is regular, although the spacetime is flat and extendible. Let $Z_\phi:=\phi^{-1}(0)$. The equality $\tau(t,x)=t$ holds if and only if $x\in Z_\phi$, and choosing $z\in Z_\phi$ we get $\tau(t,x)\ge(t^2-D^2)^{1/2}$ for $t\ge D$, so $t-\tau\to0$ uniformly for $t\to+\infty$. Along a future inextendible causal curve starting from $(t_0,x_0)$ the proper time elapsed up to the coordinate time $t$ is at most $t-t_0$, with equality on the vertical line, so the limit $L^{\textnormal{\textsc{m}}}$ of Section \ref{seccanvol} is at least $t_0$, with equality on the vertical line. Therefore
\[
b^{\textnormal{\textsc{m}}}=t,\qquad v=\p_t,\qquad \theta^b=0,\qquad S=\emptyset,\qquad w=t-\tau\ge0 ,
\]
$\textrm{dom}\, b^{\textnormal{\textsc{m}}}=M$, and $w>0$ outside $\mathbb{R}\times Z_\phi$. The slice $T_a$ is the graph of a function $\mathsf{t}_a:T^n\to[a,+\infty)$, and $\omega=i_{\p_t}\mu_g$ is the pull-back of the volume form of $h$, so the calibrated volume of a subset of $T_a$ is the $h$-volume of its projection on $T^n$, and
\[
N(a,s)=\textrm{vol}_h\big(\{x\in T^n:\ \mathsf{t}_a(x)\le a+s\}\big) .
\]
As $\mathsf{t}_a(x)=a$ if and only if $x\in Z_\phi$, and $a\le\mathsf{t}_a\le(a^2+D^2)^{1/2}\le a+D^2/(2a)$ for $a\ge D$,
\begin{align*}
N(a,0)&=\textrm{vol}_h(Z_\phi) && \textrm{for every } a>0,\\
N(a,s)&=\textrm{vol}_h(T^n) && \textrm{for } s>0,\ a\ge\max\{D,D^2/(2s)\},
\end{align*}
and $V_\infty=\textrm{vol}_h(T^n)$. For fixed $a$, $N(a,s)$ decreases to $\textrm{vol}_h(Z_\phi)<\textrm{vol}_h(T^n)$ for $s\to0$, so for small $s>0$ the function $N(\cdot,s)$ is not constant, although the calibrated congruence is parallel: in (\ref{canbal}) only the entry term is present. Thus the growth of the canonical volumes should not be read as the expansion of a region comoving with the calibrated observers. Moreover, if $Z_\phi$ is $\mathcal{H}^n$-null, for instance a point, then $N(a,0)=0$ for every $a$, while $N(a,s)\to V_\infty>0$ for every $s>0$, which shows that the condition $s>0$ in Theorem \ref{thmasym}(a) is necessary.
\end{example}

\section{An area theorem for the future cosmological horizon} \label{secareahor}

Let $N:=\p\, \textrm{dom}\, b$ be the future cosmological horizon. By \cite[Prop.\ 5.15]{minguzzi26e} it is a $C^0$ future null hypersurface and, being the boundary of a past set, it is achronal. Thus it is a past horizon in the sense of \cite[Def.\ 1]{minguzzi14d}, and the area inequality \cite[Thm.\ 13]{minguzzi14d} applies to it. By \cite[Thm.\ 3.9]{minguzzi26e}, $N$ does not depend on $\Sigma$, and the same is true for all the ingredients of this inequality, which are determined by $N$ and the metric. These are: the generators of $N$; its Alexandrov points, namely the points at which its local graphing functions are twice differentiable, at which the expansion $\theta$ of a field $k$ of semitangents of $N$ is defined; the measure $\mu_N$, defined on $N$ by $-g(k,V)\,\mu_N=i_V\mu_g$ for a timelike vector field $V$ \cite[Eq.\ (13)]{minguzzi14d}, which does not depend on $V$, since the restriction of $i_V\mu_g$ to $TN$ is linear in $V$ and vanishes for $V\in TN$, hence it is proportional to $g(k,V)$; and the area form $\sigma_N=i_k\mu_N$, which on the tangent spaces of $N$ is the volume form of the quotient metric on $TN/k$. A change of normalization $k\mapsto fk$, $f>0$, gives $\theta\mapsto f\theta$ and $\mu_N\mapsto f^{-1}\mu_N$, so that the measures $\theta\,\mu_N$ and $(\p_k\varphi)\,\mu_N$, the form $\sigma_N$ and the sign of $\theta$ do not depend on it.

What is missing for an area law is the sign of $\theta$. If $(M,g)$ is future null complete, the generators of $N$ are future complete and, under NCC, \cite[Thm.\ 15]{minguzzi14d} gives $\theta\ge 0$. In general, however, they need not be complete, and we adapt instead Hawking's argument for event horizons \cite[Thm.\ 4.1]{chrusciel01}, \cite[Thm.\ 22]{minguzzi14d}, in which the role of null infinity is played by the lifts to $\tilde M$ of the $b$-calibrated rays. Let
\[
E:=\textrm{dom}\, b\times \mathbb{R}\subset \tilde M ,
\]
an open past set of $\tilde M$, since $(y,w)\le (x,z)$ implies $y\le x$, and let $u:\tilde M\to (-\infty,+\infty]$, $u(y,w):=b(y)-w$.

\begin{lemma} \label{lemstar}
Assume that $(M,g)$ is future timelike complete. Let $\tilde C\subset \tilde M$ be a compact set such that $\tilde C\cap E\ne \emptyset$, and let $(q,z_q)$ be a minimum point of $u$ on $\tilde C$. Then $(q,z_q)\in E$ and, for every proper time parametrized $b$-calibrated ray $\alpha$ starting from $q$, the curve $\ell(t):=(\alpha(t),z_q+t)$, $t\ge 0$, is a future complete lightlike geodesic of $\tilde M$, contained in $E$, such that
\[
\ell\subset \tilde J^+(\tilde C)\backslash \tilde I^+(\tilde C).
\]
\end{lemma}

\begin{proof}
The function $u$ is continuous as a map to $(-\infty,+\infty]$ \cite[Thm.\ 5.12(iv)]{minguzzi26e}, so it attains its minimum on $\tilde C$, which is finite as $\tilde C\cap E\ne \emptyset$. Thus $q\in \textrm{dom}\, b$ and the ray $\alpha$ exists, is timelike and is defined on $[0,+\infty)$ (\cite[Prop.\ 5.6]{minguzzi26e} and \cite[Thm.\ 5.12(i)]{minguzzi26e}). As $\tilde g((\dot\alpha,1),(\dot\alpha,1))=g(\dot\alpha,\dot\alpha)+1=0$, $\ell$ is an affinely parametrized future complete lightlike geodesic, and $\ell\subset E$ as $\alpha\subset \textrm{dom}\, b$. Clearly $\ell\subset \tilde J^+(\tilde C)$. Suppose that $(y,w)\in \tilde C$ and $(y,w)\ll \ell(t)$. The projection of a timelike curve from $(y,w)$ to $\ell(t)$ is a timelike curve from $y$ to $\alpha(t)$, along which the $z$ coordinate increases less than its Lorentzian length \cite[Sect.\ 4]{minguzzi26e}, so that $z_q+t-w<d(y,\alpha(t))$. By the rushing property and calibration, $b(y)+d(y,\alpha(t))\le b(\alpha(t))=b(q)+t$. Therefore $b(y)-w<b(q)-z_q$, which contradicts the minimality of $u(q,z_q)$. Thus $\ell\cap \tilde I^+(\tilde C)=\emptyset$.
\end{proof}

The lemma is the analog of the condition $\star$ used in \cite[Thm.\ 22]{minguzzi14d}, with the difference that here it is a consequence of future timelike completeness and it holds for every compact set.

\begin{theorem} \label{thmhorexp}
Assume that $(M,g)$ is future timelike complete and that either
\begin{itemize}
\item[(i)] $(M,g)$ satisfies TCC, or
\item[(ii)] $(M,g)$ is future null complete and satisfies NCC.
\end{itemize}
Then $\theta\ge 0$ at every Alexandrov point of the future cosmological horizon $N=\p\, \textrm{dom}\, b$, hence $\mu_N$-almost everywhere.
\end{theorem}

\begin{proof}
Under assumption (ii) the generators of $N$, which are future inextendible lightlike geodesics, are future complete, and $N$ is achronal, so the claim is \cite[Thm.\ 15]{minguzzi14d}, whose proof shows that $\theta\ge 0$ at every Alexandrov point. Let us assume (i).

The set $\tilde N:=\p E=N\times \mathbb{R}$ is the boundary of an open past set of $\tilde M$, and through each of its points passes the horizontal lift of a generator of $N$, which is a lightlike geodesic of $\tilde M$; thus it is a past horizon of $\tilde M$. Its local graphing functions can be chosen independent of $z$, so its Alexandrov points are the points $(p,z)$ with $p$ an Alexandrov point of $N$, and, for the semitangent field $(k,0)$, its expansion at $(p,z)$ coincides with the expansion of $N$ at $p$, since $\p_z$ is a parallel vector field tangent to $\tilde N$. Moreover, $(\tilde M,\tilde g)$ satisfies the null convergence condition: a future directed lightlike vector of $\tilde M$ reads $(v,\zeta)$ with $v$ causal, and $\widetilde{Ric}((v,\zeta),(v,\zeta))=Ric(v,v)\ge 0$ by TCC and continuity. Observe also that $\tilde I^-(\tilde N)\subset E$, as $E$ is an open past set, and that $\tilde I^+(\tilde N)\cap \overline{E}=\emptyset$, since $x\in \overline E$ and $x\gg y\in \tilde N$ would give $y\in \tilde I^-(\overline E)=E$, which is impossible as $E$ is open and $y\in \p E$.

Suppose, by contradiction, that $\theta(p)<0$ at an Alexandrov point $p$ of $N$, and let $\tilde p:=(p,0)$. As in the proof of \cite[Thm.\ 22]{minguzzi14d}, see also \cite[Thm.\ 4.1]{chrusciel01}, the second order expansion of $\tilde N$ at $\tilde p$ provides, for every large integer $m$, a smooth spacelike codimension two surface $S_m$ with compact closure contained in the ball of radius $1/m$ centered at $\tilde p$ (for an auxiliary Riemannian metric), such that: (1) the future null normal field of $S_m$ which approximates the semitangents $(k,0)$ of $\tilde N$ has negative expansion on $S_m$; (2) $S_m$ contains a point of $\tilde I^-(\tilde p)\subset E$; (3) $\p S_m\subset \tilde I^+(\tilde N)$.

Let $\tilde q_m=(q_m,z_m)$ and $\ell_m$ be given by Lemma \ref{lemstar} applied to $\tilde C=\overline{S_m}$. By (3), $b=+\infty$ on the projection of $\p S_m$, so $u=+\infty$ there and $\tilde q_m\in S_m$. A lightlike geodesic which starts from a point of the smooth surface $S_m$ and is contained in $\tilde J^+(\overline{S_m})\backslash \tilde I^+(\overline{S_m})$ is orthogonal to $S_m$ and contains no focal point of $S_m$ \cite[Sect.\ 4.5]{hawking73}. Thus $\ell_m$ is a future null normal geodesic of $S_m$ without focal points. It is the normal of (1) for large $m$. Indeed $\tilde q_m\to \tilde p$, and the future null normals of $S_m$ at $\tilde q_m$ converge, up to positive factors, to $(k,0)$ and to a future lightlike vector $l$ at $\tilde p$ transverse to the tangent hyperplane of $\tilde N$ at $\tilde p$. The geodesic with initial vector $l$ enters the open set $\tilde M\backslash \overline E$, on the future side of $\tilde N$, immediately, while $\ell_m\subset E$, so that a limit of the initial segments of the curves $\ell_m$ is contained in $\overline E$ and cannot be tangent to $l$. Since $\ell_m$ is complete, the Raychaudhuri equation for the congruence of null normal geodesics of $S_m$ and the null convergence condition of $\tilde M$ imply that $\ell_m$ contains a focal point of $S_m$, a contradiction.
\end{proof}

Under assumption (i), TCC, rather than NCC, is needed, because the projections of the geodesics $\ell_m$ along which focusing takes place are timelike. Under assumption (ii), focusing takes place along the generators of $N$ themselves.

\begin{theorem}[Area theorem for the future cosmological horizon] \label{thmhorarea}
Assume that $(M,g)$ is future timelike complete and that either it satisfies TCC, or it is future null complete and satisfies NCC. Let $N=\p\, \textrm{dom}\, b$. For every open relatively compact subset $D\subset N$ of finite perimeter and every positive Lipschitz function $\varphi$ on $D$,
\[
0\le \int_D \varphi\,\theta\,\mu_N\le |A(\p^+D,\varphi)|-|A(\p^-D,\varphi)|-\int_D (\p_k\varphi)\, \mu_N,
\]
where $\p^\pm D$ are the future and past essential boundaries of $D$ and $A$ is the area functional of \cite[Eq.\ (26)]{minguzzi14d}. In particular, let $f$ be a locally Lipschitz Cauchy time function and let $A(t)$ be the area of the section $N\cap f^{-1}(t)$, which is well defined for almost every $t$. Then for almost every $t_1<t_2$
\[
A(t_1)\le A(t_2),
\]
and if equality holds then $N\cap f^{-1}((t_1,t_2))$ is as regular as the metric. In particular, the inequality is strict if the sections $N\cap f^{-1}(t_1)$ and $N\cap f^{-1}(t_2)$ are not homeomorphic.
\end{theorem}

\begin{proof}
The second inequality is \cite[Thm.\ 13]{minguzzi14d} applied to the past horizon $N$, and the first follows from Theorem \ref{thmhorexp}. The statement on the sections follows as in the proof of \cite[Thm.\ 23]{minguzzi14d}, applied to $D=N\cap f^{-1}((t_1,t_2))$, which is relatively compact since $f$ is a Cauchy time function and the Cauchy hypersurfaces of $M$ are compact, with \cite[Thm.\ 22]{minguzzi14d} replaced by Theorem \ref{thmhorexp}. In the equality case the singular part of the Hessian of the local graphing functions and $\theta$ vanish on $D$, and \cite[Thm.\ 17]{minguzzi14d} gives the regularity.
\end{proof}

Theorems \ref{thmhorexp} and \ref{thmhorarea} concern only the set $N$, which does not depend on $\Sigma$, and the inequalities are invariant under changes of normalization of $k$: the Cauchy hypersurface $\Sigma$ enters only the proof, through $b$. Thus the future cosmological horizon behaves as the event horizon of a black hole also in this respect: under TCC, or under NCC and future null completeness, its area cannot decrease. Past timelike completeness is not needed, so these results apply to spacetimes with an initial singularity, for instance to the horizon $\p\, \textrm{dom}\, b^{\textnormal{\textsc{m}}}$ of Section \ref{secinit}, whenever it is non-empty.

In spacetimes with regular cosmological time the slicing can be chosen canonically.

\begin{corollary}[Canonical area of the future cosmological horizon] \label{corcanarea}
Let $(M,g)$ be future timelike complete with regular cosmological time $\tau$, and assume that either it satisfies TCC, or it is future null complete and satisfies NCC. For almost every $a>0$ the area $\mathcal{A}_N(a)$ of the section $N\cap T_a$ of the future cosmological horizon $N=\p\,\textrm{dom}\, b^{\textnormal{\textsc{m}}}$ by the slice $T_a=\tau^{-1}(a)$ is well defined, and
\[
\mathcal{A}_N(a_1)\le\mathcal{A}_N(a_2) \qquad \textrm{for almost every } a_1<a_2 .
\]
The function $\mathcal{A}_N$ depends only on $(M,g)$.
\end{corollary}

\begin{proof}
The function $f=\log\tau$ is a locally Lipschitz Cauchy time function with $f^{-1}(\log a)=T_a$ (Remark \ref{remcosmvol}), and $N=\p\,\textrm{dom}\, b^{\textnormal{\textsc{m}}}$ is the future cosmological horizon (Section \ref{secinit}), so the claim is Theorem \ref{thmhorarea}.
\end{proof}

Thus, in a spacetime with regular cosmological time and non-empty future cosmological horizon, where the asymptotic calibrated volume is infinite, the geometry still selects, with no auxiliary choice, the deficit spectra $\nu_a$, whose cumulative distributions $N(a,s)$ are finite and, under the TCC, non-decreasing in $a$ (Theorem \ref{thmasym}), their slope $\sigma(a)$ at the horizon (Theorem \ref{thmslope}(b)), and the area $\mathcal{A}_N(a)$ of the horizon. For smooth slices the slope is bounded from below by the area of the section of the horizon (Theorem \ref{thmslope}(a)); we expect that $\sigma(a)\ge\mathcal{A}_N(a)$ holds as well. The slices $T_a$ are only Lipschitz, and the proof requires control of the induced metrics, of the coarea formula and of the perimeters on them, as bi-Lipschitz equivalence alone does not preserve the sharp constant.

\section{Remarks on a geometric second law} \label{secphys}

In this section we propose a physical interpretation of the balance law as a law of entropy production, a geometric second law for cosmological spacetimes. The mathematical statements are consequences of the previous results.

\paragraph{The current and its production.} Let $(M,g)$ be future timelike complete with regular cosmological time, let $b=b^{\textnormal{\textsc{m}}}$, $v=-\nabla b^{\textnormal{\textsc{m}}}$, and assume first that $\textrm{dom}\, b^{\textnormal{\textsc{m}}}=M$. For a constant $\eta>0$, with the dimensions of an entropy per unit volume, let
\[
J:=\eta\, v,\qquad S_{\rm geom}(a):=\eta\,\mathcal{V}(a)=\eta\,\textrm{Vol}_b(T_a) .
\]
For every Borel subset $E$ of a slice, $\eta\,\textrm{Vol}_b(E)$ is the flux of $J$ through $E$, since the calibrated volume is the integral of $\omega=i_v\mu_g$. By Corollary \ref{cordiv}, $\textrm{div}\, J=\eta\,\theta^b\mu_g+2\eta\,\vert\omega_+\vert\llcorner S^1$, a measure which is non-negative under the TCC, and by Corollary \ref{corhf} the increase of $S_{\rm geom}$ between two cosmological times is the total production in the intervening slab, which vanishes only if the slab is locally static, with the calibrated observers as static observers. This is the structure of a second law, with a production measure whose absolutely continuous part is due to the expansion of the canonical congruence and whose singular part is due to its creases, and with a rigidity statement for the absence of production. The production need not be positive at every time, and a plateau on an interval gives a local, not a global, static structure. In this picture the arrow of time is tied to the definition of $b^{\textnormal{\textsc{m}}}$ at future infinity: the entropy is measured by the observers who age maximally into the far future.

Monotonicity alone cannot select a formula, as every increasing function of $\mathcal{V}$ is monotone. Additivity does: a function $\Psi$ of the calibrated volume of the Borel subsets of the slices which is additive on disjoint sets and monotone satisfies $\Psi(x+y)=\Psi(x)+\Psi(y)$ on the values of the calibrated volume, which form an interval as the measure has no atoms, so it is linear, $\Psi(x)=\eta x$. If the horizon is non-empty, the same considerations apply to $\eta N(a,s)$, but by Theorem \ref{thmcanbal} its increase contains, besides the production, the term $\eta\,\Phi(s)$, which is the influx of $J$ through the moving boundary $\{w=s\}$ and not a production. A growing entropy of a subsystem need not signal production inside it.

\paragraph{Normalization and temperature.} The calibrated volume of a slice is the area of a section of the horizon $H$ of the product trick, which suggests a comparison with the Bekenstein--Hawking entropy $k_BA/(4\hbar G)$. In dimension $D=n+1$, $\hbar G$ has the dimensions of a length to the power $n-1$, while a section of $H$ is $n$-dimensional, as $H$ lies in $(M\times\mathbb{R},g+\dd z^2)$, of dimension $n+2$. The natural value is therefore $\eta=k_B/(4\hbar\tilde G)$, with $\tilde G$ the gravitational constant in dimension $n+2$. If the fields do not depend on $z$ and the auxiliary direction has length $\ell$, the dimensional reduction of the action gives $\tilde G=G\ell$, so that $\eta=k_B/(4\hbar G\ell)$, and the length $\ell$ is the new constant of the theory. Beyond Einstein gravity the entropy of a horizon depends on the action \cite{wald93}. The identification of the area with an entropy rests also on a temperature and a first law, which we are exploring in related ongoing work. A natural candidate for the temperature is suggested by the local structure at the creases. In the flat model of a crease, $b=\min\{f_+,f_-\}$ on Minkowski space, with $f_\pm$ linear and $\nabla f_\pm$ unit timelike, the horizon $H$ is the future part of a bifurcate Killing horizon of the Minkowski space $M\times\mathbb{R}$, whose bifurcation surface projects onto the crease, and on each generator the boost Killing field is $\sigma\p_\sigma$, where $\sigma$ is the proper time elapsed along the calibrated ray since its birth at the crease. Thus the horizon has unit surface gravity with respect to the boost parameter $\ln\sigma$, which suggests the local temperature $k_BT=\hbar/(2\pi\sigma)$, inversely proportional to the age of the calibrated ray; for the rest curves of the cosmological time the age is $\tau$, and this is the value $\hbar/(2\pi\tau)$ of the Milne universe.

\paragraph{Two tests.} The spatially flat radiation model of Example \ref{exwarp2}(iii) satisfies all the assumptions, with $\mathcal{V}(a)=C^3a^{3/2}\,\textrm{vol}_h(S)$. The radiation evolves adiabatically: its temperature scales as $A^{-1}$, its entropy density as $A^{-3}$, and its total entropy is constant. Thus $S_{\rm geom}$ is an additional geometric contribution, distinct from the entropy of matter. Locally, at the smooth points the entropy current $s\,v$ of a fluid comoving with $v$ has divergence $v(s)+s\,\theta^b$, which vanishes for adiabatic evolution, whereas $J$ has divergence $\eta\,\theta^b$. The compact Milne universe of Example \ref{exwarp2}(i) is flat, and $\mathcal{V}(a)=a^n\,\textrm{vol}_h(S)$ grows strictly. Thus $S_{\rm geom}$ is complementary to the measures of the clumping of the gravitational field, such as the gravitational entropy of Penrose's Weyl curvature hypothesis \cite{penrose79} or of the proposal of Clifton, Ellis and Tavakol \cite{clifton13}, which is built from the Bel--Robinson tensor and vanishes if and only if the Weyl tensor vanishes, hence on the Friedmann--Lema\^itre--Robertson--Walker models and on the compact Milne universe; it measures the growth of the global geometry along the canonical congruence. In the present form the second law requires the TCC: a positive cosmological constant violates it, as $Ric=\lambda g$, $\lambda>0$, gives $Ric(v,v)=-\lambda<0$, and recollapsing models are not future timelike complete; for weights compensating a negative $Ric(v,v)$ the weighted results of Section \ref{secmono} apply.

\paragraph{Volumes of Cauchy hypersurfaces in the literature.} The volume of Cauchy hypersurfaces enters several monotonicity statements and physical proposals, which we briefly compare with the present one.
\begin{itemize}
\item[(1)] For the vacuum Einstein flow in $3+1$ dimensions, foliated by compact hypersurfaces of constant mean curvature $H<0$ whose topology is of Yamabe type $-1$, namely which admit no metric of zero scalar curvature, Fischer and Moncrief introduced the dimensionless reduced Hamiltonian $\vert H\vert^3$ times the volume of the slice. They showed that it is strictly decreasing along the Einstein flow, in the expanding direction, unless the slice is hyperbolic with vanishing transverse traceless momentum, in which case it is constant, and they related its infimum to the $\sigma$-constant of the manifold and to the geometrization of $3$-manifolds \cite{fischer99}. A different monotonicity was obtained by Anderson \cite[Cor.\ 1.3]{anderson01}: for a vacuum spacetime in $3+1$ dimensions foliated by compact Cauchy hypersurfaces of constant negative mean curvature, the volume of the slices divided by the cube of their maximal proper time distance from an initial slice is non-increasing, and it is constant on an interval if and only if the corresponding region is a flat Lorentzian cone. The proof combines the Raychaudhuri equation with a geometric maximum principle, and the inequality persists under the TCC \cite[Prop.\ 1.2 and the remark following it]{anderson01}; in dimension $n+1$, with the $n$-th power, the ratio is then constant on an interval exactly on regions isometric to Lorentzian cones $-\dd t^2+t^2g_-$ with $Ric_{g_-}\le0$ \cite[Sect.\ 2]{anderson04}. These quantities are normalized by the scale and decrease, while $\mathcal{V}$ is not normalized and increases; in the compact Milne universe, which is the critical point of the reduced Hamiltonian, they are constant and $\mathcal{V}$ grows as $a^n$. They control the shape of the geometry, $\mathcal{V}$ its growth along the canonical observers.
\item[(2)] In the approach of Barbour, Koslowski and Mercati the arrow of time is identified with the growth of a scale invariant complexity of the shape, while the total volume of the constant mean curvature slices is the scale degree of freedom which is factored out \cite{barbour14}.
\item[(3)] In holography, Stanford and Susskind proposed that the complexity of the state dual to a black hole is proportional to the volume of the maximal slice through the Einstein--Rosen bridge, divided by the product of the gravitational constant and a length \cite{stanford14}, and Brown and Susskind formulated a second law of complexity \cite{brown18}. The volume of maximal hypersurfaces inside black holes grows linearly with the advanced time \cite{christodoulou15}. In these proposals the growing volume is interpreted as complexity rather than entropy, and it is the volume of an extremal hypersurface anchored at infinity, while here it is the calibrated volume of the slices of the cosmological time, defined through the observers who maximize the proper time to future infinity, and its growth law is a projected area theorem.
\item[(4)] Padmanabhan proposed that the expansion of the Universe is driven by the difference between the numbers of degrees of freedom on the boundary of the Hubble volume and in its bulk, through an equation for the rate of change of the Hubble volume \cite{padmanabhan12}. There the relevant volume is that of the Hubble sphere, rather than that of a Cauchy hypersurface.
\end{itemize}
In conclusion, the theorems of this paper provide a geometrically selected congruence of observers and a volume with non-negative production, whose absence forces local staticity. We regard them as the second law of a geometric entropy of cosmological spacetimes, whose first law is the subject of ongoing work.

\section{Conclusions} \label{secconcl}

The product trick of \cite{minguzzi26e} represents the graph of the Cauchy Busemann function $b$ as a Cauchy horizon, and in this paper we have shown that the regularity theory and the area theorem for horizons translate, through a simple dictionary, into statements on the volume measured by the observers who age maximally into the far future. Thanks to the SBV regularity of $\nabla b$, the resulting volume theorem is an exact balance law: the calibrated volume changes only through the expansion $-\Box b$ of the calibrated congruence and through its creases, which always act as sources. Under the TCC the calibrated volume is non-decreasing, and its constancy forces the intervening region to be locally a static product, with the calibrated observers as static observers. Weights which decrease along the calibrated rays are controlled by conditions of Bakry--\'Emery type, which for synthetic dimension $N=0$ and smooth weights express the area theorem for the horizon in a conformally rescaled metric, and the past semi-flow of \cite{minguzzi26f} localizes the monotonicity to arbitrary Borel portions of the slices transported by the calibrated observers. In a counterexample to Bartnik's conjecture no Cauchy hypersurface is contained in the future cosmological development, so the slices of every smooth Cauchy temporal function have infinite calibrated volume, and the monotonicity is expressed by the truncated and weighted volumes.

In spacetimes with an initial singularity no choice is needed. The canonical function $b^{\textnormal{\textsc{m}}}$ selects a congruence of observers, and the least asymptotic delay $w=b^{\textnormal{\textsc{m}}}-\tau$ is a Lyapunov function of its semi-flow. The deficit spectra $\nu_a$, whose cumulative distributions are the calibrated volumes $N(a,s)$ of the events of cosmological time $a$ with delay at most $s$, depend only on $(M,g)$. They satisfy an exact balance law, in which an entry term accounts for the observers whose delay falls below the bound, and under the TCC their cumulative distributions are non-decreasing in $a$ and converge to the asymptotic calibrated volume $V_\infty$, which does not depend on the bound. A finite $V_\infty$ excludes the future cosmological horizon. When the horizon is present, the truncated volumes diverge at least linearly in the bound, with a slope which is non-decreasing in time and, on smooth slices, bounded from below by the area of the section of the horizon, and this area satisfies its own area theorem under the TCC, with no assumption of past completeness.

The calibrated volume of a slice of the cosmological time differs from its induced volume by the Lorentz factor between two natural frames: the observers comoving with the cosmological time, who have aged maximally since the initial singularity, and the calibrated observers, who will age maximally into the far future. In the Friedmann--Lema\^itre--Robertson--Walker models the two frames coincide, and the calibrated volume is the familiar comoving volume. In general they differ, and the same relative boost which enlarges the calibrated volume makes the delay decrease. We find it remarkable that a second law emerges from purely geometric considerations, with a non-negative production measure and a rigidity statement for its absence, and that the observers who are naturally singled out to measure it are those who age maximally into the far future. As the calibrated volume grows also in flat spacetimes and for adiabatic radiation, the corresponding geometric entropy is distinct from the entropy of matter and from measures of gravitational clumping. It is tempting to regard it as an entropy of the cosmological spacetime itself, whose growth would provide a geometric arrow of time rooted in the asymptotic future.

Several questions remain open. The areas of the sections of the horospheres by the slices are not controlled from above: an upper bound would give the finiteness of the weighted volumes for every weight integrable at infinity, such as $\eu^{-w}$, and their Ces\`aro convergence (\ref{areaconv}) would identify the slope of the truncated volumes with the area of the section of the horizon, so that the area theorem for the horizon would be the leading term of the volume monotonicity theorem. We do not know whether the finite part of the truncated volumes is monotone (Remark \ref{remfinite}), nor whether the lower bound $\sigma(a)\ge\mathcal{A}_N(a)$ holds on the Lipschitz slices of the cosmological time. The second law obtained here requires the TCC, so it would be important to identify natural weights which compensate a positive cosmological constant, for which the weighted results of Section \ref{secmono} are available. On the physical side, the normalization of the geometric entropy, a temperature and a first law are the subject of ongoing work. Finally, it would be interesting to compute the deficit spectra in explicit inhomogeneous cosmological models, in which the two frames differ, and to investigate the observational consequences of their relative motion.

\section*{Acknowledgments}

For the writing of this paper we had the assistance of AI tools from Anthropic (Claude Opus 5.5) and OpenAI (GPT-6 Astra). This work was supported in part by a grant of access to OpenAI models through the ChatGPT for Academic Researchers program.


\end{document}